\documentstyle[12pt,epsf]{article}
\input psfig
\newcommand{\beq}{\begin{equation}}
\newcommand{\eeq}{\end{equation}}
\newcommand{\beqn}{\begin{eqnarray}}
\newcommand{\eeqn}{\end{eqnarray}}

\newcommand{\half}{{\textstyle{1\over2}}}
\newcommand{\dslash}{D\!\!\!\!/\,}
\newcommand{\Amu}{A_{\mu}}
\newcommand{\Anu}{A_{\nu}}
\newcommand{\Dmu}{D_{\mu}}
\newcommand{\Dnu}{D_{\nu}}
\newcommand{\Fmunu}{F_{\mu\nu}}
\newcommand{\Unmu}{U_{n, \mu}}

\newcommand{\sslash}{s\!\!\!/\,}
\newcommand{\psibar}{\bar{\psi}}

\def\MSbar{{\overline{\rm MS}}}
\def\CV{{\cal V}}
\def\CA{{\cal A}}
\def\CO{{\cal O}}
\def\COhat{{\widehat{\cal O}}}
\def\COhatdag{{\widehat{\cal O}}^{\dag}}

\def\qbar{{\overline q}}
\def\ubar{{\overline u}}
\def\dbar{{\overline d}}
\def\sbar{{\overline s}}
\def\Dbar{{\overline D}}
\def\Sbar{{\overline S}}
\def\Qbar{{\overline Q}}
\def\qtw{{\tilde q}}
\def\qtwbar{\overline{\tilde q}}
\def\BKhat{\widehat{B}_K}
\def\NDR{{\rm NDR}}
\def\CONT{{\em cont}}
\def\LATT{{\em lat}}
\def\lat{{\em lat}}
\def\phys{{\em phys}}
\def\eff{{\em eff}}
\def\frac#1#2{{\textstyle{#1\over#2}}}

\def\Dsl{\,\raise.15ex\hbox{/}\mkern-13.5mu D}
\def\bar#1{\overline{#1}}
\def\chibar{\bar\chi}
\def\vev#1{\langle #1 \rangle}
\def\vdag{{\vphantom{\dag}}}
\def\g#1{\gamma_{#1}}

\catcode`\@=11

\begin{document}

\newcommand{\symbolfootnote}{\renewcommand{\thefootnote}
	{\fnsymbol{footnote}}}
\renewcommand{\thefootnote}{\fnsymbol{footnote}}
\newcommand{\alphfootnote}
	{\setcounter{footnote}{0}
	 \renewcommand{\thefootnote}{\sevenrm\alph{footnote}}}

\def\abstracts#1{{
	\centering{\begin{minipage}{30pc}\tenrm\baselineskip=12pt\noindent
	\centerline{\tenrm ABSTRACT}\vspace{0.3cm}
	\parindent=0pt #1
	\end{minipage} }\par}} 

\newcommand{\bibit}{\it}
\newcommand{\bibbf}{\bf}
\renewenvironment{thebibliography}[1]
	{\begin{list}{\arabic{enumi}.}
	{\usecounter{enumi}\setlength{\parsep}{0pt}
\setlength{\leftmargin 1.25cm}{\rightmargin 0pt}
	 \setlength{\itemsep}{0pt} \settowidth
	{\labelwidth}{#1.}\sloppy}}{\end{list}}

\topsep=0in\parsep=0in\itemsep=0in
\parindent=1.5pc

\newcounter{itemlistc}
\newcounter{romanlistc}
\newcounter{alphlistc}
\newcounter{arabiclistc}
\newenvironment{itemlist}
    	{\setcounter{itemlistc}{0}
	 \begin{list}{$\bullet$}
	{\usecounter{itemlistc}
	 \setlength{\parsep}{0pt}
	 \setlength{\itemsep}{0pt}}}{\end{list}}

\newenvironment{romanlist}
	{\setcounter{romanlistc}{0}
	 \begin{list}{$($\roman{romanlistc}$)$}
	{\usecounter{romanlistc}
	 \setlength{\parsep}{0pt}
	 \setlength{\itemsep}{0pt}}}{\end{list}}

\newenvironment{alphlist}
	{\setcounter{alphlistc}{0}
	 \begin{list}{$($\alph{alphlistc}$)$}
	{\usecounter{alphlistc}
	 \setlength{\parsep}{0pt}
	 \setlength{\itemsep}{0pt}}}{\end{list}}

\newenvironment{arabiclist}
	{\setcounter{arabiclistc}{0}
	 \begin{list}{\arabic{arabiclistc}}
	{\usecounter{arabiclistc}
	 \setlength{\parsep}{0pt}
	 \setlength{\itemsep}{0pt}}}{\end{list}}

\newcommand{\fcaption}[1]{
        \refstepcounter{figure}
        \setbox\@tempboxa = \hbox{\tenrm Fig.~\thefigure. #1}
        \ifdim \wd\@tempboxa > 6in
           {\begin{center}
        \parbox{6in}{\tenrm\baselineskip=12pt Fig.~\thefigure. #1 }
            \end{center}}
        \else
             {\begin{center}
             {\tenrm Fig.~\thefigure. #1}
              \end{center}}
        \fi}

\newcommand{\tcaption}[1]{
        \refstepcounter{table}
        \setbox\@tempboxa = \hbox{\tenrm Table~\thetable. #1}
        \ifdim \wd\@tempboxa > 6in
           {\begin{center}
        \parbox{6in}{\tenrm\baselineskip=12pt Table~\thetable. #1 }
            \end{center}}
        \else
             {\begin{center}
             {\tenrm Table~\thetable. #1}
              \end{center}}
        \fi}

%

\def\fnm#1{$^{\mbox{\scriptsize #1}}$}
\def\fnt#1#2{\footnotetext{\kern-.3em
	{$^{\mbox{\sevenrm #1}}$}{#2}}}

\font\twelvebf=cmbx10 scaled\magstep 1
\font\twelverm=cmr10 scaled\magstep 1
\font\twelveit=cmti10 scaled\magstep 1
\font\elevenbfit=cmbxti10 scaled\magstephalf
\font\elevenbf=cmbx10 scaled\magstephalf
\font\elevenrm=cmr10 scaled\magstephalf
\font\elevenit=cmti10 scaled\magstephalf
\font\bfit=cmbxti10
\font\tenbf=cmbx10
\font\tenrm=cmr10
\font\tenit=cmti10
\font\ninebf=cmbx9
\font\ninerm=cmr9
\font\nineit=cmti9
\font\eightbf=cmbx8
\font\eightrm=cmr8
\font\eightit=cmti8


\begin{titlepage}
 \null
 \makebox[\textwidth][r]{UW/PT 94-15} 
 \begin{center} 
 \vspace{0.5in}
  {\Large \bf
        Phenomenology from the lattice}
  \par  \vskip 5.0em
        {\large \em  Stephen R.~Sharpe} \\
        Physics Department, University of Washington    \\ 
        Seattle, WA 98195, USA
  \par \vskip .5truein
  
	{\large ABSTRACT}\\[0.5em]
 \end{center}
\begin{quotation}
This is the written version of four lectures given at
the 1994 TASI. My aim is to explain the essentials of lattice 
calculations, give an update on
(though not a review of) the present status of calculations of
phenomenologically interesting quantities,
and to provide an understanding of the various sources
of uncertainty in the results. 
I illustrate the important issues using the
examples of the kaon B-parameter ($B_K$) and various quantities
related to $B$-meson physics.
\end{quotation}
\begin{center}
  \par \vskip .5truein
        Lectures at TASI, 1994, to appear in\\[0.5em]
{\em ``CP Violation and the Limits of the Standard Model'',}
  \\[0.5em]
        Ed. J. Donoghue, to be published by World Scientific.
  \end{center}

\vfil

\makebox[\textwidth][l]{December 1994}

\end{titlepage}

\tableofcontents
\vfill


\twelverm   
\baselineskip=14pt
\section{Why do we need lattice QCD?}
The theme of this year's TASI is CP violation. 
CP violation offers a possible window on physics beyond the
Standard Model: Will the single phase in the CKM matrix be 
sufficient to explain all the CP violating amplitudes 
that we hope to measure once B-factories, $\phi-$factories
and the next generation of $\epsilon'/\epsilon$ experiments
are up and running? 
To answer this we need to relate the phases in the CKM matrix,
which appear in the underlying quark amplitudes,
to the measurable phases in hadronic decay amplitudes.
For some B-meson decays, e.g. $B\to\psi K_s$,
we can avoid uncertainties due to hadronic structure 
by taking appropriate ratios.
For most quantities, however, the relation between CKM and measurable phases
is obscured by non-perturbative physics.
To find the relation, we need to know certain hadronic matrix elements
of fermion bilinears and quadrilinears.
One of the most practical uses of lattice QCD is the calculation of
such matrix elements.

\begin{figure}[b]
\centerline{\psfig{file=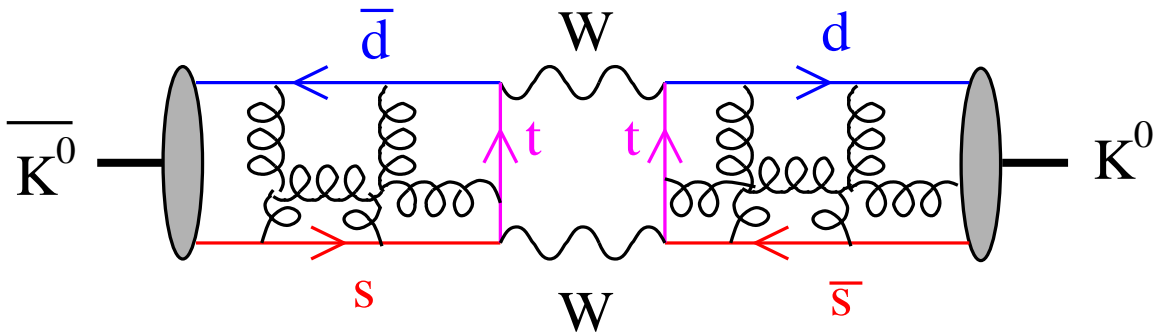,width=6truein}}
\caption{The $\bar K-K$ mixing amplitude.}
\label{fig:BKcartoon}
\end{figure}

Figure \ref{fig:BKcartoon} shows a cartoon of one of the best studied 
examples, CP-violation in the $K^0-\overline{K^0}$ mixing amplitude.
Most of the attendees of TASI were not born when this amplitude,
parameterized by $\epsilon$, was first measured. 
We understand the central ``box'' in the diagram: the large mass of 
the top quark allows us to treat it as an effective four-fermion operator,
with a coefficient proportional to 
${\rm Im}\left[ V_{ts}^2 V_{td}^2\right] $, multiplied by a
calculable QCD anomalous dimension factor.
What we do not know how to calculate analytically is the effect of
low momentum gluons and quarks ($|p|< 2$ GeV, say).
Such gluons confine the $q$-$\qbar$ pairs into kaons.
They interact with a large coupling constant, $\alpha_s(p)$,
and thus the interactions cannot be described using perturbation
theory. Non-perturbative calculations are needed. The only 
method presently available which starts from first principles and makes
no further assumptions is to simulate lattice QCD (LQCD) numerically. 
In the example of $\epsilon$, 
what the lattice must provide is the matrix element
\beq
\langle{\overline{K}|\bar s \gamma_\mu(1\!+\!\gamma_5) d\
                         \bar s \gamma_\mu(1\!+\!\gamma_5) d|K}\rangle
\equiv {16\over3} m_K^2 f_K^2 B_K \ .
\label{eq:BKdef}
\eeq
Here the decay constant is normalized so that $f_\pi=93$ MeV.

I will discuss the lattice calculation of $B_K$,
which is just a parameterization of the matrix element,
at some length below. What I want to point out here is that one needs
to know $B_K$ in order to use the experimental result for $\epsilon$
to extract a value for the ${\rm Im}(V_{td}^2)$.
A similar matrix element is needed to extract information from the
measured $B-\bar{B}$ mixing amplitude. 
Other matrix elements are needed to extract information from 
semileptonic $B$-meson decays.

Thus to probe the electroweak theory one must be able to
do non-perturbative calculations of matrix elements.
These lectures concern only lattice calculations of these matrix elements.
It is important to realize, however, that there are other methods available:
the large $N_c$ (number of colors) approach, QCD sum rules, chiral quark
model, etc. These all have the status of (more or less sophisticated) models:
they make approximations which are hard to improve upon and whose effect
is not always easy to gauge.
They often have the advantage, however, of being applicable to a wider
range of quantities than can be studied on the lattice.\footnote{%
I will discuss the limitations of lattice calculations in the following.}
What has happened in the last few years is that lattice calculations have
come of age, in the sense that for some quantities all errors are
understood and are small. Future progress will, I think, see the lattice
give reliable answers for increasingly many quantities.
This will allow us to test and improve the approximate methods,
which can then be applied with more confidence
to quantities for which lattice calculations are difficult.

Lattice calculations involve, at their core, numerical simulations,
but in order to make the various extrapolations that are needed,
considerable guidance from analytic results is required. 
Thus the lattice phenomenologist needs, in her or his tool kit, 
not only a PSC (personal super-computer),
but also expertise in
lattice perturbation theory (to match lattice onto continuum operators), 
chiral perturbation theory (to extrapolate to physical quark masses),
non-relativistic QCD or its close variants (to study heavy quarks on the
lattice), and the technology of ``improved actions'' (how to reduce
errors due to finite lattice spacing).
In the following I hope to give a flavor of how these tools are used,
and to indicate the present status of their application.

\section{Basics of Euclidean Lattice Field Theory}

I will begin with a review of the basics.
My discussion will be both sketchy and patchy.
Those wishing more detail or rigor will find both in
two recent texts on lattice field theory
\cite{Rothe,MM}. Less detailed but still useful is
Creutz's monograph \cite{Creutz}. Many results I use are standard,
and if I do not give a reference, it can be found in one or more
of these books.

The steps that are taken to get to a numerical simulation are these:
\begin{enumerate}
\item
Use the Euclidean space functional integral formulation of field theory
\beq
Z = \int [dA][dq][d\qbar] e^{-\ S_E[A,q,\bar q]} \ ,
\eeq 
where $A$, $q$ and $\qbar$ are respectively the gauge, quark and antiquark
fields.
\item
Discretize the theory on a hypercubic lattice, with spacing $a$.
In present simulations this lies in the range $0.05-0.2$ fm.
\item
Work in finite volume: $L$ points in the three spatial directions, 
$T$ points in the time direction.
The largest lattices in use
today are roughly $32^3\times64$ in size, 
which, for $a=0.1$fm, corresponds to $(3.2\,{\rm fm})^3 \times 6.4\,{\rm fm}$.
\item
Do the functional integral (now a multidimensional path integral) using
numerical Monte Carlo methods.
\item
Attempt to extrapolate $L\to\infty$ and $a\to 0$.
\end{enumerate}

Actually this is something of an idealization. Additional steps
are required in most simulations.
\begin{enumerate}
\item[3.5]
Make the ``quenched'' approximation, in which internal quark loops
are left out, keeping only valence quarks.
\item[5.5]
Extrapolate from quark masses $m_q\sim m_s/2$ down to the physical
up and down quark masses $(m_u+m_d)/2 \sim m_s/25$.
\end{enumerate}

\noindent
Many of these steps will be discussed in more detail below.

It is important to realize that,
with the exception of ``quenching'',
the approximations that are made can be systematically improved.
This improvement can occur not only because of increases in computer power
(which allows one to study larger lattices, for example),
but also due to analytical advances 
(e.g. ``improving the action'' so that one can
work with larger lattice spacings without increasing the errors due
to discretization).

An important limitation of LQCD is that the calculations are carried out
in Euclidean space. Why is this? Because the integrand
in a Minkowski path integral, $\exp(iS_M)$, is complex.
In contrast, the Euclidean integrand is, in most cases, real and positive.
Thus in the Minkowski integral there are large cancellations
between different regions of configuration space, and these make it hard to
simulate all but very small systems. This is an algorithmic, not a 
fundamental, issue. It is possible that it will be resolved in the future,
though there are no signs of this at present.
It turns out that even in Euclidean space, QCD at finite chemical potential
(i.e. finite baryon number density) has a complex action, and thus is very
difficult to simulate. 
Similarly, chiral theories have complex Euclidean actions. In fact,
the ``fermion doubling problem'' (to be discussed below) makes it difficult to
even formulate such theories on the lattice.

\subsection{Extracting information from Euclidean Correlators}

In this subsection I want to explain why LQCD is well suited to
the calculation of matrix elements involving the vacuum or single
particle states, while those involving multiple particles are much
more difficult. What can be most successfully calculated are
\begin{itemize}
\item
The energies of states, e.g. $\vev{\pi(\vec p)| H | \pi(\vec p)}$, which, 
in the continuum limit should become $E=\sqrt{{\vec p}^2 + m_\pi^2}$.
In particular, for $\vec p=0$, one directly calculates particle masses.
\item
Decay constants, e.g. $\vev{0|A_\mu|\pi}=i\sqrt2 p_\mu f_\pi$, 
where $A_\mu$ is the axial current.
\item
Single particle matrix elements $\vev{N(\vec p)| {\cal O} | M(\vec q)}$,
where $N$ and $M$ are single particle states, and $\cal O$ is a fermion
bilinear, or quadrilinear, or perhaps a gluonic operator.
\end{itemize}

I will explain why these quantities are easily calculable by showing how
they are calculated. Consider the Euclidean two-point function 
\begin{eqnarray}
C(\tau) &=& Z^{-1} \int [d\mu] \exp(-S_E) \CO^*(\tau) \CO(0) \\
&\equiv & \vev{\CO^*(\tau) \CO(0)}
\end{eqnarray}
where $d\mu$ is the measure for the quark and gluon fields,
$Z=\int [d\mu] \exp(-S_E)$ is the partition function,
and $\CO(\tau)$ is an function of the fields residing
at Euclidean time $\tau$. To study a pion at zero three-momentum,
for example, we might choose 
$\CO =\sum_{\vec x} {\overline u(\tau,\vec x)} \gamma_5 d(\tau,\vec x)$.
Recall that the fermion fields in the functional integral are 
Grassman variables.

The functional integral is constructed to give a time-ordered expectation value
\begin{eqnarray}
\label{2ptop}
C(\tau) &=& \vev{0|T[\COhatdag(\tau) \COhat(0)]|0} \\
	&=& \vev{0|e^{\hat{H}\tau} \COhatdag e^{-\hat{H}\tau}\COhat|0} \\
	&=& \vev{0|\COhatdag e^{-\hat{H}\tau}\COhat|0} \ ,
\label{2ptop3}
\end{eqnarray}
where $\hat{H}$ is the Hamiltonian operator, normalized so that
its ground state has zero energy, $\hat{H}|0\rangle = 0$, and
$\COhat(\tau)$ is the Heisenberg operator corresponding to $\CO(\tau)$
(in simple cases obtained by substituting for each field in $\CO$ the
corresponding operator).
In the second line I assume $\tau>0$, so that $\tau$-ordering, $T$,
has no effect, and I have used the Euclidean version of time translation
to shift the operator back to $\tau=0$. 
By convention $\CO(\tau\!=\!0)=\CO$.

Before proceeding, it is worthwhile understanding the conditions under
which expectation values in Euclidean functional integrals (e.g. $C(\tau)$)
can be written as time-ordered products, as in Eq. \ref{2ptop}.
Textbooks typically begin with the Minkowski space time-ordered product, 
which is then analytically continued to Euclidean space
(in the form of Eq. \ref{2ptop3}), and then written as a functional integral. 
What we want to do is run the argument the other way around.
This question was studied long ago by Osterwalder and Schrader,
who found the following \cite{OstSchrad}.
If the action $S_E$ is Euclidean invariant, and expectation values satisfy 
a property called ``reflection positivity'', plus
some other more technical conditions, 
then there exists a Hilbert space with positive norm, 
a Hamiltonian $\hat{H}$ acting on this space which is hermitian, 
and whose spectrum is bounded from below with the lowest state having
zero energy, and field operators, 
such that Eqs. \ref{2ptop}-\ref{2ptop3} hold.\footnote{%
For a nice discussion, including the definition of reflection
positivity, see Ref. \cite{MM}.}
There is then no obstruction to analytically continuing these correlators
to complex time by an inverse Wick-rotation, $\tau\to \tau'=e^{i\phi}\tau$.
For $\phi=\pi/2$, what results are Minkowski time-ordered products
\begin{equation}
\label{wick}
C(\tau\!=\!it) = \vev{0|T[\COhatdag(t) \COhat(0)]|0} \ ,
\end{equation}
where $t$ is Minkowski time.
A consequence of the initial Euclidean invariance is that
there are unitary operators acting on the Hilbert space which
implement Poincar\'e transformations. 

Once we have these time-ordered products we can, in principle,
use the LSZ reduction formalism to extract the S-matrix from the residues
of their poles.
In other words, Euclidean functional integrals with suitable properties
provide a definition of the field theory, as long as we can carry out
the analytic continuations.
It is important to realize that not all Euclidean functional integrals 
satisfy the necessary conditions. In particular, QCD in the 
quenched approximation can be written as a particular functional integral,
displayed below,
but does not satisfy reflection positivity, and does not correspond
to a well-behaved Minkowski field theory. 

With this background, let us return to the correlator $C(\tau)$.
Inserting a complete set of energy eigenstates (single and
multiparticle states), we find
\beq
\label{eq:completeset}
C(\tau) = \sum_n |\vev{n|\COhat|0}|^2 {e^{-E_n t}\over 2 E_n V} \ .
\eeq
Here I am assuming a finite volume $V$,
and using relativistically normalized states,
\beq
\vev{\vec p|\vec q} = 2 E V 
\delta_{p_x q_x} \delta_{p_y q_y} \delta_{p_z q_z} 
\stackrel{V\to\infty}{\longrightarrow}
 2 E (2\pi)^3 \delta^3(\vec p-\vec q)\ ,
\eeq
where $E^2=|\vec p|^2 + m^2$.
We could now follow the LSZ procedure:
Fourier transform to Euclidean energy, rotate to Minkowski space,
and look for poles.
We would find a series of poles corresponding to the stable states 
which couple to the operator $\COhat$, 
and various cuts for multiparticle intermediate states.
If the lightest state is a stable particle, however, then there is
no need to rotate to Minkowski space. For example, if the operator
creates a $\pi^+$ at rest, as in the example above, then we can read off
$m_\pi$ simply by looking at the exponential fall-off at large Euclidean times
\begin{equation}
\label{eq:1exp}
C(\tau)\stackrel{\tau\to\infty}{\longrightarrow}
|\vev{\pi^+(\vec p\!=\!0)|\COhat|0}|^2 \exp(-m_\pi \tau) \ .
\end{equation}
Furthermore, the coefficient of the exponential gives us a vacuum to
pion creation amplitude, which, if $\COhat=\bar u \gamma_0\gamma_5 d$,
is proportional to the decay constant $f_\pi$.
For this procedure to work it is important that there be a gap between
the lightest and next-to-lightest states.
In the present example, the latter consists of three pions at rest,
with $E= 3m_\pi$, so there is a gap for finite pion mass.

\begin{figure}[t]
\centerline{\psfig{file=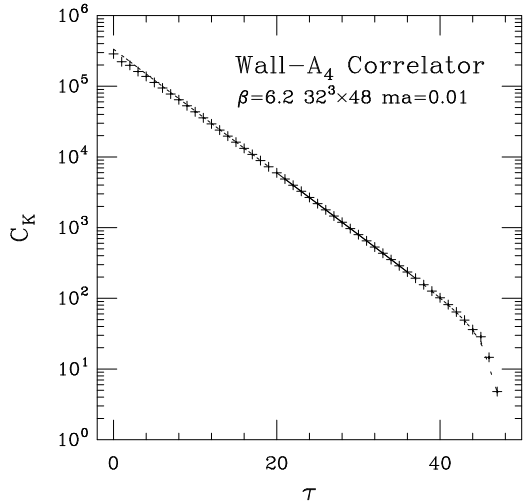,height=3.5truein}}
\vspace{-0.75truein}
\caption{``Kaon'' correlator with staggered fermions.}
\label{fig:kcor}
\end{figure}

Figure \ref{fig:kcor} shows an example of a two point function computed
numerically (from a calculation done in collaboration with Greg Kilcup 
and Rajan Gupta). It is for a particle with the quantum numbers of the kaon,
and with a mass similar to that of the kaon, but in which 
$m_s^\LATT=m_d^\LATT\approx \half m_s$.
The lattice spacing is about $a=1/12$ fm,
and the lattice size is $32^3\times48$. 
The graph shows that by $\tau\sim20$,
the data are represented almost perfectly by a single exponential.
The solid line shows a fit to such a form in the range $\tau=20-37$.
Outside this range, the dashed line shows the
extension of the fit function.
Thus you can
see the deviation of the data from a pure exponential at short times.
The curvature for $\tau>40$ is due to the boundary conditions.

The observant reader will notice an inconsistency between the data and
the expected form, Eq. \ref{eq:completeset}. $C(\tau)$ is a sum
of exponentials with positive coefficients, and
must approach its asymptotic form from above. 
This condition does not apply to the results of Fig. \ref{fig:kcor},
however, because we use different operators at times 0 and $t$.
This allows the coefficients to have either sign, and the approach
to asymptotia need not be monotonic.
One of the technical aspects of lattice
calculations which I will not go into, is the use of improved operators
(sometimes called ``sources''), designed so as to couple strongly
to the lightest state but much more weakly to higher states. 
With such operators the correlators quickly become dominated by a
single exponential. 
This allows one to use lattices which are shorter in the 
Euclidean time direction, and usually reduces the signal to noise ratio.

\begin{figure}[t]
\centerline{\psfig{file=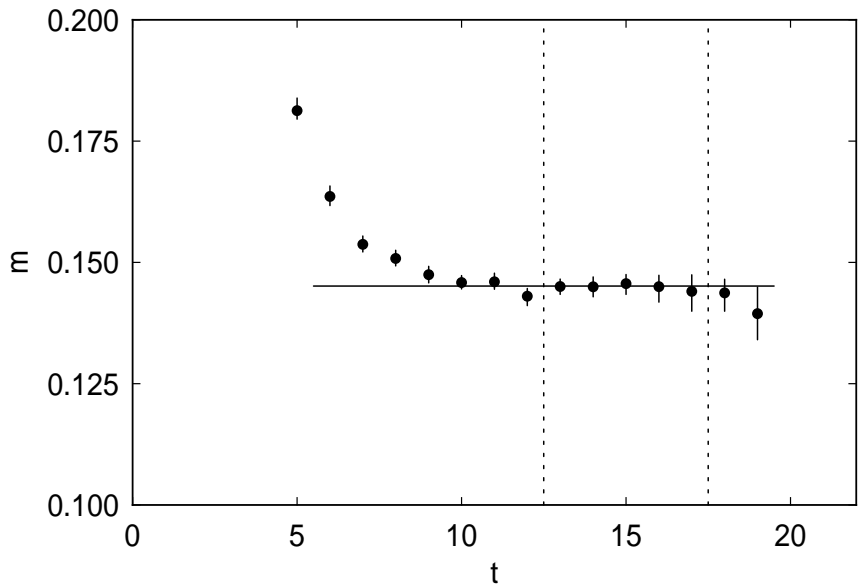,height=3truein}}
\caption{Example of an effective mass plot. The fitting range is shown
by dotted lines, the fit value by the solid line.}
\label{fig:meff}
\end{figure}

It is common for lattice practitioners to present their results in
terms of the logarithmic derivative $-d\ln[C(\tau)]/d\tau$, 
which, for $\tau\to\infty$, becomes the energy of the lightest state. 
The lattice version of this is 
$a m_{\rm eff}={\rm ln}[C(\tau)/C(\tau+1)]$, 
where $a$ is the lattice spacing.
An example of such a plot is shown in Fig. \ref{fig:meff}, 
for a calculation at similar parameters to that in Fig. \ref{fig:kcor}
(more precisely, a $30\times 32^2\times40$ lattice at $\beta=6.17$, 
$\kappa=0.1532$), taken from Ref. \cite{IBMspect}. 
(The behavior for $\tau>20$ is due to a negative parity $j=\frac32$ particle
propagating backwards in time.)
Note that, for the operators used to obtain these results, the coefficient
of the non-leading exponential turns out to be positive, as shown by
the fact that $m_{\rm eff}$ approaches its asymptote from above.

To first approximation the mass can just be read off from the graph.
It is not possible, to use the plot to give an idea of how
well the data is represented by a single exponential. This is because all
the points are correlated, fluctuating up and down nearly in unison.
One needs the full correlation matrix, and not just the ``diagonal'' errors
which are displayed, in order to test the goodness of fit.
It is true, however, that the fluctuations increase with $\tau$,
so one is always balancing the need to go to longer times,
so as to be sure one has a pure exponential, 
with the desire to work in the region where the errors are smaller.
This is a general feature of the analysis of lattice ``data''.

\bigskip
Single particle matrix elements can also be calculated directly in Euclidean
space, e.g.
\begin{equation}
C(\tau_2,\tau_1) = \vev{\Pi(\vec p,\tau_2) \CO_{K\pi}(\vec q,\tau_1) K(0)} \ .
\end{equation}
Here, the operators $K\sim \bar d \gamma_5 s$ and 
$\Pi\sim \bar u\gamma_5 d$ 
create a $K^0$ and destroy a $\pi^-$, respectively.
(I am using the loose description in which the I associate the
functions appearing as arguments in the functional integral with the
corresponding operators which appear in the operator representation.)
The operator in the ``middle'' (assuming $\tau_2>\tau_1>0$), might be,
for example, the current $\bar s \gamma_\mu (1\!+\!\gamma_5) u$.
Using the generalization of Eq. \ref{2ptop}, one finds
\begin{eqnarray}
C(\tau_2,\tau_1) 
&=& 
\vev{0|\widehat\Pi(\vec p) e^{-H(\tau_2-\tau_1)}
\widehat{\CO_{K\pi}}(\vec q) e^{-H\tau_1} \widehat K|0}  \\
&=&
\sum_{n,n'} \vev{0|\widehat\Pi(\vec p)|n} 
{e^{-E_n(\tau_2-\tau_1)} \over 2 E_n V}
\vev{n|\widehat{\CO_{K\pi}}(\vec q)|n'}
{e^{-E_{n'}\tau_1} \over 2 E_{n'} V}
\vev{n'|\widehat K|0} \ .
\end{eqnarray}
If $\tau_2-\tau_1$ and $\tau_1$ are both large enough, then we need
keep only the lightest state in the two sums.
Thus $\langle n|=\langle\pi^0(\vec p)|$,
$|n'\rangle=|K^0(\vec k)$, where $\vec k=\vec p - \vec q$.
The correlator becomes
\begin{equation}
C(\tau_2,\tau_1) \propto
\vev{0|\widehat\Pi(\vec p)|\pi^0(\vec p)}
e^{-E_\pi(\tau_2-\tau_1)}
\vev{\pi^0(\vec p)|\widehat{\CO_{K\pi}}|K^0(\vec k)}
e^{-E_K\tau_1}
\vev{K^0(\vec k)|\widehat K|0} \ .
\end{equation}
The energies and the creation and annihilation amplitudes can all be
obtained from two point correlators, and divided out.
Thus, from the three-point function one can directly extract the 
transition matrix element
$\vev{\pi^0(\vec p)|\widehat{\CO_{K\pi}}|K^0(\vec p)}$.
If $\CO\sim \bar s\gamma_\mu u$, then this is the vector form factor
which governs the semileptonic decay spectrum in $K^0\to\pi^- e^+ \nu$.
By changing the quantum numbers of the operators one can
calculate other form factors of interest, e.g. $B\to K^*\gamma$.
Using a four-fermion operator for $\CO_{K\pi}$, 
one can calculate $K-\bar K$ and $B-\bar B$ mixing amplitudes.
Most of the phenomenologically
interesting results from LQCD are for matrix elements of this type.

This exhausts the types of quantity that are simple to calculate 
in Euclidean space.
It is much more difficult, unfortunately, to calculate amplitudes involving
two or more particles in either initial or final state.
Some of the quantities we would like to calculate are
\begin{itemize}
\item $\CA(\pi\pi\to\pi\pi)$,
\item $\CA(K\to\pi\pi)$, \ and
\item $\CA(B\to\psi K_s)$.
\end{itemize}
The difficulty arise from the fact that,
in Minkowski space, these amplitudes are complex.
This follows from unitarity, as
there are on-shell real intermediate states. (The only exception is
the pion scattering amplitude at threshold, 
for which there is no phase space for intermediate states.)
But on-shell intermediate states are only possible in Minkowski space;
there are none if the external states have Euclidean momenta.
Starting from Euclidean momenta, the imaginary parts are generated 
{\em by the analytic continuation} to Minkowski space.
A simple example is the fuction $\ln(4 m_\pi^2 - (p_1+p_2)^2)$, real
in Euclidean space, but imaginary upon continuation
to physical momenta satisfying $(p_1+p_2)^2<-4 m_\pi^2$.

One way of seeing why there is no simple way of doing the calculation
directly in Euclidean space is the following. 
Consider the $K\to\pi\pi$ amplitude.
This is obtained by creating the kaon, acting with the weak Hamiltonian
to turn it into a state with the quantum numbers of two pions,
and then destroying the two pions. The physical amplitude involves the
two pions having a non-zero relative momentum.
In a Euclidean correlator, however, one does not obtain this contribution
by making a large Euclidean time separation between the weak Hamiltonian
and the pion operators. Instead, what dominates is the transition from
a kaon to two pions at rest, the latter being the lowest energy state.
Even if one uses two pion operators having relative momentum, they will
have, due to interactions, a coupling to the lowest energy state.
Thus what dominates the Euclidean correlator is an off-shell transition
amplitude $\vev{K|{\cal H}_W|\pi\pi(\vec p=0)}$, where $\vec p$ is the
relative momentum, and ${\cal H}_W$ the weak Hamiltonian.
This is not the quantity of interest.
Another way of stating the problem is that one does not create the
{\em in} and {\em out} states directly in Euclidean space.
See Ref. \cite{MaianiTesta} for a clear explanation of this point.

Thus to get the correct amplitude, both in magnitude and in phase,
one has to analytically continue. In most cases this will only be possible
if one has a model of the momentum dependence of the amplitude.
For example, for $K\to\pi\pi$ decays,
one use chiral perturbation theory, which, at leading order,
relates the decay amplitudes to calculable single particle matrix elements.
Using such a method, however, one gives up on the possibility of
a first-principles calculation, the errors in which can be systematically 
reduced. There will be an irreducible error due to the uncertainties
in the model used. For the example of $K\to\pi\pi$ decays, the uncertainties
are due to higher order terms in the chiral expansion.

Another approach is possible in the case of scattering.
L\"uscher has shown, in a beautiful series of papers \cite{Luscherscat},
how to extract scattering amplitudes indirectly, 
using the finite volume dependence of the energies of two particle states. 
Here too, however,
one must make approximations to use the results in practice.
An infinite tower of partial waves contribute to two particle energy
shifts, and one must assume that only a finite number are important.

\pagebreak
\section{Discretizing QCD}

To calculate amplitudes numerically, we must discretize QCD.
This must be done in such a way that gauge invariance is maintained,
since this invariance is required to guarantee the unitarity of the S-matrix.
How to do this was worked out long ago by Wilson.

\subsection{Continuum QCD, a brief overview}

The continuum action is given by
\beq
S_E = - \sum_{q = u, d, s, c, b , \ldots} \int_x \bar{q} (\dslash + m_q)q \ + 
\half \int_x {\sf Tr} (F_{\mu\nu} F_{\mu\nu}) \ ,
\eeq
where the integrals run over Euclidean space.
The covariant derivative is 
\beq
D_{\mu} = \partial_{\mu} - ig A_{\mu} \ ,
\eeq
in which the gauge fields are collected into a matrix $\Amu = \Amu^a T^a$,
with $T^a$ the generators of the $SU(3)$ Lie Algebra
\beq
[T^a, T^b] = i f^{abc}T^c \,,
\hspace{10mm} {\sf tr}(T^aT^b) = \half\delta_{ab} \ .
\eeq
The quark fields are color triplets, with an implicit color index.
Finally, the gauge field strength is
\beq
\Fmunu = F^{a}_{\mu \nu}T^a = \frac{i}{g} [\Dmu, \Dnu] = \partial_{\mu} \Anu
- \partial_{\nu}\Amu - ig [\Amu, \Anu] \ .
\eeq

A local $SU(3)$ gauge transformations is described by a space-time
dependent element $V(x) \in SU(3)$ ($V^{-1} = V^{\dagger}$, ${\sf det}(V)=1$):
\beqn
&&q(x) \rightarrow V(x) q(x) \quad
  \bar{q}(x) \rightarrow \bar{q}(x)V^{-1}(x)  \\
&&\Amu (x) \rightarrow V(x) \Amu (x) V^{-1}(x) + \frac{i}{g} V(x)
\partial_{\mu} V^{-1}(x) \\
&& \Fmunu(x) \rightarrow V(x) \Fmunu(x) V^{-1}(x)  \\
&& [D_{\mu} q](x)  \rightarrow V(x)  [D_{\mu} q](x) \ .
\eeqn
Given the last two lines, it is simple to see that $S_E$ is invariant.

It is useful to introduce the path-ordered integrals
\beq
L(x, y) = P exp \{ ig \int_y^x dz_{\mu} \Amu(z) \} \ ,
\eeq
which are to be thought of as going {\sl from y to x}.
The ordering is such that, for example, 
$A_\mu(x)$ is always to the left of $A_\mu(y)$.
Gauge transformation properties depend only on the end points of $L$,
and not on the path of integration
\beq
L(x, y) {\longrightarrow} V(x) L(x, y) V^{-1}(y) \ .
\eeq
They thus transport the gauge rotation from one point to another,
such that the quantity
$\qbar(x) L(x, y)q(y) $ is gauge invariant.
Another gauge invariant quantity is the trace of the path-ordered
integral around any closed path
\beq
Tr[L(x,x)]  {\longrightarrow} Tr[V(x)
L(x,x) V^{-1}(x)] = Tr[L(x,x)]
\eeq
These objects are called {\sl Wilson loops}.

\begin{figure}[t]
\centerline{\psfig{file=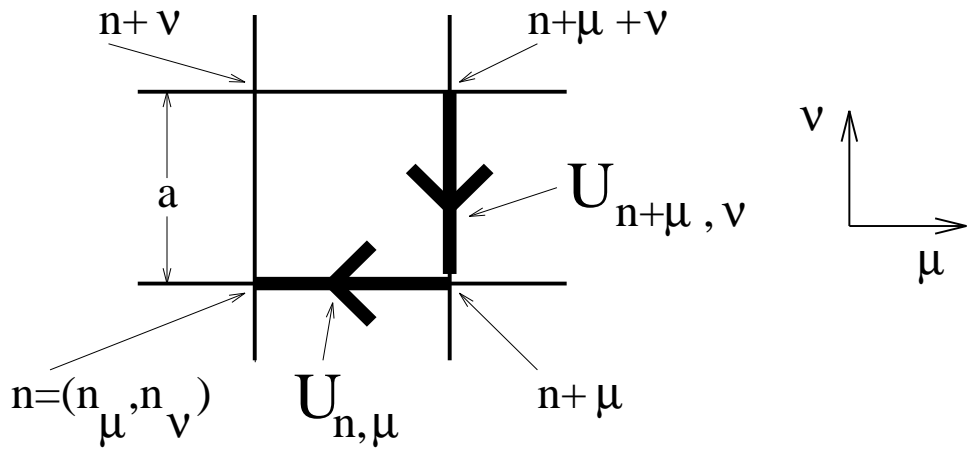,width=5truein}}
\caption{Notation for lattice quantities. $n$ is a vector of integers.}
\label{fig:lattice}
\end{figure}

With these quantities in hand, we can now construct a 
gauge invariant lattice version of QCD.
One cannot simply place the quark and gauge fields on the sites 
of the lattice and discretize the derivatives appearing in $S_E$.
Instead, the gauge fields, which transmit information about gauge
transformations from one position to another, will live on the ``links'' or
``bonds'' connecting the sites. I will choose the lattice to be hypercubical,
since this is the form most easily studied numerically, and nearly all work
has been done with it. 
The notation for the sites and links on the lattice is shown in Fig.
\ref{fig:lattice}. 

Discretizing fermions presents its own set of problems, 
not directly related with gauge invariance.
Thus I consider first the gauge part of the action.
This will be constructed from elements of $SU(3)$,
$U_{n,\mu}$, associated with the link from site $n$ to site $n+\mu$,
and corresponding to the continuum line integral along the link:
\beqn
U_{n,\mu} &\sim& L(an,an+a\hat\mu) \nonumber \\
	  &=& P \exp\left[ig\int_{n+a\hat\mu}^{n} dz_\mu A_\mu(z)\right]
\label{eq:lineintegral} \\
	  &=& 1 - iga A_\mu(n\!+\!\half\hat\mu) + O(a^2) \ . \nonumber
\eeqn
The ``$\sim$'' in the first line means ``corresponds to''. The vagueness here
is deliberate---once we put the theory on the lattice there are no gauge
fields $A_\mu$: they are replaced by the $U$'s. The expansion in the last
line is useful, however, for thinking about what the $U$'s mean,
and also for taking the classical continuum limit.

The gauge transformation properties of the $U$'s are taken to be the
same as those of the corresponding $L$'s
\beq
\Unmu \rightarrow V_n \Unmu V_{n + \mu}^{\dagger} \ ,
\eeq
where the $V_n\in SU(3)$ are gauge transformation matrices 
which live on sites. I have adopted the abbreviated notation
in which $n +\mu$ means $n+a\hat\mu$.
In correspondence with the continuum result
$L(x, y) = L(y, x)^{\dagger}$, we associate $\Unmu^{\dagger}$ to
the link from $n+\mu$ to $n$.
Note that we can multiply the $U$'s along any closed loop and take the
trace, and obtain an object which is invariant under gauge
transformations, since $V_n V_n^{\dagger}=1$. 
These are the lattice versions of Wilson loops.

We can construct a lattice version of the pure gauge action 
using the smallest Wilson loop, that around an elementary square
or ``plaquette''
\beq
P_{\mu \nu}^{\dagger} = \Unmu U_{n+\mu, \nu} U_{n + \nu,\mu}^{\dagger} 
U_{n, \nu}^{\dagger} \ .
\eeq
The geometry is illustrated here.\\
\begin{figure}[h]
\vspace{-0.1truein}
\centering
\unitlength=1pt
\begin{picture}(350,80)

\put(170, 20){\line(0,6){45}}
\put(215, 20){\line(0,6){45}}
\put(170, 65){\line(6,0){45}}
\put(170, 20){\line(6,0){45}}
\put(185, 5){$U_{n, \mu}$}
\put(185, 73){$U_{n+\nu, \mu}^{\dagger}$}
\put(148, 35){$U_{n, \nu}^{\dagger}$}
\put(220, 35){$U_{n+\mu, \nu}$}
\put(166, 40){$\wedge$}
\put(211, 40){$\vee$}
\put(190, 62){$>$}
\put(190, 17){$<$}

\end{picture}
\vspace{-0.2truein}
\end{figure}

\bigskip
\noindent
It is reasonable that such a loop is related to $F_{\mu\nu}$, because
the field strength is the curvature associated with the connection $A_\mu$.
In any case, using the correspondence given above for the $U$'s, 
and after some algebra, one finds that the classical continuum limit
of the plaquette is
\beq
P_{\mu\nu}^{\dagger} = 1 - ig a^2 \Fmunu - \frac{g^2}{2} a^4
\Fmunu^2 + ia^3 G_{\mu\nu} + ia^4 H_{\mu\nu} + 0(a^5),
\eeq
where $H_{\mu\nu}$ and $G_{\mu\nu}$ and are hermitian\footnote{%
Exercise: derive this result. Everyone should do it once!}.
Thus one can use the $\mu\nu$ plaquette as a discretized version of the 
corresponding component of the field strength, $\Fmunu$. 
If we take the trace, so as to get a gauge invariant quantity, we find
\beq
{\sf Re \ Tr} P_{\mu\nu} = N_c - \frac{g^2}{2} a^4 {\sf Tr} (\Fmunu^2)
+ 0(a^6) \,,
\label{eq:plaqexp}
\eeq
where $N_c=3$ is the number of colors.
We then have
\beq
\int d^4 x \sum_{\mu\nu} \half {\sf Tr} \Fmunu \Fmunu 
\ \sim\ \sum_{\Box} \frac{2}{g^2} ( N_c - {\sf Re Tr} \Box) \ .
\eeq
The factor of 2 arises because of the mismatch between 
the number of plaquettes per site, 6, and the number of 
terms in the sum $\sum_{\mu\nu}$, 12.

Now, it is standard (though unfortunate)
to replace the coupling constant by
$\beta=\frac{2 N_c}{g^2}$, so that
\beq
S_g = - \beta \sum_{\Box} \frac{{\sf Re Tr } \Box}
{N_c} + \mbox{\ irrelevant constant} \ .
\eeq
This is called the ``Wilson (gauge) action''.
It is important to realize that there is nothing special about using the
smallest loop to define the action. Any loop, e.g. a $1\times2$ rectangle, 
contains a term in its expansion proportional to a linear combination
of components of $(F_{\mu\nu})^2$. 
By taking an appropriate combination of loops we can
obtain the continuum action as $a\to0$.
The advantage of a small loop is that corrections proportional to powers 
of the lattice spacing are typically smaller than with a larger loop.

At this juncture it is appropriate to make a brief comment on 
``improving'' the action, i.e. reducing the errors due to discretization.
In Eq. \ref{eq:plaqexp} there are corrections of $O(a^2)$ compared
to the $F^2$ term that we want. 
Symanzik has shown how to systematically reduce these
corrections from $O(a^2)$ to $O(g^2 a^2)$ (by one loop calculations), 
and then to $O(g^4 a^2)$ (by two loop calculations,
etc. \cite{Symanzik}.
More ambitious is the program (based on using the renormalization group)
to construct an almost perfect action, i.e. one in which all terms of
$O(a^{2n})$ are almost absent \cite{perfectaction}.
The subject has lain dormant for almost a decade, but is now receiving
considerable attention, in part because progress has been made at
reducing the errors in the fermion action (which are usually of $O(a)$,
and thus larger than those in the gauge action).
To date, however, numerical
simulations leading to phenomenological results have only been carried
out using the Wilson gauge action.

To complete the definition of the theory I need to specify the measure.
Each link variable is integrated with the Haar measure over the group 
manifold. This measure satisfies
($V$ and $W$ are arbitrary group elements)
\beq
\int dU F(U) = \int dU F(UV) = \int dU F(WU) \,.
\eeq
Given this, it is simple to see that the functional integral
\beq
Z_{\rm gauge} = \int \prod_{\rm links} dU_{n,\mu} 
\exp\left({\beta \sum_\Box \frac{1}{N_c} {\sf Re\ Tr}\Box}\right)
\eeq
is gauge invariant.
What has been accomplished here is a non-perturbative, gauge invariant
regularization of pure gauge theories. What has been sacrificed is 
full Euclidean invariance: rotations and translations. The hope is that,
as one approaches the continuum limit, these symmetries are restored.

\begin{figure}[t]
\centerline{\psfig{file=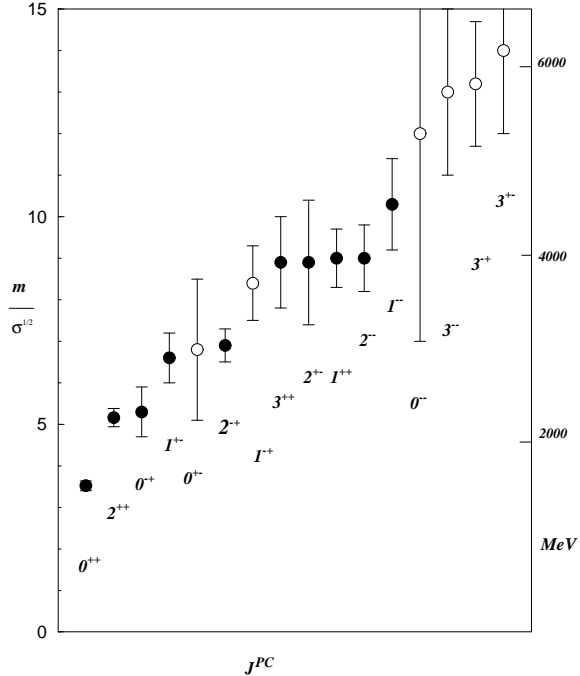,height=4truein}}
%
\caption{Spectrum of pure gauge SU(3) theory.}
\label{fig:gluespect}
\end{figure}

Pure $SU(3)$ gauge theory is not QCD.
But it is still of interest as a non-trivial field theory, 
sharing some properties in common with QCD.
In particular, its spectrum should consist of massive glueballs, in which
the gluons are confined by their self-interactions.
Considerable progress has been made in numerically simulating this theory.
I will give more details below of the methods used.
For now, let me note that one calculates
the glueball spectrum by looking at two point functions in which the
operators are Wilson loops of various shapes and sizes. By forming appropriate
linear combinations one can project onto the various representations of
the lattice rotation group, which can then be associated with certain
spin parities in the continuum limit. 
The present status of the spectrum is shown in Fig. \ref{fig:gluespect}
(Ref. \cite{gluespect}).
The masses are given in units of the square-root of the string tension,
a quantity I discuss below. For the moment, just consider the units
as arbitrary, so that all we can extract is the ratio of the masses of
different glueballs.
Clearly, we have reasonable control over the spectrum of 
this non-trivial theory, with the scalar glueball being the lightest, 
followed by the tensor and pseudoscalar.
Furthermore, there is evidence that Euclidean symmetry is being restored.
For example,
the five spin components of the tensor glueball lie in two different 
representations of the lattice cubic rotation group,
yet all have the same mass within errors.

I have not explained yet how one takes the continuum limit. 
When one does a simulation,
one picks the value of $\beta=6/g^2$, not the lattice spacing.
The output of the calculation are a set of masses in lattice units:
$a m_{\rm glue}$. One obtains a value for $a$ by comparing these
to a physically measured scale. 
In this case, there is no such scale, since the real
world is not pure gauge theory. If it were, we would use the physical
value of $m_{\rm glue}$, for a particular glueball, to fix $a$. All other
masses are then predictions. If we choose a different value for $\beta$
(i.e. for $g^2$), then we would obtain a different value for $a$.
To approach the continuum limit we adjust $\beta$ so that $a\to0$.
As we do this, the lattice mass vanishes, $a m_{\rm glue} \to 0$.
Thus correlators fall off more and more slowly, 
$C(t)\propto \exp(-a m_{\rm glue} t)$.
It is conventional in statistical mechanics to describe this in
terms of the divergence of the correlation length in lattice units:
$C(t)\propto \exp(-t/\xi)$, $\xi=1/(a m_{\rm glue})\to\infty$. 
A place at which $\xi$ diverges is called a critical point, and it is at
such points that one can take a continuum limit.

In the case of pure gauge theories, or QCD with sufficiently few
flavors that it is asymptotically free, one knows that the critical point
is at $g^2=0$, $\beta=\infty$. This is because the functional
dependence of $g^2$ on $a$ is calculable for small $g^2$ using
perturbation theory, yielding the familiar result that $g^2\propto 1/\ln(1/a)$.
Inverting this, we find that $a=(1/\Lambda) \exp(-1/2\beta_0 g^2)$
($\beta_0$ the leading order coefficient in the beta-function),
so that $a\to0$ for $g^2\to0$. What we do not know {\sl a priori}
is the proportionality constant $1/\Lambda$.
This we must determine numerically.
Given $\Lambda$, we can then predict the
variation of $a$ with $g^2$, for small enough $g^2$.
This works well for $\beta\ge6$, as long as one chooses an
appropriate scheme for the coupling constant \cite{LM}.

\subsection{Discretizing Fermions}

Discretizing the fermionic action,
\beqn
S_F = - \psibar(\dslash_E + m ) \psi \ ,
\eeqn
seems straightforward at first sight.
The subscript on the derivative is a reminder,
immediately to be dropped, that we are in Euclidean space,
with hermitian gamma matrices satisfying 
$\{ \gamma^{\mu}, \gamma^{\nu} \} = 2 \delta^{\mu \nu}$.
To discretize $S_F$, we place quark and antiquark fields on sites, 
$q(x) \to q_n$, and perform gauge transformations like so:
\beq
q_n \to V_n q_n \ ,\quad \qbar_n \to \qbar_n V_n^{\dagger} \ .
\eeq
To obtain a discrete derivative we must separate the $q$ and $\qbar$ fields,
and we make this gauge invariant using the link variables.
For example, choosing a symmetric derivative on the lattice,
\beq
\qbar D_\mu q(x) \longrightarrow \half \
\qbar_n \left[\Unmu q_{n+\mu} - U_{n-\mu,\mu}^{\dagger} q_{n-\mu}\right] \ .
\label{eq:symderiv}
\eeq
Here and in the following I have set the lattice spacing $a$ to unity.
It is left as a simple exercise to show that when one expands out the $U$'s
in terms of $A$'s one indeed finds the covariant derivative.
With this choice of $D_\mu$ the action is
\beq
- S_N(\qbar,q,U)\ =\ \sum_{n\mu} \half \qbar_n \gamma_\mu
\left[\Unmu q_{n+\mu} - U_{n-\mu,\mu}^{\dagger} q_{n-\mu}\right]
+ \sum_n m \qbar_n q_n \ .
\label{eq:naiveaction}
\eeq
For reasons about to be described, this is called the ``naive'' fermion
action. The full partition function of QCD is then
\beq
Z_{\rm QCD}\ =\ \int \prod_{\rm links} dU_{n,\mu} 
\prod_{q=u,d,s,..} [dq d\qbar]
\exp\left[ -S_g(U) -\sum_{q=u,d,s,...} S_N(\qbar,q,U) \right] \ .
\eeq
The quark measure is gauge invariant because a special unitary rotation has
unit jacobian. Thus $Z_{\rm QCD}$ is gauge invariant.

There are many choices available when discretizing derivatives
aside from that of Eq. \ref{eq:symderiv}:
the forward derivative ($\partial_\mu q \to q_{n+\mu}-q_n$), 
the backwards derivative ($\partial_\mu q \to q_n - q_{n-\mu}$), etc.
The symmetric derivative is, however, the most local choice which
preserves the anti-hermitian nature of the continuum operator $\dslash$.

The harmless looking action of Eq. \ref{eq:naiveaction}
gives rise to an infamous problem: 
in $d$ dimensions, it represents $2^d$ degenerate Dirac fermions, 
rather than one.
This sixteenfold replication in 4 dimensions
is referred to as the ``doubling (!) problem''.
Doubling has nothing to do with gauge fields, and so to discuss it
I consider a free lattice fermion. 
As discussed above, the spectrum of the states can be determined by
looking at the fall-off of the Euclidean two point function. 
More useful for analytic studies is to transform to momentum space, 
and look for poles. These occur at complex Euclidean momenta: 
$k^2 = -m^2$, where $m$ is the mass of the state.

To evaluate the two point function, i.e. the propagator,
recall the integration formulae for Euclidean fermions
(which are Grassman variables):
\beqn
Z &= & \int [ d\qbar] [ dq]\ e^{\qbar (\dslash + m)q} 
\ =\ {\sf det}(\dslash + m) \ ,\\
G(x, y) &= & - {Z}^{-1} \int  [ d\qbar] [ dq] 
e^{\qbar (\dslash + m)q} q(x) \qbar(y) 
\ =\ [\frac{1}{\dslash + m} ]_{xy}  \ .
\eeqn
To diagonalize $\dslash$ we go to momentum space.
Due to the periodicity of the lattice
one is restricted to the first Brillouin zone
\beq
q_n = \int_{-\pi}^{\pi}\frac{d^4k}{(2\pi)^4} e^{ikn}q(k) \ ,\quad
\qbar_n = \int_{-\pi}^{\pi}\frac{d^4k}{(2\pi)^4} e^{-ikn} \qbar(k) \ ,
\eeq
In terms of these fields, and using $s_{\mu} = \sin{k_{\mu}}$,
\beq
-S_N = \int_k \qbar(k) (i \sum_{\mu} s_{\mu} \gamma_{\mu} + m ) q(k) \ .
\eeq
Thus, the momentum space propagator is given by
\beq
G(k) = {1\over{i\sslash + m}} = { -i \sslash + m \over s^2 + m^2} \ .
\eeq
It is useful to reinstate factors of $a$,
so that $m = a m_\phys$ and $k = a k_\phys$.
In the continuum limit, for fixed physical quark mass, $m\to0$.
There is thus a pole near $k=0$, and we can expand 
$s_\mu = a k_{\mu,\phys} (1 + O(a^2))$, yielding
\beq
a C(k) = {-i \gamma_\mu k_{\mu,\phys} + m \over k_\phys^2 + m_\phys^2} \ .
\eeq
This has a pole at $k_\phys^2=-m_\phys^2$, representing the 
fermion that we expected to find.

Now we come to doubling. The lattice momentum function $s_\mu$
vanishes for $k_\mu=\pi$ as well as $k_\mu=0$.
In the neighborhood of the momentum $(\pi,0,0,0)$, if we
define new variables by  $k'_1=\pi-k_1$, $k'_i= k_i$, $i=2-4$, then
\beq
G(k') \approx { -i \sum_\mu k'_\mu \gamma'_\mu + m \over k'^2 + m^2} \ .
\label{eq:doubler}
\eeq
To bring the propagator into the standard continuum form,
I have introduced new gamma-matrices, 
$\gamma'_1=-\gamma_1$, $\gamma'_i=\gamma_i$, $i=2-4$,
unitarily equivalent to the standard set.
Equation \ref{eq:doubler} shows that there is a second pole, 
at $k'^2=-m^2$, which also represents a continuum fermion.
This is our first ``doubler''.

The saga continues in an obvious way.
$s^2$ vanishes if each of the four components of $k_\mu$ equals 0 or $\pi$.
There is a pole near each of these 16 possible positions.
Our single lattice fermion turns out to represent 16 degenerate states.

It is not, in fact, the replication of fermions which is the hard part
of the problem, but rather the way in which the chiralities of the states
work out. If $m=0$, then we can introduce a chiral projection
into the action $\g\mu\to \g\mu(1+\gamma_5)/2$, which in the continuum
restricts one to left-handed (LH) fields.
On the lattice, the pole near $k=0$ is LH.
The second pole I uncovered, however, actually represents a RH field. 
This is plausible, because 
$\gamma'_5=-\gamma_5$ and so $(1+\gamma_5)=(1-\gamma'_5)$.
It can be demonstrated by considering the coupling to external currents.
For each of the components of $k$ that is near to $\pi$ the chirality flips,
so that one ends up with eight LH and eight RH fermions.
It is important to note that, when one introduces gauge fields, 
all the fermions are necessarily coupled in the same way.
Thus one always obtains a ``vector'' representation of fermions,
i.e. one in which LH and RH fields lie in the same representation
of the gauge group.

How general is this result? Karsten and Smit showed that,
in {\em infinite volume}, any {\em local, antihermitian} discretization 
which maintains {\em translation invariance} will give rise to
LH and RH fermions in pairs \cite{KarstenSmit}.
There need be only one such pair:
Wilczek has given an example, using an action which breaks
Euclidean rotation invariance \cite{Wilczek}.
What are the consequences of Karsten and Smit's result?
\begin{itemize}
\item
Lattice regularization automatically takes care of the fact that
theories with anomalous representations of fermions (e.g. $SU(N_c)$ with
a single left-handed fermion) cannot be defined.
\item
It does too good a job, however. One cannot discretize a chiral theory,
i.e. one having an anomaly free, yet chiral, fermion representation.
In particular, one cannot discretize the electroweak theory.
\item
Indeed, one cannot even discretize QCD with ($n_f$) massless quarks,
in the following sense.
Such a theory should have an $SU(n_f)_L\times SU(n_f)_R$ chiral symmetry,
under which the LH and RH quarks rotate with independent phases.
But the lattice fermions are all begotten of the same lattice field,
and so cannot be rotated independently.
\end{itemize}
In summary, then, the lattice theory {\em lacks chiral symmetries}.
\smallskip

Can this be fixed up? Much effort has been devoted to this question.
There are various escapes, each with their own problems. 
Most notable are these.
\begin{itemize}
\item
One can explicitly break chiral symmetry right from the
start, and aim to recover it only in the continuum limit. This is, after all,
what one does with the rotations and translations. 
For fermions in vector representations, this is the approach originally
taken by Wilson, which I discuss in more detail below.
For chiral theories, this is the approach advocated by the Rome group,
and involves breaking the gauge symmetry at finite lattice spacing
\cite{Romechiral}.
\item
Keep the extra doublers, and divide their effects out by hand.
It turns out to be simple to reduce from 16 to 4 Dirac fermions,
the result being ``staggered'' fermions \cite{Susskind}.
I explain this when I discuss $B_K$ below.
\item
Avoid the Karsten-Smit result by using a random lattice \cite{random}.
This is very difficult to analyze, 
because even free fermions must be studied numerically.
Little progress has been made---see Ref. \cite{newrandom} for 
a recent study.
\item
Use ``domain-wall fermions'' or their descendents.
I have no time even to introduce the ideas; see Ref. \cite{domainwall}
for a review and references to the literature. It is controversial
at present whether the scheme can be implemented in a practical way.
\end{itemize}

Most present simulations use Wilson fermions, so I will briefly explain
what these are. A mechanistic way of understanding doubling is to note that
the ``forward-backward'' derivative of naive fermions is small both
for a smooth function and for one that is smooth except that it alternates
in sign. On the other hand, a discrete {\em second derivative}
will be small for the smooth function, yet large for the alternating
function. Thus Wilson suggested adding to the action a second derivative term
\beq
S_W =
\sum_{n\mu} {r \over 2} \qbar_n (U_{n,\mu} q_{n + \mu} - 2 q_n + 
U_{n-\mu,\mu}^{\dagger} q_{n - \mu}) \ ,
\label{eq:SW}
\eeq
where $r$ is a parameter.
The resulting free propagator is (I leave this as an exercise)
\beq
G(k) = { -i \sslash +(m - \frac{r}{2} \hat{k}^2) \over
	    s^2 +(m +\frac{r}{2} \hat{k}^2)^2} \ ,
\label{eq:wilsonprop}
\eeq
where $\hat{k} = 2 \sin(\frac{k_{\mu}}{2})$.  
If $k_{\mu}= \pi$, then $s_\mu= 0$ but $\hat{k_\mu} = 2$,
so that the additional pole picks up
an effective mass $m_\eff = m + 2r$.
Thus if one keeps $r$ finite in the continuum limit,
the effective mass stays finite, even if $m=m_\phys a\to0$.
Thus the effective physical mass becomes infinite. 
The same applies to all the other doublers.
Thus we expect them to decouple from the theory in the continuum limit,
leaving a single Dirac fermion.
This slightly sloppy analysis can be confirmed by studying the
position of the poles in $G(k)$.

In the presence of interactions, the doublers still decouple,
but they do cause a renormalization of the
parameters of the original Lagrangian.
This is the standard result of the Applequist-Carrazone decoupling theorem.
In particular, the quark mass $m$ gets additively renormalized.
This is not a surprise, since $S_W$ explicitly breaks chiral symmetry,
so there is nothing special about the value $m=0$.
Chiral symmetry should, however, be restored in the continuum limit,
because the Wilson term is a discretization of $a \qbar \partial^2 q$,
which vanishes when $a\to0$.

In practice one uses the value $r=1$ in simulations. One reason for this
is that the lattice theory then satisfies reflection positivity,
guaranteeing that one can construct an hermitian positive Hamiltonian
\cite{Luscher}.

\section{Simulations}

To illustrate what is involved in simulations, 
consider the pion two-point function
\beqn
\vev{\ubar \gamma_5 d(n) \dbar \gamma_5 u(p)} &=&
Z^{-1} \int [dU] \prod_q [dq] [d\qbar] e^{-S_g-S_N-S_W}
\ubar \gamma_5 d(n) \dbar \gamma_5 u(p) 
\label{eq:simulation}
\\
\nonumber
&=& 
{ -\int [dU] \prod_q {\sf det}(\dslash + m_q) e^{-S_g}
{\sf Tr}\left[
(\dslash + m_d)^{-1}_{np} \gamma_5 (\dslash+ m_u)^{-1}_{pn} \gamma_5 \right]
\over
\int [dU] \prod_q {\sf det}(\dslash + m_q) e^{-S_g}} \ .
\eeqn
Here $S_g$ is the lattice gauge action, $n$ and $p$ are lattice sites,
and $\dslash+m$ is the complete lattice Dirac operator appearing in 
the sum of naive and Wilson terms $S_N+S_W$.
Simulations are done using the form on the second line, 
i.e. after integrating out the Grassman fields. 
One is left with the functional integral over gauge fields, with measure
\beq
d\mu = [dU] \prod_q {\sf det}(\dslash +m_q) e^{-S_g} \ .
\eeq
To reduce the problem to a finite number of degrees of freedom
one uses a lattice of finite extent in Euclidean time and space.
The finiteness in time actually corresponds to putting the system
in the canonical ensemble at temperature $T=1/(N_t a)$.
For the studies I will consider, $N_t a$ is large enough that
$T\ll \Lambda_{\rm QCD}$, and the system is effectively at zero temperature.
One must choose boundary conditions on the fields.
To obtain the canonical ensemble these must be periodic (antiperiodic)
in the time direction for gauge fields (fermions).
In the spatial directions any choice can be made, though typically
periodic boundary conditions are used for all fields.
One then generates a set of ``gauge configurations'' 
(i.e. a $U$ for every link) according to the measure $d\mu$,
and calculates propagators,
$(\dslash +m)^{-1}$, on each of the gauge fields.
These are joined together into traces
such as that in Eq. \ref{eq:simulation}.
Finally, the resulting correlator is averaged over the
``ensemble'' of configurations, giving
a statistical estimate of the desired quantity.

What is the magnitude of this task? I am involved in calculations
on $V=32^3\times64$ lattices. Since each $U$ integral is over the
eight dimensional group manifold of $SU(3)$,
we are attempting to evaluate approximately an 
$ 8\times 4({\rm links/site} \times V \approx 7\times 10^{7} $
dimensional functional integral.
The Dirac operators are complex matrices of dimension
\[
[3 ({\rm color})\times 4 ({\rm spin}) \times V]^2= [2.5 \times 10^7]^2 \ .
\]
They are, however, sparse involving connections only between nearest neighbors.
Furthermore, if one fixes the site $n$ in Eq. \ref{eq:simulation},
one need only calculate a single column of the inverse 
(and a single row, but this can be obtained without extra work).
In this way one obtains the two point function from 
$n$ to all other points $p$.
When lattice practitioners talk about calculating propagators, 
they almost always are referring to such a truncated calculation.
The equation to be solved for the propagator $G$ is
\beq
\sum_p (\dslash + m)_{np} G_p = S_n \ ,
\eeq
where $S_n$ is the ``source''.
This might be $\delta_{n,n_0}$, or an extended function.

Numerical simulations, then, break up into two parts: generating
configurations, and calculating propagators. The latter is done
using a variety of standard algorithms\cite{MM}, 
conjugate gradient and minimal residue typically 
being preferred for staggered and Wilson fermions respectively.
Convergence can be improved by preconditioning. Nevertheless,
present algorithms for calculating propagators take
a number of iterations which is, for small $m$, 
roughly proportional to $1/m$. 
This slowing down is a big problem in practice,
as it forces us to work with quark masses much heavier than the
physical up and down quark masses.
Important steps towards alleviating this problem
have been taken using ``multi-grid'' algorithms \cite{multigrid}.

There are various methods for generating configurations---Metropolis, 
Langevin, Molecular dynamics, \dots---none of which I have time to discuss
in detail.
All use {\em importance sampling}, i.e. they move through the space of
possible configurations in such a way that the resulting distribution
is weighted directly according to the measure $d\mu$. If one defines
an effective action by
\beq
\exp(-S_\eff(U))
=  \prod_q {\sf det}(\dslash +m_q) e^{-S_g} \ ,
\eeq
then, as in any problem with a large number of degrees of freedom,
$S_\eff$ is very highly peaked in $U$ space. It would be utterly
hopeless to generate configurations uniformly in $U$ space, and then
include $\exp(-S_\eff)$ in the integrand.
All the methods work by moving towards the minimum of $S_\eff$ but
including some noise to keep the appropriate distribution around the
minimum.

Most of the methods will only work if $\exp(-S_\eff)$ is positive,
in which case it can be interpreted as a probability distribution on
$U$ space. The gauge action is automatically real.
In the continuum limit, for a vector representation of fermions
(which, as we have seen, is what we are forced to use on the lattice,
and, in any case, is appropriate for QCD),
the determinant of the Dirac operator is real and positive definite
(for $m>0$).
This is because the eigenvalues come in complex conjugate pairs,
with eigenvectors related by multiplication by $\gamma_5$.
On the lattice, with Wilson fermions, the determinant can be shown
to be real, but is not necessarily positive.
Thus to obtain a positive measure one must simulate degenerate
pairs of quarks, in which case the measure contains 
the square of the single quark determinant.

Almost all algorithms are based on making a change 
$\delta U$ in the gauge fields,
and then calculating the change in $S_{eff}$
\beqn
\delta S_\eff &=& - \sum_q \delta {\sf Tr\ ln}(\dslash+m_q) + 
			\delta S_g \\
&=& - \sum_q {\sf Tr}\left[ \delta U (\dslash+m_q)^{-1} \right]
    -\beta \sum_\Box \frac{1}{N_c} \delta\left({\sf Re\ Tr}\Box\right) \ .
\eeqn
To calculate the variation of the gauge action involves only a local
calculation, i.e. one involving links close to that being changed.
In contrast, the variation of the fermionic part of the effective
action involves a propagator and is non-local. Thus it is doable 
(at least one doesn't have to calculate the determinant itself!), 
but it is slow,
as it takes a number of operations that grows proportional to $V$.

The present algorithm of choice for simulating QCD is the ``hybrid Monte
Carlo'' algorithm \cite{HMCalg}. 
One can estimate that the time it takes to
generate an independent configuration is roughly
${\rm CPU} \propto V^{5/4} m^{-11/4}$. 
As the quark mass decreases, so does the pion mass, $m_\pi^2\propto m$.
Naively, it is necessary for the lattice length to exceed the
pion Compton wavelength by a factor of a few, so $L\propto 1/m_\pi$.
Using this, one finds ${\rm CPU} \propto m^{-5.25}$.
This is an asymptotic estimate, at best only roughly applicable for
today's lattice sizes and quark masses. But it shows why it
becomes rapidly more difficult to simulate lighter quarks. This is why
you rarely hear of simulations with quarks lighter than $m_s/3$,
and why progress is slow, even with an exponential growth in computer power.

There are various ways in which one can overcome this scaling law.
First, if one uses an improved action, one can get away with a larger
lattice spacing (for a given size of systematic error),
and thus a smaller number of lattice points for a fixed physical volume.
Second, a lot of the difficulty comes from the slowing down of 
propagator inversions as $m\to0$, and this should be avoidable.
This is the aim of the multigrid inversion algorithms.
Third, once the quark masses become small enough that the pions are
light, they interact weakly, and it aught to be possible to use 
chiral perturbation theory to account for errors introduced, say, by
not increasing the volume as $L\propto 1/m_\pi$.
With the presently available 10 Gigaflop machines, however,
and with the notable exception of studies of QCD at finite temperature,
systematic calculations of phenomenologically interesting quantities
are not possible in QCD. 

\subsection{Quenched QCD (QQCD)}

To make progress one must make an approximation, and what is used is
the so-called ``quenched'' approximation
\beq
d\mu_{\rm QCD} = [dU] \prod_q {\sf det}(\dslash +m_q) e^{-S_g} 
\longrightarrow d\mu_{\rm QQCD} = [dU] \exp(-S_g) \ .
\eeq
In other words, we set the fermion determinant to a constant.
This amounts to throwing away {\em internal quark loops}, 
while keeping the valence quarks, which now propagate through a
modified distribution of gauge configurations. 
For this reason, it is sometimes called the ``valence'' approximation.
Using QQCD reduces CPU requirements by a factor
of $\sim10^2-10^3$ with present values of $a$ and $m$.
Furthermore, the time to generate new configurations only grows as
${\rm CPU} \propto V^{5/4} \propto m^{-2.5}$, so that it is easier to
go to smaller quark masses.

Throwing away quark loops is a drastic approximation, the effect of which
I discuss in more detail below. Nevertheless, it is 
an approximation certainly worth studying,
because it retains the essential non-perturbative features of QCD,
confinement and chiral symmetry breaking. A quark propagating through
the ensemble of quenched gauge fields picks up an effective mass, just as it
does in QCD, and binds to form hadrons. The effective mass and the details of
the binding will differ from those in QCD, but, given the success of the
quark model in describing the hadron spectrum, one would expect the
quenched spectrum to be qualitatively similar. 

How well do we expect the QQCD to reproduce the spectrum of QCD?
Since we are stuck with QQCD for a while to come,
it is important to try and estimate such quenching errors.
Let me begin with a rough estimate.
One of the unphysical effects of quenching is that resonances in QCD, 
e.g. the $\rho$ meson, become stable states in QQCD.
This is because internal quark loops are
necessary to obtain the on-shell intermediate states 
(e.g. $\pi\pi$ in the case of the $\rho$) which
give rise to the imaginary parts of the propagators, and thus to
the width of resonances.
Discarding these intermediate states, however, affects not only the imaginary
part, but also the real part of the propagator.
In other words, not only is the width of the state changed (to zero)
but also the mass is shifted.
The most naive estimate is that $\delta m \sim \delta \Gamma=\Gamma$.
This mass shift will not be uniform in sign or magnitude, since it depends on
the available thresholds, possible cancellations, etc.
It may, in fact, be small for the $\rho$ \cite{Rhoshift}.
But it will, I expect, distort the spectrum at the 10\% level 
($100{\rm MeV}/1 {\rm GeV}$) in general.

Can one make a better estimate of the effects of quenching?
In particular cases, I think one can.
One example is quarkonia ($\bar cc$, $\bar bb$), which I touch on below.
Here I describe a somewhat more systematic method applicable 
to the properties of the pseudo-Goldstone bosons (PGBs),
the $\pi$'s, $K$'s and $\eta$.

The starting point is the formulation of QQCD
introduced by Morel\cite{Morel} 
\beq
Z_{QQCD} = \int [dU] [dq][d\qbar] [d\qtw][d\qtwbar]
e^{-S_g} e^{\qbar(\dslash+m)q} e^{-\qtwbar(\dslash+m)\qtw} \ .
\label{eq:MorelQQCD}
\eeq
Here $\qtw$ is a ghost field: a {\em commuting} spin-$\half$ variable.
I have written the partition function for only one flavor---in general
there is a ghost degenerate with each quark.
Equation \ref{eq:MorelQQCD} works because the ghost integration yields 
an inverse determinant which cancels that from the quark integration.
In other words, internal ghost loops cancel internal quark loops exactly.
This formulation shows why QQCD is a sick theory---ghosts have the
wrong connection between spin and statistics, and thus, in Minkowski
space, there will be violations of causality.
One cannot avoid the problems by 
considering correlation functions of external states composed of quarks alone,
because particles containing ghosts will appear in intermediate states.
Note, however, that, even though the Minkowski theory may be sick,
the Euclidean version is well defined. This only requires that
the eigenvalues of $\dslash+m$ have positive real parts, 
so that the bosonic functional integral converges.
This is an example of a Euclidean theory which is not reflection positive.

Morel's formulation is the starting point for the development of
``quenched chiral perturbation theory'' (QChPT) \cite{BG,Sharpelogs}.
Because of the ghosts fields, QQCD has a larger chiral symmetry than QCD, 
namely
\beq
SU(3|3)_L \times SU(3|3)_R \ .
\eeq
Here $SU(3|3)$ is the graded group of special unitary transformations
of three commuting and three anticommuting objects.
Numerical evidence suggests that this group is spontaneously broken
down to its vector subgroup, yielding not only the usual 
pseudo-Goldstone bosons (PGBs), but also
partners in which the one or both of the quark and antiquark fields
are replaced by ghosts.
One can develop an effective Lagrangian for this theory, analogous to the
usual chiral Lagrangian. Using this one can calculate the effect of
loops of PGBs, which give rise to non-analytic terms generically referred
to as ``chiral logs''. These logs are, in most cases, different in
QCD and QQCD, and one can use the difference to estimate the effect of
quenching. In particular, it should be possible to give a rough
ordering of quantities according to the size of quenching errors.
I give some examples below.

\begin{figure}[t]
\centerline{\psfig{file=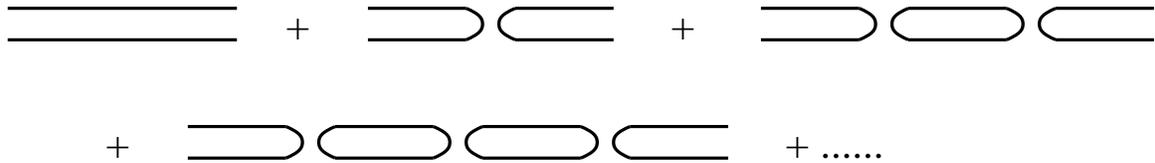,width=6truein}}
%
\caption{Contributions to the $\eta'$ propagator. Lines are quark
propagators.}
\label{fig:hairpin}
\end{figure}

There is an important qualitative difference between QChPT and ChPT
(chiral perturbation theory for QCD).
In QCD, the $\eta'$ is not a PGB, since the would-be $U(1)_A$ symmetry is
anomalous.
The $\eta'$ gets additional mass from
diagrams in which the quark-antiquark pair annihilate into gluonic
intermediate states. This ``hairpin vertex'' gets iterated to give rise
to the mass term, as shown schematically in Fig. \ref{fig:hairpin}.
But in QQCD, all terms except the first two are absent, and so the
$\eta'$ remains light, effectively a PGB. The would-be mass term
becomes a two-point vertex. The $\eta'$ must be included in the 
quenched chiral Lagrangian, along with this additional vertex.
The quenched chiral Lagrangian of Ref. \cite{BG} provides an
explicit realization of these rather handwaving diagrammatic arguments.
As explained below, the existence of a light $\eta'$
leads to a number of unphysical effects. In particular, there is an
$\eta'$ cloud surrounding all hadrons containing at least one light quark.



\pagebreak
\section{Numerical Results from quenched QCD}

In this section I present the evidence which shows that
QQCD does give a reasonable approximation to the real world.
This is essential if we are to have any hope of using QQCD to calculate
phenomenologically interesting matrix elements.
I will also discuss in more detail the size of quenching errors.

\subsection{Confinement}

There is a simple criterion for confinement in QQCD:
evaluate the potential energy $V(R)$ of an 
infinitely heavy quark ($Q$) and antiquark
($\Qbar$) as a function of the distance $R$ between them. 
The standard picture of confinement has a tube of color electric flux
joining the quark and antiquark, a tube which simply elongates as the
pair are separated. This suggests that $V(R)\propto R$ for distances
significantly larger than the width of the tube.
A better way of stating this is that the
force, $F = -dV/dR$, is expected to asymptote to a constant at large R. 
The magnitude of this constant is called the string tension, $\kappa$.
Using the force avoids the problem that $V(R)$ contains the self energies
of the quark and antiquark, which are $R$ independent,
but divergent in the continuum limit.

This criterion for confinement is not useful in QCD, because the flux tube
can be broken by the creation of a $\qbar q$ pair, leaving a $\overline Q q$
and $\qbar Q$ meson having an energy independent of $R$.
This process does not occur in QQCD because it requires internal quark loops.

To evaluate $V(R)$ one proceeds as follows.
Create the $Q \Qbar$ pair at time $\tau=0$ using the gauge invariant operator
\beq
\Qbar(\vec R) L(\vec R, 0) Q(0) \ ,
\label{eq:createQQ}
\eeq
The ordered integral in $L$ (Eq. \ref{eq:lineintegral}) can follow any path, 
or one can average over a number of paths so as
to maximize the overlap of the operator with the $\Qbar Q$ state.
Next, destroy the pair at a later time $\tau=T$ using the conjugate operator
\beq
\Qbar(0) L(0,\vec R) Q(\vec R) = \Qbar(0) L(\vec R, 0)^{\dagger} Q(\vec R) \ .
\label{eq:destroyQQ}
\eeq
In this way one has constructed a correlator which, for large $T$, should 
fall as $\exp(-V(R) T)$, where $V(R)$ is the energy of the
lightest state of the $\Qbar + Q + {\rm glue}$ system.

To evaluate the correlator we need the heavy quark propagator.
In Minkowski space,
an infinitely heavy quark just maintains its velocity, so if it
starts at rest, as here, then it remains so. That is why the final operator 
in Eq. \ref{eq:destroyQQ}
is chosen with the $\Qbar$ field at the same site as the original $Q$.
The current of a static quark does have a time component,
which couples to $A_0$, resulting in a phase for the propagator
$P\exp(-ig\int dt A_0(t))$. There is also the kinematical phase 
$\exp(-iM_Q t)$, but this does not depend on $R$ and thus does not
contribute to the force, and can be dropped.
Rotating to Euclidean space, the line integral remains a phase:
$P\exp(-ig\int d\tau A_4(\tau))$. For the antiquark, the line integral
is in the opposite direction. Combining these propagators with the line
integrals from Eqs. \ref{eq:createQQ} and \ref{eq:destroyQQ}, 
we obtain a Wilson loop, $W$.
It is roughly rectangular, having straight segments in the time direction, 
while the spatial paths (determined by the choice of $L(0,\vec R)$)
can wiggle, though they must remain in a single time-slice.
Putting this all together, we expect
\beq
\vev{W}
\stackrel{T\to\infty}{\longrightarrow}
C e^{-V(R) T} \stackrel{R\to\infty}{\longrightarrow}
C e^{-\kappa RT} \ ,
\eeq
where $C$ is a constant.
The last result is the well-known ``area-law'':
the expectation value of a rectangular $R\times T$ loop falls
exponentially with the area of the loop if there is confinement.

\begin{figure}[t]
\epsfxsize=3truein
\epsfbox{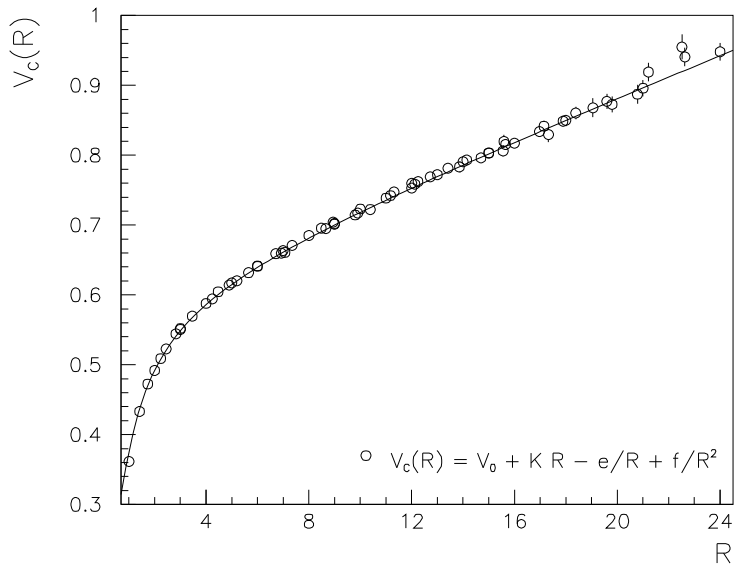}
\vspace{0.4truein}
\caption{The heavy quark potential, in lattice units.
The short distance points have been corrected for lattice artifacts
using the lattice Coulomb propagator.}
\label{fig:potential}
\end{figure}

It was established long ago, using the strong coupling expansion,
that the area law holds for large $g^2$.
This leaves the question of whether the result extends to the continuum limit,
$g^2\to0$, i.e. whether there is a phase transition at finite $g^2$ 
which breaks the analytic connection between weak and strong coupling.
Numerical evidence to date suggests that this does not happen.
For example, I show in Fig. \ref{fig:potential}
the results for the potential obtained by
Bali and Schilling \cite{Bali} at $\beta=6.4$, which corresponds to
$a\approx 0.06\,$fm. The horizontal scale is in units of $a$,
and thus extends well beyond $1\,$fm. The linear behavior of $V$ is clear, 
starting from $\sim 0.5\,$fm. 
A pleasing feature of this result is that one can see,
in the same calculation, both the long distance,
non-perturbative physics of confinement, 
and the short distance perturbative Coulomb potential. 
This is actually necessary if one is to calculate weak matrix elements,
for one must match onto the continuum using perturbation theory 
at short distances, while simultaneously including 
the long distance contributions.


\subsection{Charmonium and Bottomonium Spectra, and extracting $\alpha_S$}

I now turn to the spectrum.
After integrating out the fermions, and making the quenched approximation,
a general meson two-point correlator becomes
\begin{figure}[h]
\centerline{\psfig{file=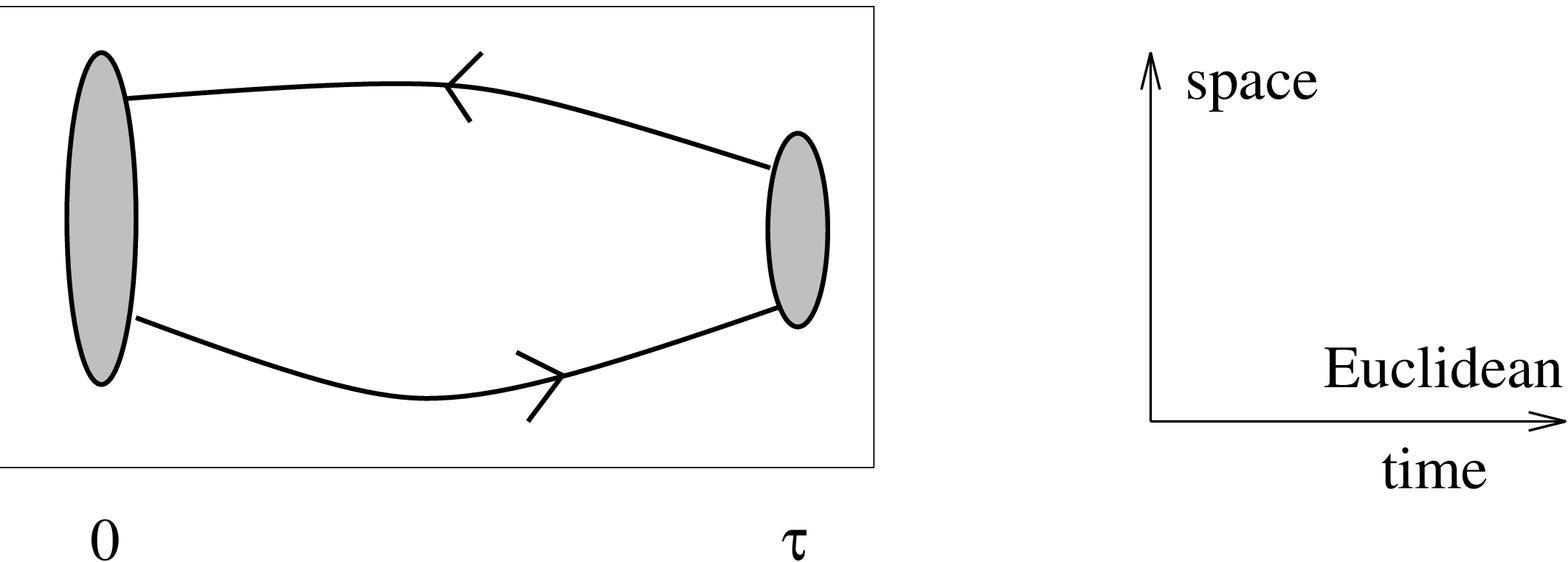,width=5.6truein}}
%
\label{fig:mesoncorrelator}
\end{figure}

\smallskip
\noindent
where the ``blob'' at $\tau=0$ represents an initial extended source,
made gauge invariant in some way, and the blob at $\tau$ represents
a similar ``sink''. The lines joining the blobs are quark propagators
in the background quenched gauge field.
If the meson is a flavor singlet (e.g. $\bar c c$), 
then there is a second diagram in which the quark and antiquark
annihilate into intermediate gluons. For heavy quark systems these
are suppressed by powers of $\alpha_s(m_q)$, and can be ignored. 
The same is not true for light quark systems;
in particular, the $\eta'$ mass comes from such diagrams.

I first discuss the $\bar c c$ system.
This has been studied on the lattice with great care by the Fermilab 
group\cite{FNAL}.
Compared to the light quark spectrum, the
$\bar c c$ system has several advantages
\begin{itemize}
\item
$\bar c c$ states are smaller than light quark hadrons, so one can
use lattices with smaller size in physical units.
\item
The CPU time needed to calculate $c$-quark propagators is less than
for light quarks. Since, in QQCD, calculating propagators consumes most
of the computer time, this allows one to use more lattices, and thus reduce
statistical errors. These errors are further decreased by that fact that
the intrinsic fluctuations of $c$-quark correlators from
configuration to configuration are typically smaller than for light quarks.
\item
The $\bar c c$ system is reasonably well described by potential models,
so one has a way in which to estimate {\rm systematic} errors,
in particular that due to quenching.
\end{itemize}
A potential disadvantage is that the charm mass in lattice units is not small,
e.g. $m_c a\sim 0.75$ at $a=0.1$ fm ($\beta\approx6$).
This might lead to significant discretization errors,
proportional to powers of $m_c a$, and thus require the use of a smaller 
lattice spacing.
This turns out not to be true, as long as one uses an improved action
\cite{Aidalat93}.

Charmonium is thus a system where all of the lattice errors can be
studied, and to a large extent controlled. 
In particular, finite volume and lattice spacing errors are small.
It turns out, as I discuss below, that similar control is possible for
the $\bar b b$ system, using non-relativistic QCD (NRQCD) to describe
the $b$ quark.
The onia are thus a good choice for accurately measuring the lattice spacing.
As I will explain, 
this can be turned into a prediction for $\alpha_s$.

The lattice spacing is obtained by comparing a quantity
measured in lattice units to its physical value.
An excellent choice for the physical quantity is the splitting between the
spin-averaged $1P$ and $1S$ levels in onio.
This is because the splitting is almost the same in $\bar c c$ and 
$\bar bb$ systems (457 and 452 MeV, respectively), so that one need
not worry if the lattice quark masses differ slightly from the experimental 
values. To extract $a$ one uses
\beq
a \ = {(am_{1P}-am_{1S})_{\rm lat} \over
		   (m_{1P}-m_{1S})_{\rm expt}}\,,
\label{eq:geta}
\eeq
where $(am_{1S,P})_{\rm lat}$ are the dimensionless masses 
one obtains from the lattice simulation.
The absence of corrections of $O(a^2)$ in this relation is a convention.
If we used other physical quantities we would obtain values of $a$ 
differing by such corrections.

This method gives accurate values of $a$ for various values of $g^2$,
or equivalently $g^2$ as a function of $a$. 
{}From the point of view of perturbation theory,
the lattice is just an ultra-violet regulator,
albeit a messy, rotationally non-invariant one.
Roughly speaking, it restricts momenta to satisfy $|p| < \pi/a$.
Thus it can be related to couplings defined in other schemes,
for small enough $a$. For example, it is related to the $\MSbar$ coupling
\beq
g^2(a) = g^2_{\MSbar}(\mu\!=\!\pi/a)\left[1-0.31 g^2 + O(g^4)\right] \ .
\label{eq:oneloopgs}
\eeq
(Neither the scheme nor the scale of the $g^2$ in the correction is determined
at this order.)
The idea is to use this result to obtain $g^2_{\MSbar}(\pi/a)$,
and then use the beta-function to run this coupling to a standard
scale, for example $m_Z$. 

An important test of the calculation is that
one should obtain the same result for $g^2_{\MSbar}(m_Z)$
starting at different values of $g^2$.
The extent to which this is not true is an indication of
the size of errors due to truncating perturbation theory in 
Eq. \ref{eq:oneloopgs}, from truncating the beta-function when running
to $m_Z$, and from discretization errors. 
It turns that the combined effect of these errors is smaller than
the statistical errors \cite{Aidalat93}.

The calculation cannot be carried out exactly as just explained.
The $g^2$ term in Eq. \ref{eq:oneloopgs} is $\sim 30\%$,
since $g^2\sim 1$ in lattice calculations.
%
%
Thus there are likely to be large corrections to Eq. \ref{eq:oneloopgs}
coming from unknown higher order terms. 
How, then, do we convert the well calibrated lattice into a result
for $\alpha_\MSbar$? 
We need a non-perturbative way of obtaining $\alpha_\MSbar$.
Such a method has been suggested by Lepage and Mackenzie \cite{LM}.

There are several parts to their method. First, we recall from 
our experience with perturbative QCD
that $\alpha_\MSbar(\mu)$ is a reasonable expansion parameter, as long as one
chooses a scale $\mu$ appropriate to the process under consideration. 
Equation \ref{eq:oneloopgs} then implies that $\alpha_\LATT(a)=g^2(a)/4\pi$ 
will be a poor expansion parameter. 
This expectation is borne out by numerical results.
For example, ratios of small Wilson loops
are not well represented by either first or second order lattice perturbation
theory---the $O(g^2)$ term is roughly half the size of the needed correction
at $g^2=1$.
If, however, one expands these quantities in terms of $\alpha_\MSbar$,
and chooses the scale according to a prescription explained
in Ref. \cite{LM}, then the leading order term does much better (because
$\alpha_\MSbar>\alpha_\LATT$), and the second order expression works very
well. See Ref. \cite{LM} for other examples.
The moral is that perturbation theory for short distance lattice quantities
is in good shape, as long as one chooses the correct expansion parameter.
It is also noteworthy that Lepage and Mackenzie have understood the source
of the large correction in Eq. \ref{eq:oneloopgs}: it comes from
extra ``tadpole'' diagrams that occur in lattice perturbation theory.

I have skipped over an important detail in the previous paragraph.
Lattice perturbation theory is rapidly convergent
for almost all quantities when expressed
in terms of $\alpha_\MSbar$ at an appropriate scale.
But how to we determine $\alpha_\MSbar$, given the need for higher order
terms in the relation Eq. \ref{eq:oneloopgs}?
Lepage and Mackenzie suggest a non-perturbative definition
in terms of the numerical result for the average plaquette
\beqn
\alpha_P &\equiv& - {3 \ln\vev{{\sf Tr}\Box} \over 4\pi} 
= {g^2 \over 4 \pi}(1 + O(g^2)) 
\label{eq:getalphaP}\\
{1\over \alpha_\MSbar(3.41 \pi/a)} &=& {1\over \alpha_P} - 0.37 + O(\alpha) \ .
\label{eq:getmsbar}
\eeqn
In other words, define an auxiliary coupling constant by the first equation,
and relate it to $\alpha_\MSbar$ using the perturbative result in
the second line. 
Note that the correction term in the second relation is small,
since $1/\alpha\sim 5-10$.
The scale $3.41 \pi/a$ takes into account a subset of the two-loop
corrections.
It is using $\alpha_\MSbar$ defined in this way that
Lepage and Mackenzie find 
lattice perturbation theory to be well behaved.
In effect this method uses the numerical data itself to sum
the leading tadpole diagrams to all orders.

In summary, using the result for the lattice spacing from Eq. \ref{eq:geta},
together with the result for the average plaquette, 
Eqs. \ref{eq:getalphaP} and \ref{eq:getmsbar} give us $\alpha_\MSbar$ at the 
known scale $q=3.41 \pi/a$. We can then run the result up to $m_Z$.

By far the largest error in the result is that due to the use of the
quenched approximation. The Fermilab group has developed a method to estimate
this error, based on the fact that the $\bar c c$ system is well
described by a potential model.
The potential in QCD differs from that in QQCD, 
but we have a reasonable idea of the form of this difference,
so we can subtract its effects.
It is useful to {\em define} a coupling constant in terms of the potential 
in momentum space
\beq
V(q) \equiv -C_F \alpha_V^{(n_f)}(q)/q^2 \ .
\eeq
The superscript gives the number of flavors in internal quark loops.
At short distances this behaves like any other coupling constant,
and in fact is close to that in the $\MSbar$ scheme 
\beq
{1\over \alpha_V(q)^{(n_f)}} \ =\ {1\over \alpha_\MSbar(q)^{(n_f)}}
 - 0.822 + O(\alpha) \ .
\label{eq:Vtomsbar}
\eeq
At long distances it must lead to a confining potential.
A useful form for $\alpha$ which interpolates between these 
two limits has been given by Richardson.
Now, since one adjusts the scale so that the $1P-1S$ splitting matches
experiment, it must be that QCD ($n_f=3$) and 
QQCD ($n_f=0$) potentials are similar at the scale
of typical momenta in low lying $\bar c c$ states, $q^*=0.35-0.7$GeV,
i.e. $\alpha_V^{(3)}(q^*)\approx \alpha_V^{(0)}(q^*)$.
But the two couplings run differently with $q$. In particular, for large
enough $q$, the QQCD coupling decreases more rapidly because there is
no fermionic screening.
This means that $\alpha_V^{(3)}(\pi/a) - \alpha_V^{(0)}(\pi/a)>0$
(assuming $\pi/a>q^*$).
The difference is calculable given an assumed form for the potential.
Using Eq. \ref{eq:Vtomsbar} we can convert this to a difference between
$\alpha_\MSbar^{(0)}(\pi/a)$ and $\alpha_\MSbar^{(3)}(\pi/a)$.
The former we have already determined, so we obtain the latter,
which we then run up to $m_Z$.

The resulting correction is substantial---a 26\% 
increase in $\alpha_\MSbar(5 {\rm GeV})$ for charmonium.
This correction itself is uncertain, due to uncertainties in the
matching scale $q^*$ and in the chosen form of the potential.
The resulting uncertainty in $\alpha_S$
is estimated to be $\sim 8\%$, slightly larger than
that coming from the neglect of higher order terms in the perturbative
relation between $\alpha_P$ and $\alpha_\MSbar$.
A clear discussion of all these issues is given in Ref. \cite{Aidalat93}.

The net result is \cite{Aidalat93}
\beq
\alpha_\MSbar^{(5)}(m_Z) = 0.110 \pm 0.008 \ .
\eeq
A similar analysis using NRQCD to study the $\bar bb$
system yields a consistent result\cite{NRQCDalp}, $0.112\pm 0.004$.
The fact that the final results agree is a test of the method of 
estimating quenching errors, which are different in the two onia.
It is appropriate that the result be included 
in the latest Review of Particle Properties \cite{RPP}.

Recently, the first results including quark loops have been obtained,
using two moderately light flavors of quarks in the 
loops\cite{NRQCDalp,NRQCDjapan}. For a review see Rev. \cite{shigemitsu}.
Results for three light flavors can now be obtained both using
the methods just described, or simply by extrapolating from $n_f=0$ 
and $n_f=2$. The two methods yield consistent results, with the latter
giving the smaller errors:\cite{NRQCDalp}
\beq
\alpha_\MSbar^{(5)}(m_Z) = 0.115 \pm 0.002 \ .
\eeq
This is a very impressive result, which is consistent with
the latest world average experimental value\cite{Webber}
$\alpha_\MSbar^{(5)}(m_Z) = 0.117 \pm 0.005$.
My only concern is whether the error fully accounts for the uncertainties 
in the extrapolation of $m_u$ and $m_d$ to their physical values.

\subsection{Light hadron spectrum}

I want to begin my discussion of the spectrum of hadrons composed
of $u$, $d$ and $s$ quarks by clarifying how, were we able to simulate QCD,
we would determined the correct lattice quark masses and extract $a$.
This must be done carefully because the measure of the functional integral
depends, through the Dirac determinant, on the quark masses.
For simplicity of presentation, I assume that $m_u=m_d=m_l$.
I will also assume that we have picked a value of
$\beta=6/g^2$ large enough that $O(a)$ errors can be ignored.
One would remove such errors in practice by simulating at
a number of values of $\beta$ and extrapolating to $a=0$.

Having chosen $\beta$, we then calculate numerically three lattice masses, e.g.
\beq
a m_{\rm proton}(am_l, am_s)\,,\ \
a m_{\pi}(am_l,am_s)\,, \ \ {\rm and}\ \
a m_{K}(am_l,am_s)
\eeq
as a function of the lattice masses $m_l a$ and $m_s a$.
We adjust these lattice masses until the two mass ratios agree with experiment
\beq
{a m_{\pi}(\bar{am_l}, \bar{am_s}) \over 
 a m_{\rm proton}(\bar{am_l}, \bar{am_s})} = {135 \over 938} \ ,\ \
{a m_{K}(\bar{am_l}, \bar{am_s}) \over 
a m_{\rm proton}(\bar{am_l}, \bar{am_s})} = {498 \over 938}  \ .
\eeq
Hopefully these equations have a solution!
With the lattice masses fixed we can now extract the lattice spacing, using
\beq
 a m_{\rm proton}(\bar{am_l}, \bar{am_s}) = 938{\rm MeV} \ .\ \
\eeq
Any other quantity (mass, decay constant, \dots) that we calculate
can now be predicted in physical units. 
Clearly, any three dimensionful quantities can be used to carry
out this program.

Knowing $a$ also allows us to extract the quark masses in physical
units. These are quark masses renormalized in the lattice scheme at the
scale $a$. They are perturbatively related to more familiar masses, such as
those in the $\MSbar$ scheme.

The same procedure applies to QQCD, but is much simpler to implement.
This is because the measure is independent of the quark masses,
so one need only generate a single ensemble for each $\beta$.
The flip-side of this simplicity is, of course, that one is throwing
away important aspects of the physics.

Unfortunately, it will be some time before this procedure can be followed
for QCD. For the moment a systematic study of the spectrum is restricted
to QQCD. The most thorough analysis is that of the IBM group,
using the GF-11 computer built by Weingarten and 
collaborators\cite{IBMspect}.
It sustains $\sim 7$ GFlops for these calculations.
I will present an outline of their results, with my major focus
being the issue of the reliability of the quenched approximation.

\begin{figure}[tb]
\centerline{\psfig{file=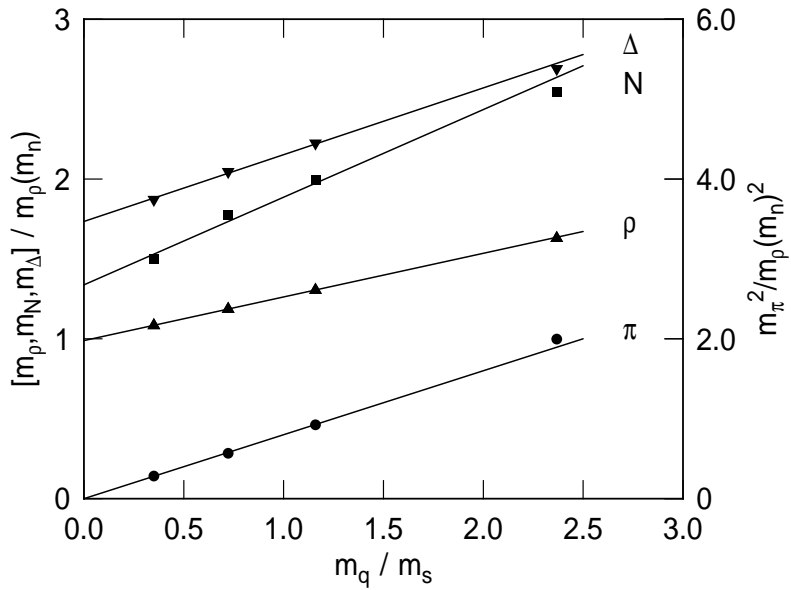,height=3truein}}
%
\caption{Mass extrapolations at $\beta=6.17$ (using sinks of size 4
on a $30\times32^2\times40$ lattice). All hadron masses are scaled
so that that $m_\rho(m_q=0)=1$. Quark masses are given in units of $m_s$.}
\label{fig:massextrap}
\end{figure}

The main features of the study are
\begin{itemize}
\item
Masses are calculated for light hadrons
(those with the quantum numbers of the $\pi$, $\rho$, $N$ and $\Delta$),
using {\em degenerate} quarks.
A range of quark masses is used, extending down to about $m_s a/3$,
where $m_s$ is the physical strange quark mass.
The results are then extrapolated to the physical up and down
quark masses. An example is shown in Fig. \ref{fig:massextrap}.
I return to the reliability of these extrapolations below. 

The use of degenerate quarks means that a direct calculation of
the strange meson and baryon masses (except for the $\Omega^-$) is
not possible. There is no fundamental obstacle to doing such a calculation;
after all, quark propagators with different masses have been calculated.
The practical problem is just that
the number of non-degenerate combinations becomes large.

The IBM group make predictions for the strange hadrons based upon the
assumption (well supported by the experimental data itself) that, in
QCD, the masses of mesons and baryons (squared masses for PGBs)
depend linearly on the masses 
of the valence quarks of which they are composed. 
This leads, for example, to the result
\beq
m_\Sigma + m_\Xi - m_N \approx m_N(m_s) \ ,
\label{eq:sigmaxiN}
\eeq
where the quantity on the right hand side is the mass the nucleon
would have if $m_u=m_d=m_s$ for the {\rm valence} quarks.
This latter quantity is easily calculated in QQCD.

\item
Finite volume effects are studied by comparing results on lattices
of different sizes.
Previous studies have shown that for $La< 1-1.5\,$fm there are significant
finite volume corrections \cite{finitevol}. 
All the lattices in the IBM study are larger than this,
so the finite volume corrections turn out to be small.

\begin{figure}[tb]
\centerline{\psfig{file=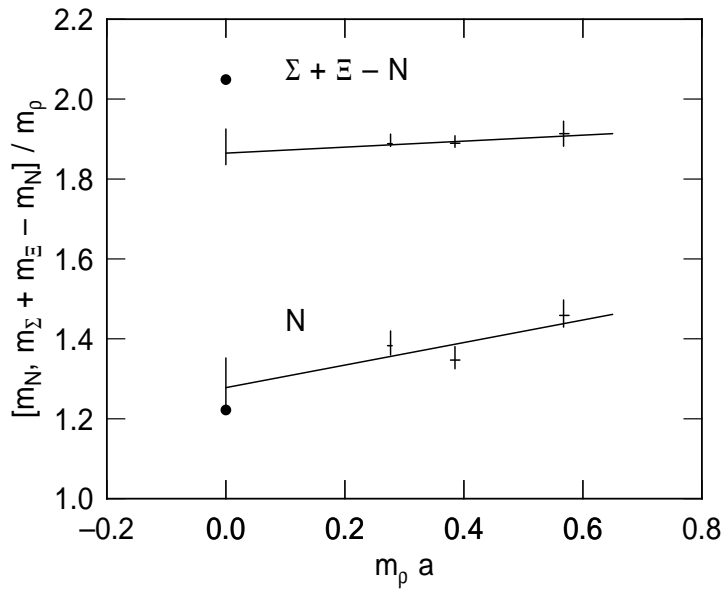,height=3truein}}
%
\caption{Extrapolation to $a=0$ (for sinks 0,1, and 2 combined).
All masses have been previously extrapolated to the continuum limit.
Dots at $a=0$ are observed values.}
\label{fig:aextrap}
\end{figure}

\item
The extrapolation to $a=0$ is done using three lattices,
having roughly the same physical volume ($L\sim 2.4$fm), 
but different lattice spacings
($\beta=5.7$, $5.93$ and $6.17$, corresponding to $a\approx
0.15$, $0.1$ and $0.07$fm).
Figure \ref{fig:aextrap} shows an example for $m_N/m_\rho$ and 
``$(m_\Sigma+m_\Xi-m_N)/m_\rho$'' (obtained using Eq. \ref{eq:sigmaxiN}). 
Since Wilson fermions are used, discretization errors are $O(a)$,
and linear extrapolation is assumed.
This plot does not include a small shift due to finite volume corrections.

\item
Statistical errors are reduced using
large ensembles of lattices, typically 200.
In addition the signal is improved by creating the hadrons
using extended sources.
Figure \ref{fig:meff} shows about the worst signal.
What is important is that the
effective mass reaches a plateau before it disappears into noise.
The fitting range is chosen by an automatic procedure, and is shown
by the dotted vertical lines.
The data sample is large enough that fits using the full correlation
matrix are stable, and errors are estimated using the bootstrap procedure.
The resulting mass value is shown by the solid horizontal line,
and has statistical errors of about $2\%$. 
\end{itemize}
The final results are shown in Fig. 10.
The predictions all agree with the experimental numbers to within $6\%$, 
or less than 1.6 standard deviations. QQCD seems to work extremely well!
Indeed Ref. \cite{IBMspect} interprets this success as indicating both that
QCD is the theory of the strong interactions, and that
the quenched approximation works well, at least for masses.

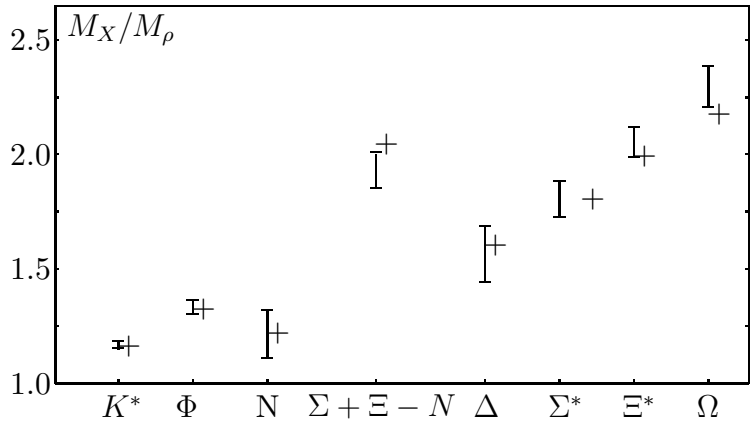
\begin{figure}[bth]
\begin{center}
\setlength{\unitlength}{.006in}
\begin{picture}(565,400)(0,0)
\put(10,50){\line(0,1){330}}
\multiput(10,100)(0,50){6}{\line(1,0){3}}
\multiput(612,100)(0,50){6}{\line(1,0){3}}
\put(6,50){\makebox(0,0)[r]{1.0}}
\put(6,150){\makebox(0,0)[r]{1.5}}
\put(6,250){\makebox(0,0)[r]{2.0}}
\put(6,350){\makebox(0,0)[r]{2.5}}

\put(20,360){\makebox(0,0)[l]{$M_X / M_\rho$}}

\put(10,50){\line(1,0){605}}
\put(10,380){\line(1,0){605}}
\put(615,50){\line(0,1){330}}

\multiput(65,50)(65,0){3}{\line(0,1){3}}
\put(50,30){\makebox(0,0)[l]{$K^*$}}
\put(115,30){\makebox(0,0)[l]{$\Phi$}}
\put(185,30){\makebox(0,0)[l]{N}}
\put(290,50){\line(0,1){3}}
\put(230,30){\makebox(0,0)[l]{$\Sigma+\Xi-N$}}
\multiput(385,50)(65,0){4}{\line(0,1){3}}
\put(375,30){\makebox(0,0)[l]{$\Delta$}}
\put(440,30){\makebox(0,0)[l]{$\Sigma^*$}}
\put(505,30){\makebox(0,0)[l]{$\Xi^*$}}
\put(570,30){\makebox(0,0)[l]{$\Omega$}}

\put(85,82.8){\makebox(0,0)[r]{+}}
\put(65,84){\line(0,1){3.2}}
\put(65,84){\line(0,-1){4.2}}
\put(60,87.2){\line(1,0){10}}
\put(60,80.8){\line(1,0){10}}

\put(150,115.4){\makebox(0,0)[r]{+}}
\put(130,117){\line(0,1){6}}
\put(130,117){\line(0,-1){6}}
\put(125,123){\line(1,0){10}}
\put(125,111){\line(1,0){10}}

\put(215,94.4){\makebox(0,0)[r]{+}}
\put(195,93.2){\line(0,1){20.8}}
\put(195,93.2){\line(0,-1){20.8}}
\put(190,114){\line(1,0){10}}
\put(190,72.4){\line(1,0){10}}

\put(310,259.4){\makebox(0,0)[r]{+}}
\put(290,235.4){\line(0,1){14.8}}
\put(290,235.4){\line(0,-1){14.8}}
\put(285,252.2){\line(1,0){10}}
\put(285,220.6){\line(1,0){10}}

\put(405,170.8){\makebox(0,0)[r]{+}}
\put(385,163){\line(0,1){24.4}}
\put(385,163){\line(0,-1){24.4}}
\put(380,187.4){\line(1,0){10}}
\put(380,138.6){\line(1,0){10}}

\put(490,210.6){\makebox(0,0)[r]{+}}
\put(450,211.2){\line(0,1){16}}
\put(450,211.2){\line(0,-1){16}}
\put(445,227){\line(1,0){10}}
\put(445,195){\line(1,0){10}}

\put(535,249.2){\makebox(0,0)[r]{+}}
\put(515,261){\line(0,1){13}}
\put(515,261){\line(0,-1){13}}
\put(510,274){\line(1,0){10}}
\put(510,248){\line(1,0){10}}

\put(600,285.4){\makebox(0,0)[r]{+}}
\put(580,309.2){\line(0,1){17.8}}
\put(580,309.2){\line(0,-1){17.8}}
\put(575,327){\line(1,0){10}}
\put(575,291.4){\line(1,0){10}}

\end{picture}
\end{center}
\caption{Final results for the light hadron spectrum.
Masses are measured in units of $m_\rho$.
Experimental numbers are denoted by crosses.}
\end{figure}

\subsection{Quenching errors in the light hadron spectrum}

The IBM study is an impressive piece of work, one which sets the standard
for future simulations. Only a few other
quantities ($\alpha_s$, $m_b$ and $B_K$) have been studied so thoroughly.
I am not, however, fully convinced of the conclusions,
i.e. that the spectra of QQCD and QCD differ by less than 6\%.
As I will explain, there are reasons to expect typical deviations to be
10-20\%.
There are then three possibilities:
(i) the reasons I will present are not valid;
(ii) the reasons are valid, but the deviations turn out fortuitously to
be smaller for the light hadron spectrum; and
(iii) the assumptions made in order to do the various extrapolations in
the IBM study are not valid, and the actual answers differ 
more substantially from the physical spectrum.
I am biased, so I expect the answer to be a combination of (ii) and (iii),
but only further simulations will resolve the issue.


The two extrapolations which might need further refinement are those
in lattice spacing and in quark mass.
As for the former, one expects corrections 
of the form $1 + a\Lambda_1 + (a\Lambda_2)^2 + \dots$,
where $\Lambda_1$ and $\Lambda_2$ are non-perturbative scales.
Linear extrapolation has been assumed ($\Lambda_2=0$).
But if $\Lambda_2\sim m_\rho$, then the quadratic term would be a
25\% correction at the largest values of $a$ used in the extrapolations,
and could not be ignored.
illustrated by 
For example, for the nucleon data in Fig. \ref{fig:aextrap}, 
such a quadratic term could lead to an extrapolated value
$m_N/m_\rho\approx 1.35$, in contrast to the result, $1.28$, from a linear fit.
Such large values for $\Lambda$ are not unreasonable---they have been seen
in the calculation of $B_K$ with staggered fermions \cite{Sharpelat93}.

The chiral extrapolations, such as those of Fig. \ref{fig:massextrap},
have been done linearly using the lightest three mass points.
Looking at the nucleon data, there is some evidence of negative curvature,
as has been also seen in other simulations.
As explained below, we expect there to be an $m_\pi^3$ term with a
negative coefficient. Including such a term would lead to a slightly
lower extrapolated value.

My point here is not that the extrapolations are necessarily wrong,
but that there are reasons to examine them more carefully,
in particular by accumulating data at more values of $a$, and
at smaller quark masses.

\begin{figure}[tb]
\centerline{\psfig{file=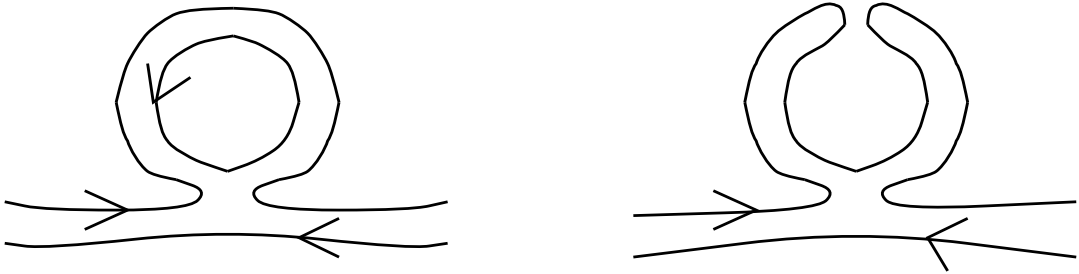,width=5.6truein}}
%
\caption{Quark diagrams leading to pion and $\eta'$ clouds around mesons.}
\label{fig:pioncloud}
\end{figure}

Let me now explain the reasons why I expect the spectra of QQCD and QCD
to differ by 10-20\%.
The essential point is that hadrons in QQCD have very different
``pion'' ($\pi$, $K$, $\eta$ and $\eta'$) clouds than in QCD.
To understand this, consider Fig. \ref{fig:pioncloud}, 
which shows ``quark-flow'' diagrams
for processes leading to pion loops. The justification for the use of
such diagrams is discussed in Refs. \cite{BG,Sharpelogs}.
The first diagram is present in QCD, but absent in QQCD.
The second is present in both theories, but in QCD it involves the
flavor singlet $\eta'$, which is heavy and thus does not contribute
significantly to the long-distance cloud. In QQCD, by contrast,
the $\eta'$ remains light, as discussed above, and this diagram leads
to the appearance of an $\eta'$ cloud around hadrons.
So what happens is that the pion cloud surrounding mesons in QCD 
is replaced by an $\eta'$ cloud around quenched. 
These two clouds are not related---the $\eta'$ cloud is 
simply a quenched artifact.
To the extent that particle properties in QCD depend upon the composition
of the cloud, they will be altered in QQCD.

The extent of the alteration can be investigated using 
(quenched) chiral perturbation theory.
One of the most striking results is that the chiral limit is more singular
in the quenched theory.
For example, the QChPT result for the pion mass is\cite{BG,Sharpelogs}
\beq
m_\pi^2 = 2 \mu m_q 
\left[ 1 - 2 \delta \ln(m_\pi/\Lambda) 
+ O(m_\pi^2 \ln(m_\pi/\Lambda')) \right] \,.
\label{eq:qupimass}
\eeq
Here $\delta$ is a constant proportional to the $\eta'$ two point
vertex in the right-hand diagram of Fig. \ref{fig:pioncloud}.
The term proportional to $\delta$ diverges as $m_\pi\to0$. 
This is contrast to the corrections in QCD, which are proportional to
$m_\pi^2 \ln(m_\pi)$, and vanish in the chiral limit.
Thus the $\eta'$ loops of QQCD introduce a sickness that is entirely an 
artifact. Indeed, the chiral expansion ceases to make sense once $m_\pi$
gets small enough that the ``$\delta$ term'' is comparable to 1.\footnote{%
For a pion made of degenerate quarks, one can resum the leading
terms,\cite{Sharpelogs,Sharpelogs2} 
i.e. those proportional to $(\delta \ln m_\pi)^n$.
It is not known how to do this in general.}
In fact, a general feature of QChPT is that
there are corrections which diverge as one or more quark masses are
sent to zero. This suggests that one cannot hope to use QQCD below a 
certain quark mass. This ``critical'' mass will likely
depend on the quantity being calculated. 

To estimate the critical mass one need to know $\delta$.
In QCD, the two-point vertex leads to the major part of $m_{\eta'}$,
the remainder due to quark masses.
Assuming the vertex is the same in QQCD as in QCD, one finds 
$\delta\approx0.2$.
In principle, one need not appeal to QCD, but rather can calculate the
$\eta'$ two-point function in QQCD and directly extract $\delta$.
This is a difficult quantity to measure, requiring propagators from
many sources. A calculation has been recently completed\cite{etaprimevertex},
however, albeit at a rather large lattice spacing ($a\sim 0.15$ fm).
The result, $\delta\approx0.15$, is close to the QCD-based estimate.
The smallness of $\delta$ justifies the use of perturbation theory,
and implies that critical masses turn out to be small.

\begin{figure}[tb]
\vspace{-0.25truein}
\centerline{\psfig{file=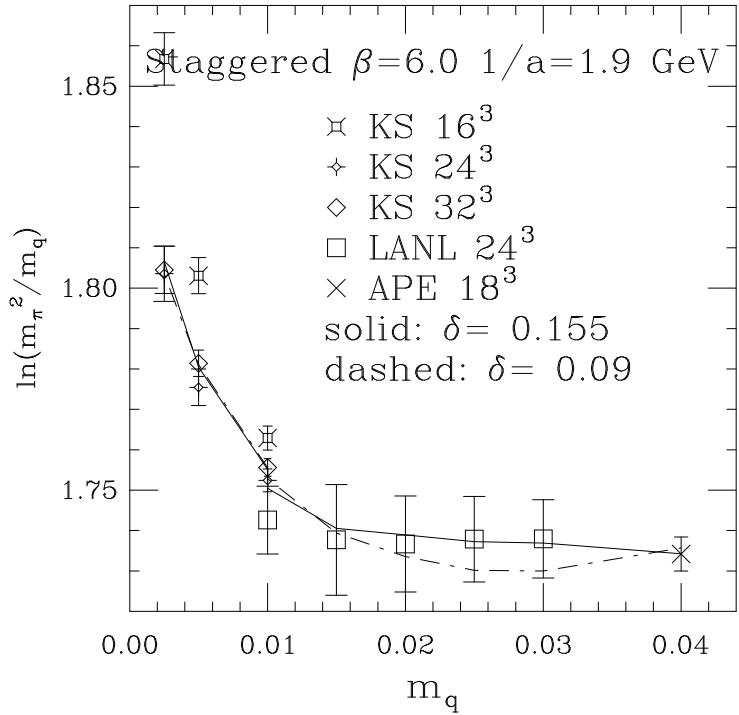,height=5truein}}
\vspace{-0.9truein}
\caption{Searching for the $\eta'$ cloud. All masses are in lattice units.
The two fits lead to the values of $\delta$ shown.}
\label{fig:qupimass}
\end{figure}

There is now some numerical support for QChPT.
Figure \ref{fig:qupimass} shows a test of Eq. \ref{eq:qupimass}.
I have taken this from the recent review of Gupta \cite{Rajanlat94},
which provides more details of the fits.
QChPT predicts that the ratio $m_\pi^2/m_q$ should diverge at small quark
masses. This can be tested using staggered fermions, for which
the quark mass is multiplicatively renormalized, and one knows where $m_q=0$.
The data at small quark masses are from Ref. \cite{Kim}, and come from
three lattice volumes. There is a pronounced finite volume effect betwee
$16^3$ and $24^3$ lattices, but no such effect from $24^3$ to $32^3$.
Thus the $32^3$ data is close to the infinite volume limit.
A technical point, which is crucial numerically, is that the mass appearing
in the logarithm in Eq. \ref{eq:qupimass} is that of a pion which is
not an exact PGB on the lattice. The best fit (solid line) gives a value
$\delta \approx 0.16$---other reasonable fits give different values,
but all are definitely non-zero.

A less striking, but equally important, test comes from decay constants
($f_\pi$, $f_K$ etc.)
Thes are also expected to diverge for mesons composed of non-degenerate
mesons when one of the quark masses vanishes.
The world's data for such decay constants is consistent with the expected
QChPT form, if $\delta=0.10(3)$\cite{Rajanlat94}, consistent with the
two other determinations.
To be complete, I should note that there is as yet no evidence 
for a divergence in $m_\pi^2/m_q$ for Wilson fermions \cite{Donlat93}.
It is, however, much more difficult to do the fit, since the quark mass
is additively renormalized, and one does not know {\sl a priori} 
where $m_q=0$.

Thus I have some confidence that QChPT makes sense, and is applicable
to present simulations.
What does it imply for baryon masses?
In QCD the chiral expansion of baryon masses is 
$m_N = m_0 + c m_\pi^2 +c' m_\pi^3 + O(m_\pi^4)$, 
where $m_\pi$ represents any of the PGB masses, and
$c,c'$ are constants. The non-analytic
$m_\pi^3\propto m_q^{3/2}$ term is due to loops of PGBs,
so-called ``chiral loops''. Its coefficient is known in terms of
the PGB couplings to the nucleon, e.g. $g_{\pi NN}$. 
The analytic corrections of $O(m_\pi^4)$, by contrast, involve additional,
unknown, parameters. But in the chiral limit, the non-analytic term is
enhanced relative to the analytic term by one power of $m_\pi$,
so one has some predictive power.
This is advantageous compared to mesons masses, for which
the chiral logs of size $m_\pi^4\ln m_\pi$ are enhanced only by
a logarithm over the analytic terms of $O(m_\pi^4)$.

In QQCD it turns out that for baryons, unlike for mesons, some,
though not all, of the chiral loops involving PGBs remain\cite{LS}.
Thus there is still an $m_\pi^3$ term, though with a different coefficient.
There are also new terms, due to $\eta'$ loops, which are proportional
to $\delta \times m_\pi$, 
and thus enhanced in the chiral limit. 
These are the analogs of the divergent terms in $m_\pi^2$,
and are pure artifacts.
Jim Labrenz and I have calculated the chiral expansion for octet and decuplet
baryon masses in QChPT\cite{LS}.
To give an example of the results, I use the QCD values for $g_{\pi NN}$
and other constants, and include only intermediate octets. We then find
(all masses in GeV)
\beq
m_N = m_0 - 0.35 (\delta/0.15) m_\pi + 3.4 m_\pi^2 - 1.5 m_\pi^3 + 
O(m_\pi^4 \ln m_\pi) \ .
\label{eq:quenchedN}
\eeq
The numerical values are only to be used as rough guides, since the
coupling constants in QQCD will not be the same as in QCD.
The most important points are qualitative: when plotting $m_N$ vs. $m_\pi^2$
there should be curvature at larger masses due to the $m_\pi^3$ term.
and there should be a peculiar 
behavior at small masses due to the $m_\pi$ term,

\begin{figure}[tb]
\vspace{-0.25truein}
\centerline{\psfig{file=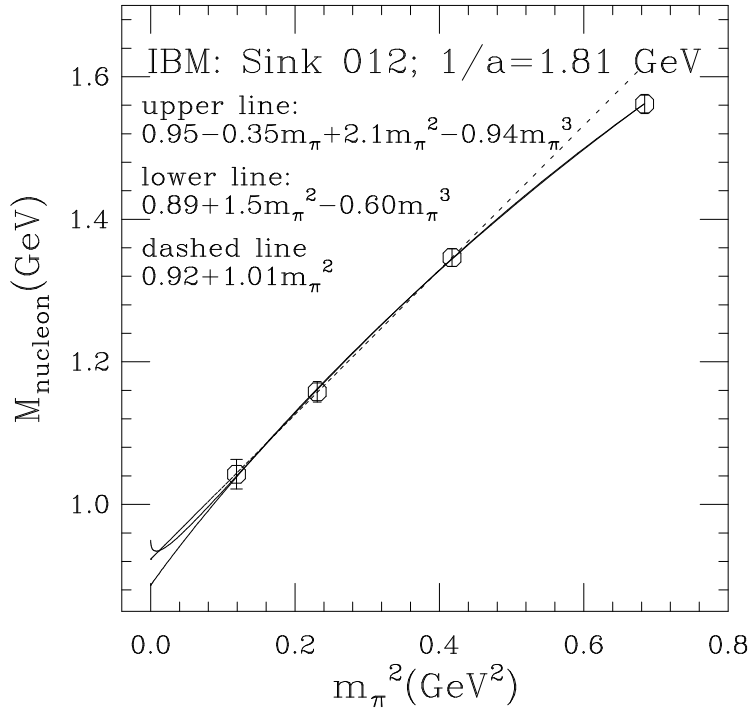,height=5truein}}
\vspace{-0.9truein}
\caption{Fits to the IBM data at $\beta=5.93$. All masses in GeV.}
\label{fig:baryfit}
\end{figure}

These results are illustrated in Fig. \ref{fig:baryfit}.
This shows the nucleon mass from Ref. \cite{IBMspect} at $\beta=5.93$.
I have done three types of fits: a fit of $m_0 + b m_\pi^2$ to the lightest
three points; and fits of Eq. \ref{eq:quenchedN} to all four points 
with $\delta$ fixed to be $0$ and $0.15$, but with all three other
coefficients free. The results of the fits are given on the plot.
All fits are reasonable, but the best are the two with curvature.
This is qualitative support of an $m_\pi^3$ term,
but is by no means definitive. 
The observation of curvature depends on heaviest mass point, and this is at
sufficiently large $m_\pi^2$ that higher order terms in the chiral
expansion could also be significant and give curvature \cite{Donlat93}.
The data provides no evidence for or against an $m_\pi$ term---as the 
Figure shows, the ``hook'' which appears for $\delta=0.15$ is too small to
be important unless one goes to very small quark masses.

In this instance, the variation in $m_0$ (the intercept) is quite small:
the linear fit to $m_\pi^2$ gives $m_0=0.92\,$GeV, 
while the fits with $\delta=(0,0.15)$ give $(0.89,0.95)\,$GeV.
Roughly, then, adding the $m_\pi^3$ term reduces $m_N$ by 3\%,
while the adding the $m_\pi$ term increases $m_N$ by 7\%.
These are small effects, but they are important given that they are
larger than the typical statistical errors.
It is not clear what to do about the $m_\pi$ term. Since it is an
artifact of quenching, it may be best to extrapolate ignoring it.
But one should certainly include an $m_\pi^3$ term.

I now return to the reason for this digression into QChPT.
We want to make an estimate of the effects of quenching on baryon masses.
We do this assuming that QCD and QQCD differ only because the contributions
of chiral-loops are different. In particular,
we assume that all coupling constants in QChPT are the same as in ChPT.
We apply this method to ratios of baryon masses, and to $m_N/f_\pi$.
Considering ratios removes changes in overall scale between QQCD and QCD.
We find that there are 10-30\% differences between ratios in 
the two theories\cite{LS}. 
This is certainly hand-waving---the coupling constants 
could conspire to make the two theories agree more closely on the various
ratios. But it indicates the magnitude of the typical effect due
to the difference in the physical composition of hadrons in QQCD and QCD.
It is because of this general argument that I expect the final quenched
spectrum to differ more substantially than 6\%.
We are in the process of extending this analysis to other data sets and
to the decuplet baryons. 

For more extensive reviews of QChPT,
see Refs. \cite{Rajanlat94,MaartenZakopane}.

\section{Anatomy of a calculation: $B_K$}

I now turn to applications of QQCD where we do not know the experimental
results in advance. We have already seen one example---$\alpha_S$; from 
now on I will focus on matrix elements of the electroweak 
effective Hamiltonian.
These so-called ``weak matrix elements''
govern weak decays and transition amplitudes.
I begin with a detailed discussion of the calculation of $B_K$, not only
because it is dear to my heart, but also because it clearly illustrates 
all aspects of such calculations.

$B_K$ arises when calculating CP-violation in $K-\bar K$ mixing,
which is parameterized experimentally by $\epsilon$.
This mixing is caused by box diagrams such as that in 
Fig. \ref{fig:BKcartoon}.
Using the renormalization group (RG), we integrate out the top quark, 
Z and W bosons, and then bottom quark, as well as gluons with momenta
exceeding the renormalization scale $\mu$.
We lower $\mu$ down to a scale at which we can match onto the
lattice calculation, $\mu\approx\pi/a \sim 5-10$ GeV.
At this stage the $\Delta S=2$, CP-violating part
of the effective Hamiltonian is, to good approximation
\beq
{\cal H}_{\rm eff}(\mu) \propto  G_F^2 {\rm Im}\left[ V_{ts}^2 V_{td}^2\right] 
c(\mu) \left[
\bar s \gamma_\mu(1\!+\!\gamma_5) d\
                      \bar s \gamma_\mu(1\!+\!\gamma_5) d \right] \,.
\eeq
Here $c(\mu)$ is a perturbatively calculable coefficient function,
known at present to two-loop order.
The non-perturbative part of the problem is the evaluation
of $\vev{\bar K|{\cal H}_{\rm eff}(\mu)|K}$, which 
is parameterized by $B_K(\mu)$ (Eq. \ref{eq:BKdef}). 
To evaluate this,
we switch to the lattice renormalization scheme,
matching the continuum four-fermion operator with a corresponding
lattice operator.
We then evaluate the matrix element of the lattice operator
using the numerical methods of lattice QCD,
and finally convert the result back into one for $B_K(\mu)$.
Combined with $c(\mu)$ and the other constants one obtains an expression
for $\epsilon$, from which one can extract the value of
${\rm Im}\left[ V_{ts}^2 V_{td}^2\right]$. 

I would like to stress a point which occasionally gets overlooked.
What the lattice calculation gives us, once we match back
onto the continuum, is $B_K(\mu)$ in a continuum scheme of our choice
(e.g. naive dimensional regularization)
at a scale $\mu$ which should be near to $\pi/a$. We can choose $\mu$ to
be a standard scale, e.g. 2 GeV. This result
contains all sorts of lattice related errors, discussed in detail below.
For phenomenological applications, we must combine it with $c(\mu)$, the
result of a perturbative calculation. 
This quantity has errors due to truncation of the RG equations,
and due to the uncertainty in the value of $\alpha_s$ 
(or equivalently the value of $\Lambda_{\rm QCD}$).
These errors have nothing to do with the lattice calculation.\footnote{%
There is a small correlation in the errors, since the 
lattice-to-continuum matching coefficients depend on $\alpha_s$, but this
is a minor effect.}
Since $c(\mu)\propto \alpha_s^{-6/25}(\mu)$ at {\em leading order},
it has become standard to quote results for
$\BKhat = \alpha_s^{-6/25}(\mu) B_K(\mu)$.
I do not like this practice, for various reasons.
The most important is that it mixes up errors from different sources.
In addition, $\BKhat$ is only $\mu$ independent at leading order,
so it is not what one actually uses in a two-loop phenomenological analysis.
And, finally, using $\BKhat$ amounts to running $B_K$
down to a very low scale, that at which $\alpha=1$, where I, at least,
have little intuition for the physics.
I propose, instead, that we quote $B_K(\mu)$ for a standard
scheme and scale, just as we do for $\alpha_S$.
Other methods of calculation (large $N_c$, QCD sum rules, \dots)
give $B_K$ at different scales, and will have
to be run to the standard scale. If the change in scale is small, however,
the uncertainties introduced by the running will also be small.
In this way we can make comparisons between models
without the overall common errors in $c(\mu)$.
To do phenomenology, we
can take $c(\mu)$ from the one of the standard RG analyses.

With that off my chest, let me return to the issue of how we calculate
$B_K$, and in particular, how we estimate and reduce the errors.
The sources of errors are these.
\begin{itemize}
\item
Numerical method. Statistical errors are now at the 1-2\% level.
\item
Matching continuum and lattice operators. With staggered fermions,
the errors from neglecting two- and higher loop terms in the
matching are small, 1-2\%.
\item
Making sure the result has the correct behavior in the limit
$m_K\to 0$. This is much simpler using staggered than Wilson fermions.
\item
Extrapolating to the physical kaon (containing a highly non-degenerate
$\bar s$ and $d$) from the lattice kaon (with degenerate, or nearly
degenerate, quarks of mass $\approx m_s/2$).
\item
Finite volume effects. It turns out that these can both be estimated
theoretically to be very small ($< 0.5\%$), and are numerically 
observed to be smaller than the statistical errors on lattices of
size $1.6-2.4$fm across.
\item
Errors due to the use of the quenched approximation.
\item
Extrapolating to $a=0$---possibly using improved actions.
\end{itemize}
I discuss the most important of these in turn.

\subsection{Numerical method}

\begin{figure}[tbh]
\centerline{\psfig{file=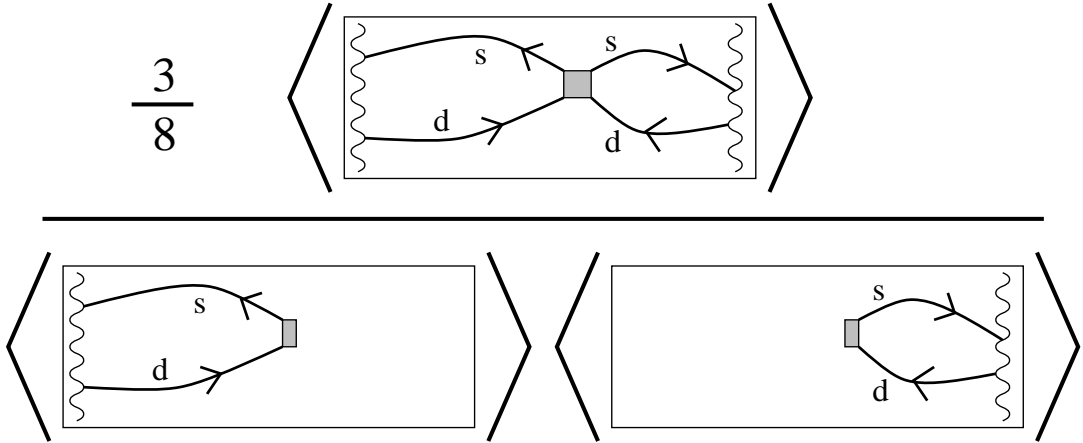,width=5.6truein}}
%
\caption{Schematic depiction of the method used to calculate $B_K$.}
\label{fig:BKmethod}
\end{figure}

Staggered fermions give the most accurate numerical results for $B_K$,
and I will describe how we (Rajan Gupta, Greg Kilcup and I) do the
calculation. For more details, see Refs. \cite{book,hmks}.
We begin by expressing $B_K$ as a ratio
\beq
B_K = {\vev{\overline{K}(\vec p=0)|\bar s \gamma_\mu(1\!+\!\gamma_5) d\
                 \bar s \gamma_\mu(1\!+\!\gamma_5) d|K(\vec p=0)} \over
	{8\over3}
	\vev{\overline{K}(\vec p=0)|\bar s \gamma_4\gamma_5 d|0} \
	\vev{0|\bar s \gamma_4\gamma_5 d|K(\vec p=0)} } \,,
\label{eq:BKasratio}
\eeq
Each of the operators in this expression is a shorthand for the lattice
operator which results from matching with the continuum.
Figure \ref{fig:BKmethod} shows schematically how
we did the calculation in Ref. \cite{BKprl}. Since then we have changed
the method slightly, but not essentially \cite{BKlat90}. 
The vertical and horizontal directions represent, respectively,
space (3-dimensional in practice) and Euclidean time. 
The expectation values indicate the functional integral
over configurations weighted, in QQCD, by the gauge action. 
In practice this means an average over some number of configurations
generated with the correct measure.
The lines with arrows are quark propagators, 
and the boxes represent the lattice bilinear and quadrilinear operators. 
Finally, the wavy lines at the edge are ``wall sources''.
These create the quark (or antiquark)
with equal amplitude across the entire timeslice, and thus ensure that
the mesons which are created have $\vec p=0$. Gauge invariance is 
maintained by first fixing the source timeslices to Coulomb gauge.
The boundary conditions are periodic in space, but Dirichlet in time,
so that the propagating quarks can bounce off the ends of the box, but
not propagate through them.

\begin{figure}[tb]
\vspace{-0.2truein}
\centerline{\psfig{file=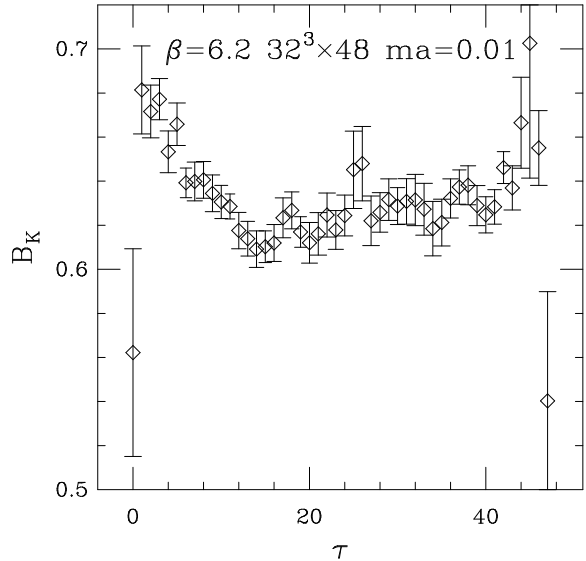,height=4.truein}}
\vspace{-0.8truein}
\caption{Data for $B_K$ (with tree-level matching). The quark mass is such
that the lattice kaon is slightly heavier than the physical kaon.
Sample of 23 configurations.}
\label{fig:BKresults}
\end{figure}

We use wall sources because they give us the freedom to insert
any operator we wish for the ``boxes'' in the diagram. 
For example, the operators
can involve quarks and antiquarks at slightly different lattice sites.
This freedom turns out to be crucial for staggered fermions, 
because (as explained below) the operators we want to use
involve $\bar q$ and $q$ at different positions.
More generally, it allows us to test that different discretizations of
the continuum operator yield the same results.
The disadvantage of wall sources is that (in the way we implement them
\cite{hmks}) they create not only kaons, but also $K^*$'s and other
excited strange mesons. These give contributions which fall off like
$\exp[-(m_{K^*}-m_K) \tau]$, where $\tau$ is the distance from the source.
Thus if one is far enough away from both sources only the kaon contributes
to the matrix element. 
I give an example of our data in Fig. \ref{fig:BKresults}.
This shows the ratio of Eq. \ref{eq:BKasratio} as a function of the timeslice
on which the operator resides. The operator has been summed over all
space, which significantly improves the signal, ans is another advantage
of the wall sources.
The desired signal is independent of $\tau$, since for all $\tau$ a particle
of mass $m_K$ (either a $K_0$ or a $\bar{K_0}$) propagates the length
of the lattice. 
Edge effects due to excited states and particles
bouncing off the boundary are apparent, but there is a ``plateau''
covering a considerable
number of timeslices from which to extract the signal. We improve
the statistics further by averaging over this plateau, although,
since the results at different times are correlated, the improvement is not
as significant as one would naively expect.

\begin{figure}[tb]
\vspace{-0.2truein}
\centerline{\psfig{file=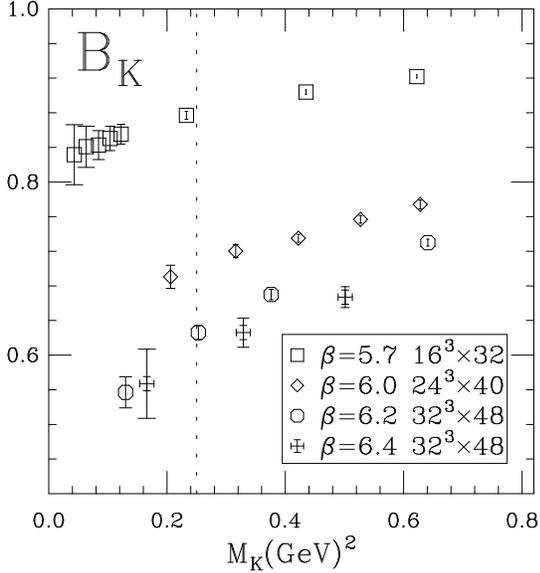,height=4truein}}
\vspace{-0.8truein}
\caption{Results for $B_K$ using tree-level matched operators.
The lattice spacings are roughly $0.2$, $0.105$, $0.08$ and $0.06$ fm
as one descends the plot.}
\label{fig:BKbare}
\end{figure}

To give you a feel for how the analysis proceeds, 
I show in Fig. \ref{fig:BKbare}
the results at four lattice spacings, and for a variety of values of $m_K$.
These are for operators matched to the continuum only at tree level.
The kaon mass has been converted to physical units using the hadron
spectrum to set the scale \cite{Sharpelat91}.
The dashed vertical line shows the value of the physical kaon mass.

The statistical errors in the results are small,
in part due to the use of a ratio to calculate $B_K$.
The small errors allow us to clearly observe the following features:
\begin{itemize}
\item
$B_K$ has a smooth dependence on $m_K^2$, and, apparently,
a finite chiral limit. 
\item
There is {\em no need to extrapolate to get to the physical kaon mass}.
Quenched particles with the kaon mass can be simulated directly on
present lattices. Of course, this is a cheat, since
the results in the figure are mostly for degenerate quarks, and
one must extrapolate to the non-degenerate case. 
\item
There is a clear and significant dependence on lattice spacing.
\end{itemize}

\pagebreak
\subsection{Matching continuum and lattice operators}

Lattice operators are chosen so that, at tree level, 
they have the same matrix elements as the continuum operators,
for momenta much lower than the cut-off ($p\ll\pi/a$).
With Wilson fermions it is easy to find such operators.
A continuum bilinear or four-fermion operator is discretized into a
lattice operator having exactly the same form, 
with all fermion fields on the same lattice site.
The tree-level matrix elements are the same as those of the continuum
operator for all the momenta that are available on the lattice.

\begin{figure}[tb]
\centerline{\psfig{file=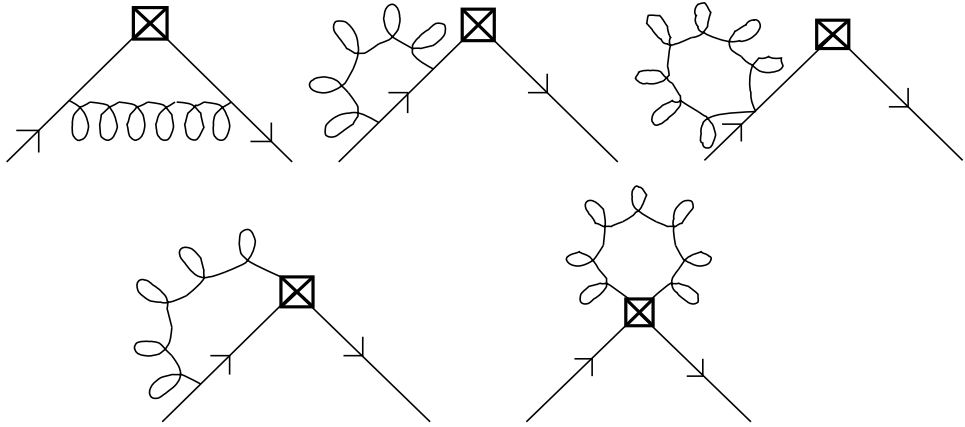,width=5truein}}
%
\caption{Diagrams contributing to one-loop matrix elements needed for 
matching.}
\label{fig:1loop}
\end{figure}

The matrix elements of lattice and continuum operators do not, however,
agree when one includes loops.
Examples of one-loop diagrams are shown in Fig. \ref{fig:1loop}.
Lattice propagators and vertices differ from their continuum 
counterparts when the momentum in the loop is of $O(1/a)$.
We have already seen this difference for fermion propagators---compare 
Eq. \ref{eq:wilsonprop} with the continuum propagator.
It is true also for gluon propagators ($D^{-1}=\sum_\mu 4 \sin(k_\mu/2)^2$
versus $k_\mu^2$), and the quark-gluon vertex.
To do the one-loop matching, one must add parts to the lattice operator,
proportional to $g^2$, so as to make the matrix elements agree.
These additional terms are finite, because they come from short distances
and the lattice integrals have an ultraviolet cut-off.
Since the momenta involved are $\sim \pi/a$, 
the corrections should be reliably calculable using perturbation theory,
as long as $a$ is small enough.

Let me mention a subtle point that one tends to forget when doing 
the matching calculations.
It is not sufficient that perturbation theory be reliable
at the scale $\pi/a$. One actually needs the stronger condition
that it be reliable down to $(0.5-1)/a$.
This is because, to do the matching, one must, in principle,
compare matrix elements with external Euclidean
momenta satisfying $|p| \gg \Lambda_{\rm QCD}$.
This is necessary to avoid
the infra-red region where perturbation theory breaks down.
But one also must have $pa$ small enough that lattice artifacts in
the matrix elements are small. Typically this remains true for $pa < 0.5-1$.
In practice, at one-loop one uses a gluon mass to regulate the
infra-red divergences, and sets the external momenta to zero.
This is adequate, because in the matching calculation, the infra-red
contributions are identical and cancel.
This point is emphasized in Ref. \cite{nonpertsub}.

One-loop continuum calculations are straightforward.
The corresponding lattice integrations are, however, a mess, and
are evaluated numerically.
I can personally attest that,
soon after beginning such a calculation, one asks oneself the questions:
Is this matching really necessary?
Can't we just use the lattice regularization throughout?
Unfortunately, the answers are yes and no, at least for the moment.
The point is that the electroweak theory has a chiral representation of
fermions, so its discretization is problematic.
The chirality is not an obstacle to discretizing a left-handed operator 
such as that in $B_K$ (Eq. \ref{eq:BKasratio}), 
because only the even-parity part of the operator contributes, 
which is the average of left-handed and right-handed operators.
Incidentally, it is possible to directly match from the electroweak
theory including weak bosons to the lattice effective Hamiltonian,
and then run down to a low scale on the lattice.

The result of one-loop matching takes the general form
\beq
\CO^{\CONT}_i(\NDR, \mu) = \CO^{\LATT}_i  + {g^2\over 16\pi^2} 
\sum_j \left(\gamma_{ij}^{(0)} \ln(\pi/\mu a) + c_{ij}\right)  \CO^{\LATT}_j
+ O(g^4) + O(a)\ .
\label{eq:contlatt}
\eeq
Here $i$ is a set of continuum operators which mix with each other
under the continuum RG. To define these operators, which in general contain
$\gamma_5$, one must pick a renormalization scheme as well as a 
renormalization scale. One of the standard schemes is naive dimensional
regularization (NDR). The set of lattice operators which are required for
matching is, in general, larger than the set of continuum 
operators. \footnote{%
The appearance of extra operators is not peculiar to the lattice.
Even in the continuum, when one uses dimensional regularization, extra
``evanescent'' operators are needed at intermediate stages of
the calculation, associated with the additional $-2\epsilon$ dimensions.}
Thus the finite matrix $c_{ij}$ is rectangular.
The anomalous dimension matrix $\gamma^{(0)}_{ij}$ is, however, square.
It governs the dependence of the continuum operators on $\mu$.

Let me give a concrete example, relevant to the present subject.
In order to simplify the allowed Wick contractions 
consider four-fermion operators composed of four distinct flavors
\beqn
\CO_{LL} &=& \half \left[
\psibar_1 \gamma_\mu (1\!+\!\gamma_5)\psi_2\
\psibar_3 \gamma_\mu (1\!+\!\gamma_5)\psi_4
+ (2 \leftrightarrow 4) \right] \\
{\cal S} &=& \half \left[
\psibar_1 \psi_2\
\psibar_3 \psi_4
+ (2 \leftrightarrow 4) \right] \\
{\cal P} &=& \half \left[
\psibar_1 \gamma_5 \psi_2\
\psibar_3 \gamma_5 \psi_4
+ (2 \leftrightarrow 4) \right] \ , {\rm etc.}
\eeqn
The result of matching for Wilson fermions is \cite{martinelli}
\beqn
\CO_{LL}^\CONT &=& \left[ 1 + 
{g^2\over 16\pi^2} (-4 \ln({\mu a\over\pi}) - 54.753)
\right] \CO_{LL}^\LATT \\
&&\mbox{}\ + {g^2\over 16\pi^2} \left[
c_s {\cal S} + c_p {\cal P} + c_t {\cal T} + c_v {\cal V} + c_a {\cal A} 
\right] + O(g^4) \,.
\eeqn
The most important feature of this result is the appearance of
lattice operators having all possible tensor structures.
In the continuum, the matrix elements of $\CO_{LL}$ are constrained
by chiral symmetry to vanish as $m_K^2$ in the chiral limit.
This is not true for the matrix elements of $\CO_{\cal S}$, $\CO_{\cal P}$,
etc, which couple to both LH and RH quarks.\footnote{%
The only exception is ${\cal V}+{\cal A}$, which has the same 
positive parity part as $\CO_{LL}$, and does have vanishing matrix elements
in the chiral limit.}
This problem is due to the breaking of chiral symmetry by the Wilson
fermion action---even though the breaking is $O(a)$ at tree-level, divergent
loops give factors of $1/a$ leading to finite contributions at one-loop.
The consequence is that if matching is not done exactly, 
to all orders in $g^2$, 
matrix elements will not have the correct continuum chiral behavior.
In addition, there will be lattice artifacts, proportional to $a$, 
with the wrong chiral behavior.
Here we have the doubling problem coming back to haunt us.
This proves to be a significant obstacle in practice, and, because of this,
it is preferable to calculate $B_K$ using staggered fermions. 
Probably the only hope for similar accuracy with
Wilson fermions is to use non-perturbative matching\cite{nonpertsub}.


I want to make two general comments about matching. 
As we have seen, lattice calculations must be combined with continuum
coefficient functions obtained from RG equations. To be consistent,
if one uses 1-loop matching, one must use 2-loop RG equations.
To see this, recall the solution to the 2-loop RG equation for the
case of an operator which does not mix.
Running from a heavy scale $m_H$ down to $\mu$, one finds
\beq
c(\mu) = c(m_H) 
\left[ g^2(m_H) \over g^2(\mu) \right]^{\gamma^{(0)}/2\beta_0}
\left[ 1 + {g^2(m_H) - g^2(\mu) \over 16 \pi^2} 
\big({\gamma^{(1)}\over2\beta_0} - {\gamma^{(0)}\beta_1 \over 2\beta_0^2}\big) 
+ O(g^4) \right] \ ,
\label{eq:coeffunction}
\eeq
where $\gamma^{(n)}$ and $\beta_n$ are the $(n\!+\!1)$-loop 
contributions to the anomalous dimension and $\beta$-function, respectively.
The term in the rightmost parenthesis proportional to
$g^2(\mu)$ is of the same form as a one-loop matching correction.
To be consistent, one must include this term in $c(\mu)$.
Since it contains $\gamma^{(1)}$ and $\beta_1$, it comes from two-loop running.

My second comment is that the perturbative matching can be done equally
well in QCD and QQCD. Indeed, at one-loop the two theories give the
same results for fermionic operators, since fermion loops do not enter.
The issue does arise, however, when picking the scale
at which to evaluate $c(\mu)$. The product of $c(\mu)$ calculated in QCD,
with $B_K(\mu)$ evaluated in QQCD, is not independent of $\mu$ because
$\beta_0$, $\beta_1$ and $\gamma_1$ receive contributions from quark loops
and thus differ in the two theories. One must simply guess a value of $\mu$
for which it seems reasonable that QQCD will do the best job of imitating QCD.

\subsection{Staggered fermions and chiral behavior}

I now give a lightning review of the essentials of staggered fermions.
For more details see the two texts, or my articles explaining in detail
how and why one uses staggered fermions to calculate matrix elements
involving PGB\cite{book}.

We begin with naive fermions
\beq
- S_N = \sum_{n\mu} \half \psibar_n \gamma_\mu
\left[\Unmu \psi_{n+\mu} - U_{n-\mu,\mu}^{\dagger} \psi_{n-\mu}\right]
+ \sum_n m \psibar_n \psi_n \ ,
\eeq
and perform a change of variables, known as 
``spin-diagonalization'' \cite{kawamoto},
\beq
\label{eq:Gammaeq}
\psi(n) = \g{n}^{\vdag}\chi(n)\ ,\ \  
\bar\psi(n) = \bar\chi(n) \g{n}^{\dag}\ ,\ \
\g{n}^{\vdag} = \g1^{n_1} \g2^{n_2} \g3^{n_3} \g4^{n_4} \,.
\eeq
Note that $\g{n}$ depends only on ${\rm mod}_2(n_\mu)$.
The result is
\begin{equation}
\label{staggaction}
- S_N = \sum_n \left[ \bar\chi_n \sum_\mu 
\frac12\eta_\mu(n)\left( \Unmu \chi_{n+\mu} 
                       - U_{n-\mu,\mu}^{\dagger}\chi_{n-\mu}\right)
+ m \bar\chi_n \chi_n \right] \,.
\end{equation}
The gamma matrices have been replaced by the phases (thus
the name of the transformation) 
\begin{equation}
\eta_\mu(n) = (-)^{\sum_{\mu<\nu} n_\nu} \,.
\end{equation}
Note that the four spinor components of $\chi$ 
are only coupled by gluon exchanges.
Staggered fermions result from simply deleting three of the four components of
$\chi$, leaving a one component fermion on each site.
This reduces the number of continuum fermions from 16 to 4.

Staggered fermions represent four degenerate flavors in the continuum limit,
but at finite lattice spacing the $SU(4)$ flavor symmetry is broken.
The only continuous symmetry of the action Eq. \ref{staggaction} is 
that corresponding to fermion number
\begin{equation}
\chi(n) \to \exp[i\theta_V] \chi(n) \ ,\quad
\bar\chi(n) \to \bar\chi(n) \exp[-i\theta_V] \ .
\end{equation}
In the massless limit, the continuum symmetry enlarges to 
$SU(4)_L\times SU(4)_R$, while the lattice action has an additional symmetry
\begin{equation}
\chi(n) \to \exp[i(-)^n\theta_A] \chi(n)\ ,\ 
\bar\chi(n) \to \bar\chi(n) \exp[i(-)^n\theta_A] \ ,
\end{equation}
where $(-)^n=(-)^{n_1+n_2+n_3+n_4}$.
This turns out to be a flavor non-singlet axial symmetry in the
continuum limit, and we refer to it as $U(1)_A$. 
It is very important as it guarantees that mass is multiplicatively
and not additively renormalized (so that setting the lattice parameter
$m$ to zero really means massless quarks).
It also guarantees that $B_K$ has the correct chiral behavior 
\cite{toolkit,wius,book}.
The remaining symmetries are 
translations, rotations, reflections and charge conjugation---the
discrete residue of flavor and Poincar\'e symmetries.

To use staggered fermions, we must learn how to
identify the continuum spin and flavor transformation properties of 
fields constructed from $\chi$ and $\chibar$.
To do this requires some notation.
We denote quark fields in a continuum theory with four degenerate fermions
using upper case letters, e.g. $Q_{\alpha,a}$. 
These have a spinor index (here $\alpha$) and a flavor index 
(here $a$), both running from 1 to 4.
Quark bilinears in this theory are of the general form
\beq
\bar Q_{\alpha,a} \gamma_S^{\alpha\beta} \xi_F^{ab} Q_{\beta,b} 
\ \equiv \ \Qbar (\g{S}\otimes\xi_{F}) Q\ .
\label{eq:contbilin}
\eeq
$\gamma_S$ determines the spin of the bilinear, $\xi_F$ the flavor.
The Dirac matrices are labeled using the notation of 
Eq. \ref{eq:Gammaeq}, except that now $S_\mu$ is one of the 16 ``hypercube'' 
vectors having components which are either 0 or 1.
It is convenient to label the possible flavors in a similar way
(rather than using the generators of the group $SU(4)$), 
using the complex conjugate matrices $\xi_F = \gamma_F^*$ \cite{daniel}.

With staggered fermions, the discretization of the continuum field $Q$ is
spread over the 16 positions of a $2^4$ hypercube\cite{kluberg}.
Divide the lattice into such hypercubes (choosing one of the sixteen
alternatives), labeled by the positions of their corners, $y=(0,0,0,0)$, 
$(2,0,0,0)$, $(2,2,0,0)$, etc. Label positions within hypercubes by
a hypercube vector $A_\mu$. Then the $4\times4$ matrix field
\beq
\chi_Q(y)_{\beta,b} = \frac18 \sum_A \chi(y+A) (\gamma_A)_{\beta,b} \ ,
\label{eq:Qlat}
\eeq
corresponds to the continuum field $Q$ as $a\to0$.
In particular, the discretization of the continuum bilinears 
is obtained by substituting $\chi_Q$ for $Q$ in Eq. \ref{eq:contbilin}.
In general, these involve $\chibar$ and $\chi$ fields at different
sites on the underlying lattice, and so must be made gauge invariant.
This is usually done either by joining quark and antiquark by an
appropriate string of link matrices, or by fixing to a smooth gauge,
such as Landau gauge.
In the new notation the axial symmetry is
\beq
\chi_Q \rightarrow \exp\left[i \theta_A (\g5\otimes\xi_5)\right] \chi_Q \ ,
\eeq
which shows that it is a flavor non-singlet.

To write the action in this new notation requires extending it to
include derivatives. There are various choices. To study the
corrections due to discretization, it is convenient to 
keep the invariance of the action under translations by a single lattice
spacing as transparent as possible. A notation with this feature is given in
Ref. \cite{SP}. The action is simply
\beq
\sum_y \chi_Q(y)[(\g\mu\otimes 1)D_\mu + m] \chi_Q(y) \,,
\eeq
which looks just like the continuum action.
The complications of the lattice are hidden in the definition
of $D_\mu$, which mixes up $\chi(y)$ with the field in adjacent
hypercubes. For this reason the action is $SU(4)$ invariant only
in the continuum limit. The recovery of this invariance can be made
more explicit using a form of the action written entirely in terms of
hypercube fields \cite{kluberg}, but I do not display this result here
as it will not be needed in the ensuing discussion.

We are now in a position to understand how to find a lattice operator
which matches onto the continuum operator needed to calculate $B_K$, i.e.
\beq
\CO_\CONT = \sbar_a \g\mu d_a\ \sbar_b\g\mu d_b
          + \sbar_a \g\mu\g5 d_a\ \sbar_b\g\mu\g5 d_b 
\label{eq:contop}
\eeq
(color indices now explicit).
Staggered fermions show their dark side in such a matching calculation. 
We choose to introduce a different staggered fermions for each continuum quark.
This means that the lattice theory does not become QCD in the continuum limit,
but rather a related theory in which there are 
{\em four versions of each quark}.
Thus the matching really proceeds in two steps \cite{book}. 
(1) Find operators in the extended theory whose matrix elements
involve the same {\em contractions} as does 
$\CO_\CONT$ between a physical $K$ and $\bar K$.
In this step we divide out the effect of the extra fermions by hand. 
(2) Find the lattice operators which match onto the operators 
in the extended theory needed in step (1).

\begin{figure}[hbt]
\centerline{\psfig{file=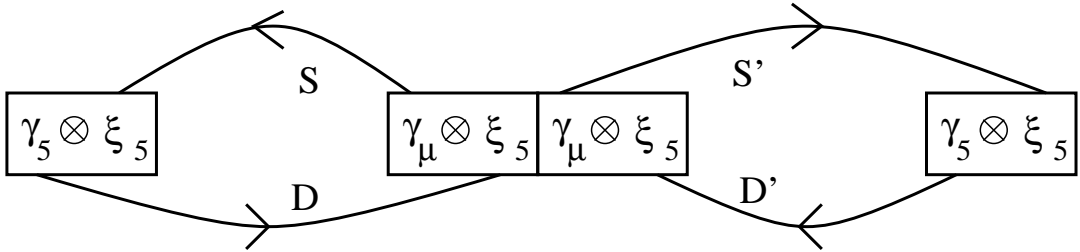,width=5.6truein}}
%
\caption{The two spinor loop contraction of the vector operator.}
\label{fig:twoloops}
\end{figure}

Step (1) can be done in many ways. Our choice is to use the following
operator in the extended theory (for a different choice see Ref. 
\cite{leeklomfass})
\beqn
\CO'_\CONT &=& \CA^I + \CA^{II} + \CV^I + \CV^{II} \\
\CA^I &=& 
   \Sbar_a (\g\mu\g5\otimes\xi_5)D_b\ \Sbar'_b (\g\mu\g5\otimes\xi_5)D'_a \\
\CA^{II} &=& \frac{1}{16}
   \Sbar_a (\g\mu\g5\otimes\xi_5)D_a\ \Sbar'_b (\g\mu\g5\otimes\xi_5)D'_b \\
\CV^I &=& \frac{1}{16}
   \Sbar_a (\g\mu\otimes\xi_5)D_b\ \Sbar'_b (\g\mu\otimes\xi_5)D'_a \\
\CV^{II} &=& \frac{1}{16}
   \Sbar_a (\g\mu\otimes\xi_5)D_a\ \Sbar'_b (\g\mu\otimes\xi_5)D'_b \ .
\eeqn
This must be sandwiched between an initial kaon created by the source
$\Dbar' (\g5\otimes\xi_5) S'$, and a final $\bar K$ destroyed by
$\Dbar (\g5\otimes\xi_5) S$.
The primes are a device to restrict Wick contractions so that
the spin and flavor indices are contracted in two-loops, as
sketched in Fig. \ref{fig:twoloops}.
This is necessary because Fierz transformations are completely different
in the four-flavor theory from those in QCD.
It is simple to convince oneself that the Wick contractions of $\CO'_\CONT$
in this particular matrix element are the same as those of $\CO_\CONT$
between a physical $K$ and $\bar K$, the factors of 16 canceling the
contributions from the extra quarks.

To carry out this first matching step, we could just as well use
any other flavor $\xi_F$ in place of $\xi_5$, 
as long as we change the flavor in both the operator and the external sources.
Our choice of flavor $\xi_5$ is only important once we have matched
onto lattice operators. The point is that the $U(1)_A$ symmetry
has flavor $\xi_5$, so that the lattice PGB arising from 
chiral symmetry breaking has this flavor. Only the matrix elements
involving this pion satisfy the Ward Identities which are needed to show that
the chiral behavior is the same as for the continuum operator \cite{toolkit}.

Step (2) must be done perturbatively.
At tree level, the result
is an operator of exactly the same form, written in terms of
the lattice fields $\chi_S$, $\chi_D$, $\chi_{S'}$ and $\chi_{D'}$.
At higher orders, flavor symmetry breaking in the lattice action 
means that matching introduces many of the 65,536 operators of the form
\beq
\chibar_S (\g{S}\otimes\xi_{F}) \chi_D \ 
\chibar_{S'} (\g{S'}\otimes\xi_{F'}) \chi_D\ !
\eeq
Fortunately, at one-loop, we need only consider the subset of these
operators with flavor $\xi_5$. The contributions of operators with
other flavors are suppressed either by powers of $a$ or $g^2$. 
Higher loop calculations would likely be a nightmare,
as many more operators must be included at intermediate stages.

\begin{figure}[tb]
\centerline{\psfig{file=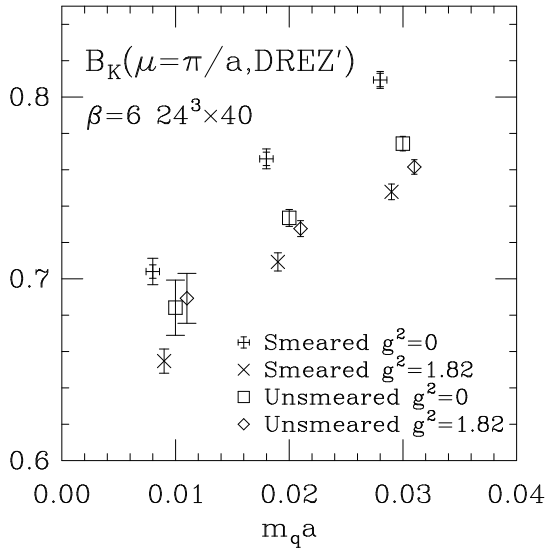,height=3truein}}
%
\caption{Tree-level ($g^2=0$) vs. one-loop ($g^2=1.82$)
matching for Landau gauge $2^4$ (unsmeared) and 
$4^4$ (smeared) operators.}
\label{fig:pertcorr}
\end{figure}

The one-loop result for operators made gauge-invariant by fixing
to Landau gauge is \cite{Ishi,SP} 
\beqn
\CO'_\CONT(\NDR,\mu) &=& \CO'_\LATT
\left( 1 + {g^2\over 16\pi^2} [4\ln{\pi\over\mu a} - {4\over3}]\right) \\
&&\mbox{} + {g^2\over 16\pi^2} \left[ 
10.7 \CA^I_\LATT + 1.7 \CA^{II}_\LATT -10.1 \CV^I_\LATT - 7.5\CV^{II}_\LATT
\right]  \ .
\eeqn
The calculations have also been
done for two other types of operator: gauge invariant $2^4$ operators
with gauge links joining the $\chibar$ and $\chi$ fields\cite{Ishi}, and
``smeared'' Landau gauge operators, which are spread across a $4^4$ hypercube
in such a way that their tree-level matrix elements have no terms
of $O(a)$\cite{PS}.
The corrections turn out to be small,
particularly for the $2^4$ and $4^4$ Landau gauge operators.
For the former, this is apparent from the above result.
Recalling that $g^2/(16\pi^2)\approx1/80$, 
typical corrections are seen to be $\sim 10\%$.
In fact, there are cancellations, so that the overall correction is
only a few percent. The corrections are nevertheless significant, as they are
larger than the statistical errors. This is illustrated in Fig.
\ref{fig:pertcorr}, which shows an example of
the difference between tree-level and one-loop matching for 
$B_K$\cite{Sharpelat93}.

When using the matching equation, one must choose a value of $g^2$,
i.e. choose both a scheme and a scale. (Note that the scale in $g^2$
{\em need not be the same as} $\mu$.)
When making these choices one is, in effect, guessing the higher order terms.
As discussed above, a physical coupling scheme such as $\MSbar$ 
leads, in general, to more rapid convergence.
As for the scale, a reasonable choice would be to use the BLM 
prescription\cite{BLM} (the simpler scheme of Ref. \cite{LM} does
not apply to a matching calculation). 
Unfortunately, this requires knowledge of the parts of the
two-loop graphs involving fermion loops, 
which have not yet been calculated.
So one is left to make reasonable guesses.
We make two choices: $\mu=\pi/a$, the lattice cut-off,
and $\mu= 2 GeV$, the scale at which we quote our result. We use
the average of the results from these two choices as our central value,
and half the range as an estimate of the systematic error. The resulting
error is small, $\sim 2\%$. 

\subsection{Chiral perturbation theory}

ChPT allows one to predict the expected behavior of $B_K$.
At leading order, the operator $\CO_\CONT$ (Eq. \ref{eq:contop}) 
is represented in the effective theory of PGB's by
\beq
\frac43 f^4 B (\Sigma \partial_\mu \Sigma^{\dagger})_{ds}
(\Sigma \partial_\mu \Sigma^{\dagger})_{ds} \ .
\eeq
(For the notation see, for example, the lectures of de Raphael.)
$f$ is the pion decay constant at leading order (in the normalization
that $f_\pi=93$ MeV),
while $B$ is the leading order value for $B_K$. 
ChPT does not give any information on $B$, but does determine the
form of the chiral expansion, and the normalization of the non-analytic
contributions from chiral loops. One finds 
(with $m_u=m_d$)\cite{Sharpelogs}
\beq
B_K = B \left[ 1 - (3 + {\epsilon^2\over3}) y\ln y + b y + c y \epsilon^2 
+ O(y^2) \right] \ ,
\label{eq:bkchpt}
\eeq
where $y=m_K^2/(4\pi f)^2\approx 0.2$ is the usual chiral expansion parameter,
while $\epsilon=(m_s-m_d)/(m_s+m_d)$ measures the non-degeneracy of the
$s$ and $d$ quarks. $b$ and $c$ are unknown constants.
The important features of this result are
\begin{itemize}
\item
$B_K$ is predicted to have a finite chiral limit. This means
(given the definition Eq. \ref{eq:BKdef}) that the matrix element
of $\CO_\CONT$ vanishes as $m_K^2$ in the chiral limit.
This can be traced back to the fact that the operator contains only
left-handed fields. It is this behavior which is maintained on the
lattice when using staggered fermions, but not with Wilson fermions.
\item
The dependence on $\epsilon$ is non-leading, suppressed by $y\ln y$.\footnote{%
That the dependence is quadratic in the mass difference follows from
the ``CPS'' symmetry introduced in Ref. \cite{bernardetal}.}
Thus there should not be a large change in $B_K$ when
changing from a theory in which $m_s(\lat)=m_d(\lat)=m_s(\phys)/2$, 
to one in which $m_s(\lat)=m_s(\phys)$ and $m_d(\lat)\approx0$.
A rough estimate is given by assuming the chiral log gives the entire
effect
\beq
{B_K(\epsilon=1) - B_K(\epsilon=0) \over B_K(\epsilon=0)} \approx 0.03-0.05 \ .
\eeq
The range comes from varying $f$ between $f_\pi$ and $f_K$, 
and varying the cut-off in the log from $m_\rho$ to $4\pi f_\pi$.
\item
To really calculate this ratio one needs to know $c$, 
one of the large number of higher order constants that
appear when one extends the chiral Lagrangian to include the 
electroweak effective Hamiltonian. One might have hoped that $c$ could
be measured from another weak decay, but this turns out not to be the case
\cite{wylerpc}.
\end{itemize}

What happens in the quenched approximation? The result becomes
\beq
B_K^{(0)} = B^{(0)} \left[
1 - (3+\epsilon^2) y\ln y + b^{(0)} y + c^{(0)} y\epsilon^2 +
\delta \left\{{2-\epsilon^2 \over 2 \epsilon} 
\ln\left({1 - \epsilon \over 1 + \epsilon}\right) + 2 \right\} \right] \ .
\label{eq:bkquenched}
\eeq
All except the last term is contained in Ref. \cite{Sharpelogs}.
The term proportional to $\delta$ is a new result \cite{Zhang}.
The most noteworthy feature of Eq. \ref{eq:bkquenched} is that, 
for degenerate quarks, the form is identical to that for QCD.
(The term proportional to $\delta$ vanishes when $\epsilon=0$.)
In particular, the
chiral logarithmic term has the same coefficient in both theories.
Because of this, I would argue that the QQCD result for
$B_K$ is likely to be more reliable than that for quantities for which
the chiral logs differ between QCD and QQCD, e.g. $f_\pi$ and $m_N/f_\pi$.
This does {\em not} necessarily mean that $B_K$ will be the same in QQCD and
QCD, because the constants $B^{(0)}$, $b^{(0)}$ and $c^{(0)}$ are 
different from their QCD counterparts. 
But, in some sort of average sense, quantities for
which the chiral logs are same in the two theories are likely to be more
similar. How it is that the chiral logs are the same for $B_K$ is explained in
Ref. \cite{Sharpelogs}.

The QQCD and QCD results differ, however, for non-degenerate quarks. 
Indeed, the term multiplying $\delta$ diverges in the limit
$\epsilon\to1$. This is similar to the divergence noted above in $m_\pi^2$.
Thus there is no sense in 
using slightly non-degenerate quarks in QQCD and trying to extrapolate to 
$\epsilon=1$: the $\delta$ contribution is a quenched artifact,
and the logarithmic term proportional to $\epsilon^2$ 
differs in the two theories.
Thus I think that the best approach is use QQCD to study $B_K$ for
degenerate quarks, and wait until we have QCD itself under control
before studying the physical kaon. For QCD an extrapolation from
small $\epsilon$ may be possible.

\subsection{Errors due to quenching}

The only way to really know the effect of quenching is to calculate
$B_K$ in QCD. An important step in this direction has been made by
two recent calculations which include the fermion determinant
\cite{BKtsukuba,BKkilcup}.
Both calculate $B_K(\epsilon=0)$ for theories with two degenerate quarks,
the masses of which range from $m_s$ down to $m_s/2$,
Although these theories are a fair distance from QCD---their ``pion'' clouds
are composed of particles with the masses of kaons---these results
do provide very useful information about $B_K$ in QCD.

These calculations are done on $16^3$ lattices at $a\approx 1$fm, 
roughly the smallest volume and lattice spacing that it is reasonable to use.
This illustrates one way in which QQCD will remain useful as QCD calculations
improve. We can use quenched calculations to study the dependence of
quantities on the volume, lattice spacing and quark masses, and map out 
the regions in parameter space where the systematic errors are small. 
We then do initial QCD calculations at the edge
of the allowable parameter space, and use the quenched results to
estimate systematic errors.
Later on we check these errors directly in QCD.

The results of the two simulations are simple and striking:
they agree with quenched results at the same lattice spacing, 
within $\sim 5\%$ errors.
This agreement is both for the slope of $B_K$ vs $m_K^2$
and for the intercept. 

What is the significance of this result?
First let me sound a note of caution. This is a result at only
one lattice spacing.
As shown in Fig. \ref{fig:BKbare}, $B_K$ has a substantial dependence
on the lattice spacing in QQCD.
The agreement at $a\approx 0.1$ fm could be fortuitous, 
with the lattice spacing dependence being
different in QCD and QQCD.

Assuming this is not so, let us analyze the result in terms of
the chiral expansions Eqs. \ref{eq:bkchpt} and \ref{eq:bkquenched}.
The result for two flavors, and degenerate quarks, has exactly
the same form as for QCD and QQCD \cite{Sharpelogs}.
Thus what the two simulations show is that
\beq
B^{(2)} \approx B^{(0)} \ \ \ {\rm and}\ \ \
b^{(2)} \approx b^{(0)} \,.
\eeq
Extrapolating linearly in the number of flavors, this suggests that
\beq
B \equiv B^{(3)} \approx B^{(0)} \ \ \ {\rm and}\ \ \
b \equiv b^{(3)} \approx b^{(0)} \,,
\eeq
in which case
\beq
B_K(\epsilon=0) \approx B_K^{(0)}(\epsilon=0) \,.
\eeq
This result is exactly what the hand-waving of the previous section
suggested, but it is perhaps surprising how small the difference
between QCD and QQCD is. Of course, it needs to be checked 
by adding a third quark.

What these simulations do not address is the value of $c$---the coefficient
which determines the dependence of $B_K$ on $m_s-m_d$.
As explained above, the chiral logs suggest that the extrapolation
to $\epsilon=1$ will raise $B_K$ by $\sim 5\%$.

\subsection{Symanzik's improvement program}

The final source of systematic error comes from working at
finite lattice spacing. 
If we calculate at a variety of different spacings, but always match
to the same continuum renormalization point, we expect
\beq
B_K(\NDR,\mu)_a = B_K(\NDR,\mu)_\CONT 
\left[1 + a \Lambda_1 + a^2 \Lambda_2^2 + O(a^3) \right] \ .
\label{eq:aextrap}
\eeq
The $\Lambda_n$ are non-perturbative
scales characterizing the discretization errors.
In fact, we will see that the first correction term is absent 
($\Lambda_1=0$) if the
calculation is done with staggered fermions in the manner described above.
This is in contrast to Wilson fermions, for which the $O(a)$ term is
present, and diverges like $1/m_K^2$ in the chiral limit.

Let me sketch how one arrives at Eq. \ref{eq:aextrap} using
perturbation theory \cite{ELCimprove}. One compares matrix elements of
continuum and lattice operators between quark states,
for some set of external momenta, 
but now keeping terms proportional to powers of $a$.
At tree level, powers of $a$ come only from the discretization of the 
operator. At higher orders, one also has to worry about corrections to
propagators and vertices.
For example, the staggered fermion propagator involves
\beq
\sin(k_\mu a) = a k_\mu(1 - a^2 k_\mu^2/6 + \dots) \ ,
\eeq
(no implicit sum on $\mu$).
Note that there are only even powers of $a$; the same is 
true for the vertices, but not, in general for the operators.
For Wilson fermions the corrections to propagators, vertices,
and operators are all, in general, of $O(a)$.

At $O(g^2)$, one can show, by power counting, that the $O(a)$ terms
can multiply logarithmically divergent integrals.
Thus one gets corrections of the form $p a g^2 \ln(am)$,
where $p$ is some combination of the external momenta or masses,
and $m$ a similar combination providing the infra-red cut-off to the
integral. Since $g^2\propto 1/\ln(a)$ at small $a$, this term is no
smaller than tree-level contributions. Similarly $n-$loop terms proportional
to $a g^{2n} \ln(am)^n$ are of the same size.
The conclusion is that the entire $O(a)$ term in Eq. \ref{eq:aextrap}
is of the form $a \Lambda_1$, where $\Lambda_1$ is a non-perturbative scale.
A similar result holds for the $a^2$ and higher order terms.

Symanzik has laid out a systematic program for removing discretization
errors \cite{Symanzik}. 
It has been proved to work for scalar theories, and there is no
reason to doubt that it works also for gauge theories.
I will provide only a thumbnail sketch. There are two parts.
\begin{enumerate}
\item
Add all possible dimension 5 terms to the action,
\beq
S_4 \longrightarrow S_4 + a \sum_i c_i(g^2) S_{5i} \ ,
\eeq
and adjust their coefficients so that some set of physical quantities
does not have $O(a)$ corrections.
This is to be done order by order in perturbation theory,
$c_i = \sum_{n=0}^{\infty} c_{in} g^{2n}$.
At tree-level, one adjusts $c_{i0}$ so as to cancel the $O(a)$ terms
in $S_4$. The strong result of Symanzik is that this removes 
terms of the form $a (g^{2}\ln a)^n$ for all $n$, at least for
physical quantities. 
What remains are terms of size $a g^2$, which are removed by
determining the $c_{i1}$, etc.
A very important constraint is that the
$S_{5i}$ have the same symmetries as $S_4$.
This is because one is trying to cancel lattice artifacts due to $S_4$, 
and these are of restricted form due to translation, rotations,
parity and any other symmetries.
\item
If one wants to ``improve'' the matrix elements of operators, one must
not only improve the action, but also improve the operators themselves.
For example, for the dim-6 operators appearing in $B_K$, the improvement
is made by 
\beq
\CO_6 \longrightarrow c(g^2) \CO_6 + \sum_j d_j(g^2) \CO_{7j} \ .
\eeq
Tree-level improvement removes $a (g^{2}\ln(a))^n$ terms, leaving
$a g^2$, etc. The operators that one has to add are all those
with the same symmetries as $\CO_6$. If this is a local operator it will
have fewer symmetries than $S_4$, and thus typically there will be more
possibilities for $\CO_{7j}$ than for $S_{5j}$.
\end{enumerate}

I now sketch the application of this program to staggered fermions
\cite{Sharpelat93,Sharpeimproved}.
As noted above,
the staggered fermion Lagrangian can be written in a compact form
\beq
S_4 =\sum_y
 \chibar(y) (\g\mu\otimes I)D_\mu\chi(y) + m \chibar(y) (I\otimes I)\chi(y) 
\,.
\eeq
The advantage of this notation is that the only bilinears which transform
as singlets under translations are those with $\xi_F=I$.
This and other symmetries restrict the possible dimension 5 operators to be
\beq
\chibar (I\otimes I) D_\mu D_\mu \chi \ ,\ \ \
\chibar (\sigma_{\mu\nu}\otimes I) F_{\mu\nu} \chi \ \ \ {\rm and}\ \ \
m^2 \chibar (I\otimes I) \chi \ .
\eeq
All these three terms are, however, forbidden by the $U(1)_A$ symmetry.
(To see this the mass must be treated as a spurion field which rotates
under the symmetry.)
There are thus {\em no dimension 5 operators available to improve the action}.
This in turn must mean that there are no $O(a)$ corrections to on-shell
quantities, because, if there were, then from the nature of the improvement
program, they could be removed by the addition of $S_5$.
This is true to all orders in perturbation theory, and I shall assume it
also holds non-perturbatively.

Now I turn to the operators needed for the lattice calculation of $B_K$.
Let $\CO^6_\LATT$ be a vector composed of 
all lattice operators which appear when
we match to the continuum operator we are interested in. As discussed above,
this is a long vector. Let $\CO^6_\CONT$ be the corresponding
vector of continuum operators.
The dimension 7 operators that we must subtract from
$\CO^6_\LATT$ will be all those with the same symmetries. The symmetry
group is much smaller now because these are quasi-local operators;
translations and rotations reduce to those transformations which
map the $2^4$ hypercube into itself---the hypercubic group.
The transformation properties of staggered fermion bilinears
under this group has been studied by Verstegen \cite{Verstegen}.
The net result is that one ends up with an extremely
long vector of dimension 7 operators, which I call $\CO^7_\LATT$. 
I now assume that Symanzik's results apply,
so that it is possible to find an improved lattice operator whose matrix
elements, in lattice states, agree with those of the continuum
operators in continuum states, up to corrections of $O(a^2)$
\beq
\vev{c(g^2)\CO^6_\LATT + d(g^2) a \CO^7_\LATT}
= \vev{\CO^6_\CONT} (1 + O(a^2)) \,.
\label{eq:dim6eq}
\eeq
Here $c(g^2),d(g^2)$ (square and rectangular matrices, respectively)
are perturbatively calculable coefficients,
obtained by removing $O(a)$ terms in a set of matrix elements. 
The claim is that this improvement then applies to all other matrix elements.

To demonstrate the desired result I need to relate $\CO^7_\LATT$ to its
continuum counterpart. But dimension 7 lattice operators mix 
with dimension 6 operators when one includes loops---the 
ultraviolet divergences give contributions
$\propto 1/a$ which cancel the $a$ which comes with the 
operator.\footnote{%
There would be mixing with yet lower dimension operators, 
if any were available, but to make a $\Delta S=2$ operator requires at least
four-fermion fields.}
Thus to obtain a continuum dimension 7 operator one needs to subtract
out some dimension 6 lattice operators. I assume the form
\beq
\vev{\tilde c(g^2) \CO^6_\LATT + \tilde d(g^2) a \CO^7_\LATT}
= 
\vev{a \CO^7_\CONT}(1 + O(a)) \ .
\label{eq:dim7eq}
\eeq
Notice that this is only good up to corrections of $O(a)$ and not $O(a^2)$n.
Combining Eqs. \ref{eq:dim6eq} and \ref{eq:dim7eq} one finds
\beq
\vev{\CO^6_\LATT} = {1 \over c - d \tilde d^{-1} \tilde c}
\vev{(\CO^6_\CONT - d \tilde d^{-1} a \CO^7_\CONT)} (1 + O(a^2)) \ .
\eeq
The product of matrices multiplying the matrix element on the r.h.s. is
just a perturbative matching matrix, part of which
has been calculated to one-loop, as described above.

The final part of the argument is messy but straightforward. One simply
enumerates the operators $\CO^7_\LATT$, using the results of Refs. 
\cite{Verstegen,SP}. An example is
\beq
\chibar_{S'} \sum_\mu (\g\mu\g5 \otimes \xi_5) D_\mu \chi_{D'}\ \ 
\chibar_S (I \otimes \sum_\nu \g\nu) \chi_D \,.
\eeq
The continuum version has the same form, but with
$\chi_s\to S$, etc. The important result is that in none of the
operators do both bilinears have flavor $\xi_5$. 
Thus, when we insert the operators between states of flavor $\xi_5$,
their matrix elements {\em vanish identically in the continuum},
where flavor is a good symmetry.
In other words
\beq
\vev{\CO^6_\LATT}_{\xi_5} = {1 \over c - d \tilde d^{-1} \tilde c}
\vev{(\CO^6_\CONT}_{\xi_5} (1 + O(a^2)) \ .
\eeq
Thus the corrections to {\em all} of the lattice dimension 6, $\Delta S=2$
operators in these matrix elements are of $O(a^2)$.
This is true whatever order in perturbation theory one does the matching.
Truncating the matching calculation at one-loop just gives rise to
errors of $O(g^4)$, and not of $O(a g^2)$.
This result greatly simplifies the extrapolation to the continuum limit,
since one knows not to include linear terms in $a$.
I do not know whether there are terms of $O(a^3)$, though I suspect
that they can be ruled out by similar considerations.

With Wilson fermions one actually has to do the calculations, rather than
just discuss formal manipulations. Fortunately, there are many fewer
operators because there are no additional flavors. The action, together with
various interesting operators, have been improved at tree level,
and the one-loop matching coefficients have been calculated.
This reduces the discretization errors to $O(g^2 a)$. The 
higher order terms in perturbation theory coming from tadpole
diagrams have also been calculated, which appears to remove the
bulk of the $O(a g^{2n})$ terms. 

\subsection{Status of results for $B_K$}

\begin{figure}[tb]
\vspace{-0.2truein}
\centerline{\psfig{file=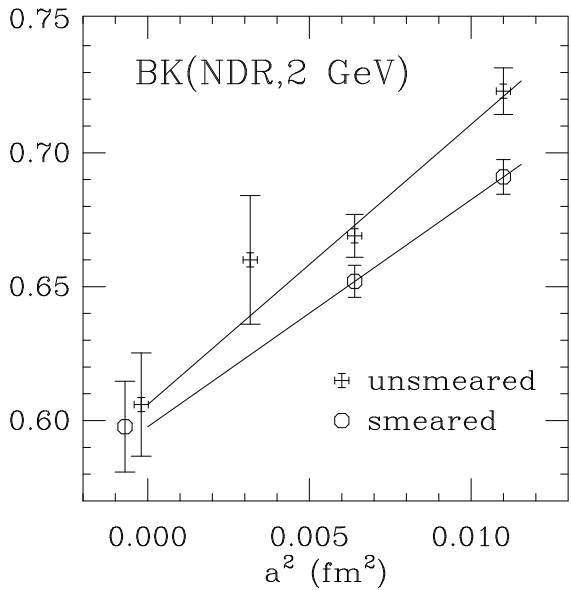,height=4truein}}
\vspace{-0.8truein}
\caption{Extrapolating quenched $B_K$ to the continuum limit.}
\label{fig:BKfinal}
\end{figure}

In Fig. \ref{fig:BKfinal} I show how our results extrapolate to the
continuum limit \cite{Sharpelat93}.
Based on the discussion of the previous subsection,
we assume quadratic extrapolation---the data itself
is not good enough to differentiate quadratic from linear dependence on $a$.
(The results at $a\sim 0.2$ fm are too far from the continuum limit to be
usefully included in the fit.)
We find that the scale characterizing discretization errors
is quite large, $\Lambda_2\sim 1$ GeV.
For this graph, we have fixed the lattice spacing using $m_\rho$,
and taken $g^2$ in the $\MSbar$ scheme at $\mu=\pi/a$.
The results for the two different types of operator disagree slightly at
the largest lattice spacing, $a\sim 0.1$ fm, but are consistent when
extrapolated. This is as expected: two different operators will, in general,
have different discretization errors. To quote a result we average
that for the two operators and use half the difference as an estimate
of the systematic error.

To estimate the other systematic error in our extrapolated result, we repeat
the calculation using $f_\pi$ to determine $a$,
and repeat it using matching coefficients evaluated with
the coupling scaled down to $\mu=2$ GeV.
There are also some small systematic errors, having to do with our numerical
method, which I will not discuss.
The final result is
\beqn
B_K(\NDR, 2{\rm GeV}) &=& 0.606 \pm 0.020\ ({\rm statistical}) \\
&&\mbox{} \pm 0.014 \ ({\rm choice\ of}\ g^2) \\
&&\mbox{} \pm 0.009 \ ({\rm choice\ of\ lattice\ spacing}) \\
&&\mbox{} \pm 0.004 \ ({\rm choice\ of\ operator})  \,.
\label{eq:BKfinal}
\eeqn
In the past I have combined the systematic errors in quadrature,
but this is not really justified.
What is important and striking is that the statistical errors are small,
and the systematic errors smaller still.
Let me also mention that the Japanese staggered group have also calculated
$B_K$ with $2^4$ Landau gauge and gauge-invariant operators, at
$\beta=6$ and $6.3$ and find results consistent with those shown above
\cite{BKtsukuba}. So the numerics appear under control.

This result is for {\em degenerate quarks in the quenched approximation}.
The error in $B_K$ due to these approximations is not known, but,
as discussed above, is likely to be small, conservatively $\sim 10\%$.

\subsection{Other matrix elements from QQCD}

Many other weak matrix elements are being calculated using lattice methods,
although the calculations are not as advanced as that for $B_K$.
I have no time to discuss these in any detail.
We have been attempting for years to calculate the matrix elements
leading to the CP-conserving and CP-violating kaon decay amplitudes
\beq
\vev{K|\CO_{1-8}|\pi\pi} \,.
\eeq
$\CO_i$ are the operators appearing in the effective electroweak Hamiltonian.
Of particular interest are $\CO_{6-8}$, which give the largest
contribution to $\epsilon'$. Since there are two particles in the final
state, these calculations are done indirectly.
ChPT is used to relate the matrix element to the simpler, and doable,
$\vev{K|\CO|\pi}$ and $\vev{K|\CO|0}$. 
To raise the level of these calculations to anything approaching that
of $B_K$ will require either considerably more more statistics 
and/or an improved method.

There are by now extensive results for various decay constants
($f_\pi$, $f_K$, \dots) and for the semileptonic
form factors $K\to \pi e\nu$, $D\to (\pi,K) e \nu$. These have been
calculated successfully using Wilson fermions, since chiral
symmetry does not give important constraints.
The statistical precision has reached the 10\% level, and
the perturbative corrections are understood. 
What remains to be done
is the extrapolation to the continuum limit, 
and for some estimate of the quenching errors to be made.


\section{Heavy mesons on the lattice}

B mesons will be the major focus of experimental studies of CP violation
in the next decade. Some measurable quantities are directly related to
elements of the CKM matrix, without hadronic uncertainties. 
Others are not, and here the lattice can contribute. Examples include
\begin{itemize}
\item
$f_B$ and $B_B$, which allow an extraction of $V_{td}$ and $V_{ts}$ from
measurements of $B_d-\overline{B}_d$ and $B_s-\overline{B}_s$ mixing.
\item
The form factors for $B\to (D,D^*) e \nu$ (which in the heavy quark limit
are related to the Isgur-Wise function). These will allow an extraction of
$V_{bc}$.
\item
The form factors for $B\to (\pi,\rho) e \nu$, which will allow an extraction
of $V_{bu}$.
\item
The matrix element for $B\to K^* \gamma$, allowing extraction of $V_{ts}$.
\item
The quark mass $m_b$. This is a fundamental parameter of QCD, and an input
into (or prediction of) models of physics beyond the standard model.
\end{itemize}
The hope is that not all the results for CKM elements will
agree, thus giving us an indication of new physics.

My discussion in this section will focus mainly on theoretical issues.
Numerical results have not yet settled down---for example there are still
significant discrepancies between results for $f_B$ in the limit that the
$b$ is static. For an excellent review of the status of numerical results
see Ref. \cite{shigemitsu}.

An important point to keep in mind is that the properties of hadrons containing
one heavy quark are determined by non-perturbative, long distance physics,
just as for light quark hadrons.
In other words, it is no easier to calculate $f_B$ than $f_\pi$.
Similarly, the problems introduced by quenching have the same character
as for light quark hadrons---$B$-mesons have PGB clouds, which are
absent in QQCD, and are replaced by an $\eta'$ cloud.
Quenched chiral perturbation theory for heavy-light mesons has been
developed by Booth \cite{Booth}, and by Zhang and I \cite{Zhang}. 
Typical estimates of the effect of quenching on decays constants are $10-20\%$.

There are, however, two simplifications. The first is that the constraints
of chiral symmetry, central for kaons,
are much less important for $B$-mesons. Thus Wilson fermions are
the discretization of choice for the light quark.
The complications associated with staggered fermions are 
probably not worth the effort, though not all agree
with this assessment\cite{NRQCDjapan}!

The second and most important simplification concerns the dynamics of
the $b$-quark. As discussed in the lectures of Grinstein, when $m_Q\to\infty$,
the heavy quark travels at constant (Minkowski) velocity, and acts as a
spinless color-triplet source. Furthermore, quark and anti-quark dynamics
are decoupled. These observations are formalized in
heavy quark effective theory (HQET), a methodology which allows
one to systematically expand quantities in powers of $1/m_Q$.
Standard examples of the results of HQET are
\begin{itemize}
\item 
$m_{B^*} = m_{B} + O(m_B^{-1}) $\ ,
\item
The large mass behavior of the decay constant of a heavy-light meson $P$ 
in which the $b$-quark is replaced by a fictitious quark with a 
different mass
\beq
\phi_P \equiv f_P \sqrt{m_P}
\left[{\alpha(m_P)\over\alpha(m_B)}\right]^{2/\beta_0}
=\phi_\infty \left[ 1 + {A\over m_P} + {B\over m_P^2} + \dots \right] \ .
\label{eq:phiP}
\eeq
\item
The four form-factors involved in $B\to (D,D^*) e\nu$ are all related
to a single function, the Isgur-Wise function $\xi(v\cdot v')$.
\end{itemize}
What the lattice can do is to calculate the unknown functions in
these expressions, and to check the importance of the non-leading corrections.

To calculate $\phi_\infty$, or $\xi(v\cdot v')$, or other quantities
which are present in the $m_Q\to\infty$ limit, one must discretize the HQET. 
There are some subleties associated with the continuation to Euclidean space,
but these appear to be understood \cite{mandulaogilvie}.
Things are simplest when the heavy quark is at rest. 
As mentioned in the discussion of the confining potential, 
the propagator for such a static quark is proportional to 
the product of time directed $U$ matrices
\beq
G(\vec n,\tau; \vec m, 0) = \delta_{\vec n, \vec m}\,
U_{(\vec m,0),4}\, U_{(\vec m,1),4}\, \dots U_{(\vec m,t-1),4} \ .
\eeq
When combined with the usual light antiquark propagator, this makes a
heavy meson propagator from which one can extract $\phi_\infty$.
The fact that the heavy quark does not move increases the 
statistical noise---a light quark propagator wanders all over the lattice
and thus averages out some of the fluctuations in the gauge fields.
To reduce the noise one has to work hard at producing good sources.

In order to estimate the size of the corrections to the heavy quark
limit, one must put ``dynamical'' heavy quarks 
(i.e. quarks moving at a variable velocity)
directly on the lattice. There is a potential obstacle
to doing so: at present lattice spacings the $b$-quark Compton wavelength 
exceeds the lattice spacing, $m_b a \sim 2-3$. 
$b$-quark propagation thus includes large lattice artifacts.
Two approaches have been adopted to remove these artifacts.

\begin{figure}[tb]
\centerline{\psfig{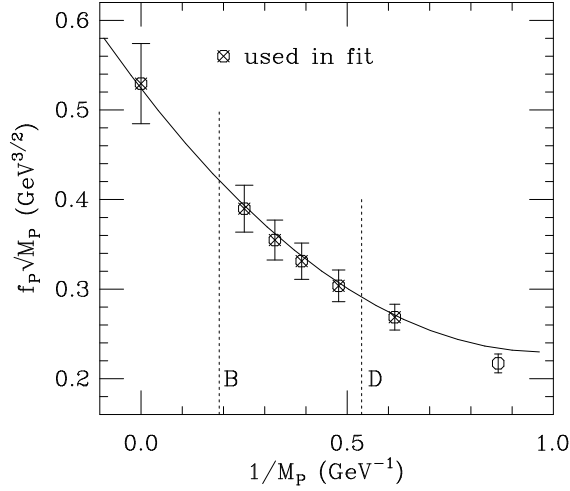}}
\caption{Static and Wilson fermion results for $\phi_P$.}
\label{fig:FB}
\end{figure}

The first is to work at small enough quark masses that lattice
artifacts are small, and then extrapolate up to the $b$ quark.
Crudely speaking this requires $m_q a < 1$, implying $m_q < 3.5$GeV
for present lattices. Using an improved action probably increases the
maximum mass that can be used. This method has been used for
$f_B$, $B\to K^* \gamma$ and the Isgur-Wise function.
I show in Fig. \ref{fig:FB} results for $\phi_P$ from Ref. \cite{BLS}.
The point at $1/M_P=0$ comes from a calculation with a static heavy
quark, while the other points come from dynamical (Wilson) heavy quarks.
These are results from $\beta=6.3$, corresponding to $1/a\approx 3.2\,$GeV.
The fit to the form of Eq. \ref{eq:phiP} is good, and one can read off 
$f_B\sqrt{m_B}$ from the intercept with the vertical line marked ``$B$''.

This is not quite the whole story.
The two points with the smallest non-zero values of $1/M_P$,
have $m_Q a\approx M_P a > 1$, and are likely to be
afflicted by substantial lattice artifacts. 
Indeed, an approximate way of removing these artifacts
has been used to correct all except the static point, 
and the corrections are substantial (as large as $\sim 50\%$) for the points
at small $1/M_P$. I explain the source of this correction below.
The bulk of it is reasonable,
but there remains a systematic uncertainty in these large mass points.
Fortunately, this uncertainty does not have much impact on $f_B$,
because the curve is pinned down from both sides of the $B$ mass.
Even if were to discard the two points with the smallest values of $1/M_P$, 
the result for $f_B$ would be similar.
The present status has been summarized by Soni \cite{soniglasgow},
who quotes $f_B=173\pm 40$\,MeV 
(in the normalization where $f_\pi=135\,$MeV).

The alternative approach is to use yet another effective theory to
describe the $b$-quark, namely non-relativistic QCD (NRQCD) \cite{NRQCD}.
If $m_Q >> \Lambda_{\rm QCD}$, then in its couplings to low momentum
gluons, the quark is non-relativistic.
The couplings to high-momentum gluons must be treated relativistically, 
but these can be accounted for using perturbation theory.
The NRQCD Lagrangian is
\beqn
\nonumber
{\cal L}_{\rm NRQCD} &=& \psi^{\dagger} D_{\tau} \psi 
	+ {1\over 2 m_Q} \psi^{\dagger} \vec D^2 \psi \\
\label{eq:NRQCD}
&&\mbox{} + c_1(g^2) {g\over 2m_Q} \psi^{\dagger} \vec\sigma \cdot \vec B \psi
+ c_2(g^2) {g \over 8 m_Q^2} \psi^{\dagger} 
\vec\sigma\cdot(\vec D \times \vec E - \vec E \times \vec D) \psi \\
\nonumber
&&\mbox{} + c_3(g^2){1\over 8 m_Q^3} \psi^{\dagger} (\vec D^2)^2 \psi
- c_4(g^2){i g\over 8 m_Q^2} \psi^{\dagger}
(\vec D \cdot \vec E - \vec E \cdot \vec D) \psi + \dots \,,
\eeqn
where $\psi$ is a two-component field. The coefficients $c_i$ are obtained by
perturbative matching with QCD. At leading order they are all unity.

The crucial point is that, when one discretizes NRQCD, 
the errors are no longer determined by $m_Q a$, 
but rather by $\vec p a$, where $p\sim \Lambda_{\rm QCD}$
is a typical momentum.
Thus, in heavy-light systems, discretization errors for heavy
and light quarks are comparable.
In the the $\bar b b$ system the typical momentum is somewhat
higher, $p\sim M v\approx 0.1 M\approx 0.5$GeV, but still satisfies
$p a < 1$ for present lattices.

NRQCD teaches us something about heavy Wilson quarks.
The point is that, as long as $m_Q >> \Lambda_{\rm QCD}$, the quarks are
non-relativisitic, and must be describable by a non-relativistic 
effective Lagrangian. This will have the same form as Eq. \ref{eq:NRQCD}
(since this contains all terms), but
the effect of discretization errors will be to change the coefficients
$c_i$ substantially from those in NRQCD\cite{Kronfeldlat92}.
The coefficients can be determined using perturbation theory.
By suitably changing the normalization of the fields
one can get the leading term in ${\cal L}_{\rm NRQCD}$ with
the correct normalization, while by changing the 
definition of $m_Q$ one can obtain the second term up to perturbative
corrections. After these corrections, results 
for heavy Wilson quarks should only have errors of 
$O(\Lambda_{\rm QCD}/m_Q)$ (from the incorrectly normalized
third term in ${\cal L}_{\rm NRQCD}$), and not of $O(m_Q a)$.
These are the corrections that have been applied to 
points in Fig. \ref{fig:FB}.

Nevertheless, the action is not that of NRQCD.
Kronfeld and Mackenzie have suggested studying
heavy quarks using a modified Wilson action \cite{kronmack}. 
Their scheme interpolates between the standard Wilson fermion action 
for light quarks and NRQCD for heavy quarks. For my purposes here, however,
there is no important distinction between this and the NRQCD approach,
and I will focus on the latter, for which more results are available.

NRQCD is a non-renormalizable theory, because ${\cal L}_{\rm NRQCD}$
contains operators of dim-5 and higher.
It must be formulated with a UV cut-off, 
which is here provided by the lattice, $\Lambda \approx 1/a$.
As mentioned above in the discussion of $B_K$,
loop diagrams mix the higher dimension operators with those of lower
dimension. Thus, for example, there are contributions to wave-function
renormalization proportional to $F(g^2) \Lambda/m_Q=F(g^2)/(m_Q a)$. 
These are calculable order by order in perturbation theory.
Since in practice one can only work to finite order, it is clear that
one cannot take $a$ too small, for then the uncalculated corrections
will become large. 
This is not a practical problem at present lattice spacings.

This does bring up an important issue that, in my view,
remains to be fully resolved.
Even if one could calculate $F(g^2)$ to all orders in perturbation theory, 
there might be non-perturbative terms proportional to 
$\exp[-n/(2 \beta_0 g^2)]\propto (\Lambda_{\rm QCD} a)^n$
($\beta_0$ is the first coefficient in the QCD $\beta$-function).
If there are such terms with $n=1$, they give non-perturbative
contributions to wave-function renormalization $\propto \Lambda_{\rm QCD}/M$,
which are not calculable. This in turn means that there is an
ambiguity in the $A/M_P$ contribution to $\phi_P$ (Eq. \ref{eq:phiP})
of size $\propto \Lambda_{\rm QCD}/M_P$. But $A\sim \Lambda_{\rm QCD}$,
so the ambiguity is of the same size as the quantity we wish to extract.
This argument was first given in Ref. \cite{nonpertterms}.

There is no dispute over whether such terms can
exist (examples are infra-red renormalon ambiguities),
but what is controversial is their size \cite{Lepagelat91}.
These are non-perturbative effects at short-distances, and the lore
is that these are small. This is true of the contributions of short
distance instantons, for example. 
How can this dispute be resolved? One can think of these terms as
non-perturbative contributions to the matching between QCD and NRQCD.
Thus one should compare the results of perturbative matching to those
of non-perturbative matching---the latter requiring either physical
matrix elements, or quark and gluon states in a particular gauge.
Such a program has been initiated in Ref. \cite{nonpertsub}.
An alternative comparison is obtained by calculating physical quantities
in different ways. If the answers agree, the non-perturbative ambiguities
are likely small. The example of $m_b$ is discussed below.
I think it is important to study this issue further.

The aim of the NRQCD program is to directly simulate $b$-quarks.
The first step in this program is to make sure that one obtains a
good description of the $\bar bb$ system. The ordering of the terms
in Eq. \ref{eq:NRQCD} is according to decreasing importance in
the $\bar bb$ system \cite{NRQCD}.
Terms in the first line
are of $O(m_Q v^2)$, where $v$ is the velocity of the heavy quark,
and give the spin-averaged splittings (e.g. the 1P-1S splitting discussed
a long way above). The second and third lines contain terms of $O(m_Q v^4)$,
the former being the leading contribution to the hyperfine splittings,
the latter correcting the spin-averaged splittings.
Since $v^2\sim 0.1$, the first line gives fine structure to 10\%, 
the next line hyperfine structure to 10\%, and the third improves the
accuracy of fine structure to 1\%.

\begin{figure}[tb]
\begin{center}
\setlength{\unitlength}{.02in}
\begin{picture}(130,120)(10,930)
\put(15,935){\line(0,1){120}}
\put(140,935){\line(0,1){120}}
\multiput(13,950)(0,50){3}{\line(1,0){4}}
\multiput(14,950)(0,10){10}{\line(1,0){2}}
\put(12,950){\makebox(0,0)[r]{9.5}}
\put(12,1000){\makebox(0,0)[r]{10.0}}
\put(12,1050){\makebox(0,0)[r]{10.5}}
\put(12,1057){\makebox(0,0)[r]{GeV}}
\put(15,935){\line(1,0){125}}
\put(15,1055){\line(1,0){125}}



\put(27,930){\makebox(0,0)[t]{${^1S}_0$}}
\put(25,943){\circle*{3}}
\put(30,942){\circle{3}}

\put(52,930){\makebox(0,0)[t]{${^3S}_1$}}
\multiput(43,946)(3,0){7}{\line(1,0){2}}
\put(50,946){\circle*{3}}
\put(55,946){\circle{3}}
\put(60,946){\makebox(0,0){$\Box$}}

\multiput(43,1002)(3,0){7}{\line(1,0){2}}
\put(50,1004){\circle*{3}}
\put(50,1004){\line(0,1){3}}
\put(50,1004){\line(0,-1){3}}
\put(55,1003){\circle{3}}
\put(55,1004){\line(0,1){2}}
\put(55,1002){\line(0,-1){2}}
\put(60,1009){\makebox(0,0){$\Box$}}
\put(60,1011){\line(0,1){7}}
\put(60,1008){\line(0,-1){7}}

\multiput(43,1036)(3,0){7}{\line(1,0){2}}
\put(50,1034){\circle*{3}}
\put(50,1034){\line(0,1){12}}
\put(50,1034){\line(0,-1){12}}
\put(55,1029){\circle{3}}
\put(55,1030){\line(0,1){13}}
\put(55,1028){\line(0,-1){13}}

\put(92,930){\makebox(0,0)[t]{${^1P}_1$}}

\multiput(83,990)(3,0){7}{\line(1,0){2}}
\put(90,987){\circle*{3}}
\put(90,987){\line(0,1){2}}
\put(90,987){\line(0,-1){2}}
\put(95,990){\circle{3}}
\put(100,987){\makebox(0,0){$\Box$}}
\put(100,989){\line(0,1){3}}
\put(100,986){\line(0,-1){3}}

\multiput(83,1026)(3,0){7}{\line(1,0){2}}
\put(90,1032){\circle*{3}}
\put(90,1032){\line(0,1){7}}
\put(90,1032){\line(0,-1){7}}
\put(95,1025){\circle{3}}
\put(95,1026){\line(0,1){4}}
\put(95,1024){\line(0,-1){4}}
\put(100,1024){\makebox(0,0){$\Box$}}
\put(100,1026){\line(0,1){4}}
\put(100,1023){\line(0,-1){4}}

\put(120,930){\makebox(0,0)[t]{${^1D}_2$}}
\put(120,1020){\circle*{3}}
\put(120,1020){\line(0,1){7}}
\put(120,1020){\line(0,-1){7}}
\put(123,1031){\makebox(0,0){$\Box$}}
\put(123,1033){\line(0,1){6}}
\put(123,1030){\line(0,-1){6}}

\end{picture}
\end{center}
\caption{Spin-averaged $\Upsilon$ spectrum from 
(a) NRQCD ($n_f=0$): filled circles;
(b) NRQCD ($n_f=2$): open circles; and
(c) UK-NRQCD ($n_f=0$): boxes.
Experimental results are the dashed horizontal lines.}
\label{fig:bbspect}
\end{figure}
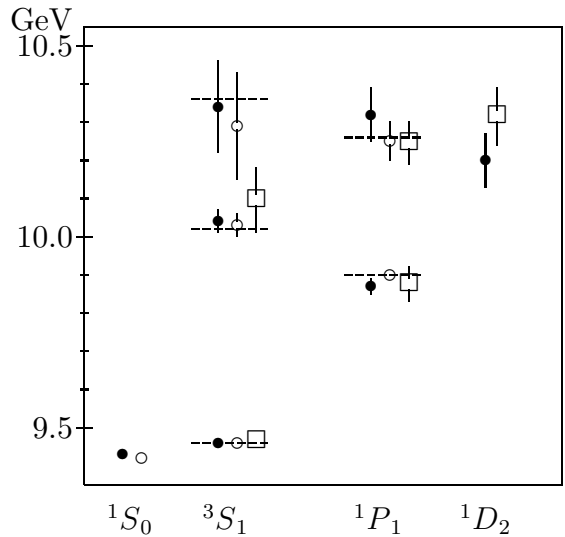

All these terms have now been included in the simulations
\cite{NRQCDalp,NRQCDmb}.
Propagators can be calculated in a single pass, because the time
derivative can be discretized as a forward difference.
Thus the calculations are much faster than for light quark propagators.
The status of the calculation of the spin-averaged spectrum is shown in Fig.
\ref{fig:bbspect}. NRQCD refers to Ref. \cite{NRQCDmb},
UK-NRQCD to Ref. \cite{UKNRQCD}.
Notice that simulations with two 
(moderately light) flavors of dynamical fermions have now been done.
The spectrum is little changed with $n_f=2$ from that in QQCD; both are in
good agreement with the experimental data.
The lattice spacing has been adjusted to fit the $1S-1P$ and $2S-2P$
splittings (from which $\alpha_s$ can be determined as discussed above).
Reasonable agreement for spin splittings is also obtained.

What about $m_b$? Spin-averaged splittings are quite insensitive to $m_b$;
a reasonable value has been used in the simulations.
An accurate determination of $m_b$ can be made in two ways.
First, one can use $m_{\Upsilon}a = 2 m_b a + E_{\rm NR}a-2 E_0 a$,
i.e. adjust $m_b$ until twice its mass plus the binding energy
(determined from the simulation) equals the physical $\Upsilon$ mass.
The measured binding energy must be corrected for the self energy
of the quark ($E_0 a$). This correction
can be calculated in perturbation theory,
$E_0 a = c_1 g^2 + O(g^4)$. It is an example of a quantity which
might contain the previously discussed uncalculable non-perturbative terms
$\propto \Lambda_{\rm QCD} a$. These would lead to an uncertainty
in the extracted value of $m_b$ of size $\Lambda_{\rm QCD}$.

The second method is to measure the kinetic energy of the $\bar b b$ states,
 e.g.
\beq
a E_\Upsilon(\vec p) = a E_{\Upsilon,\rm NR} + 
{(a\vec{p})^2 \over 2 a M_{\Upsilon,kin}} 
+ \dots \,.
\eeq
One adjusts the bare lattice mass $m_b^0 a$ until
$a M_{\Upsilon,kin}$ agrees with the physical mass in lattice
units, $a M_\Upsilon$. One then uses perturbation theory to match
the lattice mass to the pole mass, $m_b = (1 + c' g^2 + O(g^4) ) m_b^0$.
In this method an unknown non-perturbative term is truly
a small correction.

Both methods lead to consistent results, with errors of $\sim 200 MeV$.
Thus a difference between the two methods of approximately $\Lambda_{\rm QCD}$,
which is what would be expected from non-perturbative terms,
is not ruled out. More accurate data is needed to
resolve the above mentioned dispute.
Combining the results from the two methods, the final quoted values 
for the pole mass are \cite{NRQCDmb}
\beqn
m_b(n_f=0) &=& 4.94 \pm 0.15 {\rm GeV}\\
m_b(n_f=2) &=& 5.0 \pm 0.2 {\rm GeV}\,.
\eeqn

The next stage is to apply these methods directly to B-mesons.
This is just beginning.
It is my hope that the issue of non-perturbative ambiguities can be
resolved, and the lattice can, in due course,
give results for a number of transition
amplitudes including $1/M_B$ corrections.

\section{A final flourish}

I hope to have convinced you that lattice calculations are now 
reliable enough to make significant contributions to phenomenology. 
As time progresses, the importance of these calculations will increase.
To give an idea of what might happen I have played the following game.
The lattice results for $B_K$ and $f_B\sqrt{B_B}$ constrain,
respectively, $Im(V_{td}^2)$ and $|V_{td}|^2$ 
(ignoring small charm quark contributions).
Recalling that $V_{td}=A\lambda^3(1-\rho-i\eta)$, we can convert
these into constraints on $\rho$ and $\eta$.
I want to imagine how these constraints might look, say, five years hence.

\begin{figure}[tb]
\vspace{-0.2truein}
\centerline{\psfig{file=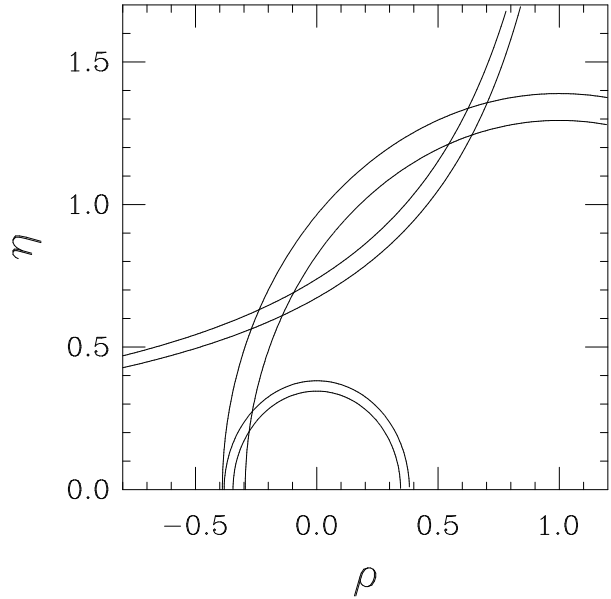,height=4truein}}
\vspace{-0.4truein}
\caption{Possible future constraints on $\rho$ and $\eta$.}
\label{fig:rhoeta}
\end{figure}

It is my guess that by that time we will have made enough progress
with simulating QCD that the error in $B_K$ will be roughly the
same as the present error in the quenched result, Eq. \ref{eq:BKfinal}.
For purposes of illustration, I will also take the same central value.
I think that the errors in $f_B$ and $B_B$ will be reduced substantially,
and I assume $f_B\sqrt{B_B}=200\pm 10\,$MeV,
instead of the present $200\pm40\,$MeV.
Let me further assume that the experimental errors in $V_{cb}$ and
$V_{ub}$ will drop significantly.
I take $|V_{cb}|=0.038$, $|V_{ub}/V_{cb}|=0.080\pm 0.004$,
and $x_d=0.67\pm0.02$ (this is the measure of $B-\bar B$ mixing).
Finally, I assume $m_t=172\,$GeV.
The resulting constraints on $\rho$ and $\eta$ are shown in
Fig. \ref{fig:rhoeta}. The parameters are supposed to lie between
the three pairs of curves 
(the hyperbolae are from $K-\bar K$ mixing,
the large circles from $B-\bar B$ mixing,
and the small circles from $|V_{ub}/V_{cb}|$).
The reduction in errors has moved the members of each pair much closer
than at present, and there is no solution for $\rho$ and $\eta$!
It will be most interesting to see how things develop in reality.

\section{Acknowledgements}
Many thanks to Rajan Gupta for comments and for providing plots,
and to Maarten Golterman for helpful discussions.

\section{References}

%
\def\PRL#1#2#3{{\it Phys. Rev. Lett.} {\bf #1} (#2) #3}
\def\PRD#1#2#3{{\it Phys. Rev.} {\bf D#1} (#2) #3}
\def\PLB#1#2#3{{\it Phys. Lett.} {\bf #1B} (#2) #3}
\def\NPB#1#2#3{{\it Nucl. Phys.} {\bf B#1} (#2) #3}
\def\NPBPS#1#2#3{{\it Nucl. Phys.} {\bf B ({Proc. Suppl.}) #1} (#2) #3}

\def\brookhaven{Proc. {``Lattice Gauge Theory '86''}, 
             Brookhaven, USA, Sept. 1986, NATO Series B: Physics Vol. 159}
\def\china{Proc. CCAST Symposium {``Lattice Gauge Theory Using 
    Parallel Processors''}, Beijing, China, June 1987, Gordon and Breach 1987}
\def\seillac#1{Proc. {``Field theory on the Lattice'}, 
             Seillac, France, Sept. 1987, \NPBPS{4}{1988}{#1}}
\def\fermilab#1{Proc. {``LATTICE 88''}, 
             Fermilab, USA, Sept. 1988, \NPBPS{9}{1989}{#1}}
\def\capri#1{Proc. {``LATTICE 89''}, 
             Capri, Italy, Sept. 1989, \NPBPS{17}{1990}{#1}}
\def\ringberg#1{Proceedings of the Ringberg Workshop,
``Hadronic Matrix Elements and Weak Decays'', Ringberg, Germany,
(4/88), \NPBPS{7A}{1989}{#1}}
\def\talla#1{Proc. {``LATTICE 90''}, 
             Tallahassee, USA, Oct. 1990, \NPBPS{20}{1991}{#1}}
\def\tsukuba#1{Proc. {``LATTICE 91''}, 
             Tsukuba, Japan, Nov. 1991, \NPBPS{26}{1992}{#1}}
\def\amsterdam#1{Proc. {``LATTICE 92''}, 
             Amsterdam, Netherlands, Sep. 1992, \NPBPS{30}{1993}{#1}}
\def\dallas#1{Proc. {``LATTICE 93''}, 
             Dallas, USA, Oct. 1993, \NPBPS{34}{1994}{#1}}

\end{document}